\documentclass[aps,pra,reprint,floatfix,superscriptaddress, nofootinbib,longbibliography,onecolumn,notitlepage,12pt,tightenlines]{revtex4-1}
\pdfoutput=1
\usepackage[ascii]{inputenc}
\usepackage[bookmarksnumbered,hypertexnames=false]{hyperref}

\usepackage{bbm,braket,microtype,mathrsfs,amsmath,amssymb,xcolor,amsthm,mathtools,graphicx,enumitem,booktabs}
\usepackage[capitalize]{cleveref}
\crefname{section}{Sec.}{Secs.}
\usepackage[T1]{fontenc}
\usepackage[USenglish]{babel}
\usepackage{verbatim}
\usepackage{placeins}
\usepackage{subfigure}
\crefname{thm}{theorem}{theorems}
\crefname{lem}{lemma}{lemmas}
\crefname{cor}{corollary}{corollaries}
\crefname{rem}{Remark}{Remarks}
\crefname{dfn}{definition}{definitions}
\DeclarePairedDelimiter{\abs}{\lvert}{\rvert}

\DeclareMathOperator{\tr}{tr}
\DeclareMathOperator{\Tr}{Tr}
\DeclareMathOperator{\Moeb}{Moeb}
\DeclareMathOperator{\Wg}{Wg}
\DeclareMathOperator{\U}{U}
\DeclareMathOperator{\NB}{NB}
\DeclareMathOperator{\dist}{dist}

\renewcommand{\vec}{\mathbf}
\renewcommand{\Re}{\mathfrak{Re}}
\renewcommand{\Im}{\mathfrak{Im}}
\newcommand{\vu}{\vec u}
\newcommand{\vv}{\vec v}

\newcommand{\CC}{\mathbb C}
\newcommand{\ZZ}{\mathbb Z}
\newcommand{\RR}{\mathbb R}
\newcommand{\EE}{\mathbb E}

\newcommand{\NN}{\mathbb N}
\newcommand{\ot}{\otimes}

\newcommand{\Hil}{\mathcal H}

\newcommand{\Norm}{\mathrm{N}}
\newcommand{\hf}{\frac12}
\newcommand{\ssection}[1]{\smallskip\phantomsection\addcontentsline{toc}{section}{#1}\textit{#1.---}}

\newcommand{\PFour}[4]{\begin{psmallmatrix}1&2&3&4\\#1&#2&#3&#4\end{psmallmatrix}}
\allowdisplaybreaks[4]
\newcommand{\SNErase}[1]{}

 \def\cj{\mathcal{J}}
 \def\tfd{\mathrm{TFD}}
 \def\hf{ \frac{1}{2} }
 \def\la{\label}
 \newcommand{\lp}{\left (}
 \newcommand{\rp}{\right )}
 \newcommand{\lb}{\left [}
 \newcommand{\rb}{\right ]}
 \newcommand{\ev}[1]{\left \langle #1 \right \rangle}
 \def\nref#1{(\ref{#1})}
 \newcommand{\eqn}[1]{\begin{equation}\begin{split} #1 \end{split}\end{equation}}
 \newcommand{\be}{\begin{equation}}
 \newcommand{\bea}{\begin{eqnarray}}
 \newcommand{\eea}{\end{eqnarray}}
 \newcommand{\beq}{\begin{equation}}
 \newcommand{\ee}{\end{equation}}
 \newcommand{\eeq}{\end{equation}}

\makeatletter
\def\l@subsubsection#1#2{}
\makeatother

\begin{document}
\title{Quantum Gravity in the Lab: \texorpdfstring{\\}{} Teleportation by Size and Traversable Wormholes, Part II}
\author{Sepehr Nezami}
\affiliation{Institute for Quantum Information and Matter, Caltech, Pasadena, CA 91125, USA}
\affiliation{Google, Mountain View, CA 94043, USA} 
\affiliation{Department of Physics, Stanford University, Stanford, CA 94305, USA}
\author{Henry W. Lin}
\affiliation{Google, Mountain View, CA 94043, USA}
\affiliation{Physics Department, Princeton University, Princeton, NJ 08540, USA}
\author{Adam R. Brown}
\affiliation{Google, Mountain View, CA 94043, USA}
\affiliation{Department of Physics, Stanford University, Stanford, CA 94305, USA}
\affiliation{Sandbox@Alphabet, Mountain View, CA 94043, USA}
\author{Hrant Gharibyan}
\affiliation{Institute for Quantum Information and Matter, Caltech, Pasadena, CA 91125, USA}
\author{Stefan Leichenauer}
\affiliation{Google, Mountain View, CA 94043, USA}
\affiliation{Sandbox@Alphabet, Mountain View, CA 94043, USA}
\author{Grant Salton}
\affiliation{Amazon Quantum Solutions Lab, Seattle, WA 98170, USA}
\affiliation{AWS Center for Quantum Computing, Pasadena, CA 91125, USA}
\affiliation{Institute for Quantum Information and Matter, Caltech, Pasadena, CA 91125, USA}
\affiliation{Department of Physics, Stanford University, Stanford, CA 94305, USA}
\author{Leonard Susskind}
\affiliation{Department of Physics, Stanford University, Stanford, CA 94305, USA}
\affiliation{Google, Mountain View, CA 94043, USA}
\affiliation{Sandbox@Alphabet, Mountain View, CA 94043, USA}
\author{Brian Swingle}
\affiliation{Brandeis University, Waltham, MA 02453, USA,  University of Maryland, College Park, MD 20742, USA}
\author{Michael Walter}
\affiliation{Korteweg-de Vries Institute for Mathematics, Institute for Theoretical Physics, Institute for Logic, Language and Computation \& QuSoft, University of Amsterdam, The Netherlands}

\vspace*{5mm}

\begin{abstract}
\noindent 
In \cite{firstpaper} we discussed how quantum gravity may be simulated using quantum devices and gave a specific proposal---teleportation by size and the phenomenon of size-winding. Here we elaborate on what it means to do `Quantum Gravity in the Lab' and how size-winding connects to bulk gravitational physics and traversable wormholes. Perfect size-winding is a remarkable, fine-grained property of the size wavefunction of an operator; we show from a bulk calculation that this property must hold for quantum systems with a nearly-AdS$_2$ bulk.
We then examine in detail teleportation by size in three systems: the Sachdev-Ye-Kitaev model, random matrices, and spin chains, and discuss prospects for realizing these phenomena in near-term quantum devices. 
\end{abstract}
\maketitle
\clearpage
\tableofcontents

\clearpage
\section{Introduction}

\subsection{Quantum gravity in the lab}

The holographic principle~\cite{tHooft:1993dmi,Susskind:1994vu, Maldacena:1997re} is an important guidepost in our quest to understand quantum gravity. Its best-understood incarnation is the AdS/CFT correspondence, where holography means that there is an exact equivalence or duality between two descriptions of the same system: theories of quantum gravity on the one hand and lower-dimensional theories that feature quantum mechanics but not gravity on the other. Since some phenomena are complicated from one viewpoint but simple from the other, this provides a great opportunity for intellectual arbitrage. This opportunity has not gone unexploited, with an enormous amount of work done in the context of the AdS/CFT, for example. 

In a previous paper~\cite{firstpaper} we explored a subject that we called `quantum gravity in the lab'. We defined this as being when the non-gravitational side of the duality is physically realized as a strongly coupled quantum system in a low-energy physics lab; the emergent holographic gravity dual is then the `quantum gravity in the lab'. In~\cite{firstpaper} we went on to explore a particular example of quantum gravity in the lab in the context of size-winding and holographic teleportation \cite{gao2017traversable,maldacena2017diving}. The bulk of this paper will be given over to a more in-depth examination of these `teleportation by size' examples. 

We emphasize that `quantum gravity in the lab', as we have defined it, must involve holography. There are therefore a number of explorations that, despite involving all three of quantum mechanics, gravity, and labs, nevertheless fall beyond our scope. For example, any experiment that has anything to do with the gravitational field that stops the lab equipment from floating off into space \cite{goldman1986experimental,page1981indirect} is beyond our definition of `quantum gravity in the lab'. While understanding quantum effects for the `naturally occurring' gravitational field is a key long-term goal, `quantum gravity in the lab' is not directly interested in any effect that involves the $G$ that Newton discovered. In the language of AdS/CFT, the quantum hardware lives in the `boundary', and we are not interested in gravity in the boundary. We are interested in the emergent gravity in the bulk. 

\subsection{Quantum gravity in the lab and the NISQ era}

The `quantum gravity in the lab' program does not need to wait for large error-corrected quantum computers. Progress can be made even in the Noisy Intermediate-Scale Quantum (NISQ) era. 

For most of the long-term goals of the quantum computing community, e.g.\ simulating quantum chemistry, we think that the NISQ era offers the start of a road that eventually leads to useful applications and new science. These payoffs may not be realized in the NISQ era, but there are nevertheless many good reasons to start walking the relevant road. In some ways, quantum gravity in the lab is similar to other long-term programs that hope to eventually shed light on profound open problems in science.

However, the near-future prospects for quantum gravity in the lab could be better than this. In the NISQ era, it is necessary to foster the growth of quantum technologies by proposing scientific and industrial applications that require a small number of qubits, and a lesser degree of fine-tuning or error correction. We argue that this is also an area where the ideas of quantum gravity in the lab come into play, as recent developments in the theory of quantum gravity have shown that many random-looking and not fine-tuned systems possess interesting gravitational duals. Examples of such systems are the Sachdev-Ye-Kitaev (SYK) model, which has a Hamiltonian with random couplings, and some random matrix ensembles that are dual to low-dimensional gravity models. In particular, such systems are generically better understood when they have a large number of degrees of freedom, so this can make intermediate-scale experiments more interesting, as they can directly probe corrections that are otherwise hard to calculate. Moreover, the NISQ era offers at least one certainty: that we can learn a great deal about the strong coupling dynamics of various toy quantum models. The issues at play in that context, including questions of chaos and thermalization, are all closely connected to deep issues in quantum gravity.

\subsection{Overview of results of this paper}

The remainder of this paper is devoted to a detailed study of one example of quantum gravity in the lab, building on our first paper~\cite{firstpaper}, in which we proposed a new mechanism for teleporting quantum information (``teleportation by size'') inspired by traversable wormholes in gravity and introduced the concept of size-winding. In this paper, we present a comprehensive set of examples in which teleportation by size and size-winding can be implemented and studied in detail, such as the SYK model, a random matrix model, and a chaotic spin chain. We use these examples to illustrate general features of the teleportation by size protocol, and to give a clear illustration of the size-winding phenomenon at low temperatures.

In~\cref{sec:survey}, we review the basic definitions and concepts introduced in our first paper. In particular, Fig.~\ref{fig:Wormhole_Circuit} shows the teleportation by size protocol. Readers familiar with the ideas from \cite{firstpaper} should feel free to skip this section. 

In~\cref{subsec:examples}, we study in detail several example systems. In~\cref{sec:subsecGGH}, we discuss teleportation by size at infinite temperature and then exhibit a weak version of the size-winding phenomenon at low temperatures that enhances the teleportation capacity. In~\cref{sec:spchm}, we present analytical results on state transfer in chaotic spin chains and again establish a link between size distribution and teleportation. In~\cref{sec:brownian}, we study Brownian evolution where the Hamiltonian changes in time and analytically study the growth of size operators. In~\cref{sec:exsyk}, we present explicit results for teleportation by size and size-winding in the SYK model at finite temperature. We further discuss the large-$q$ limit of the model and the potential connection to stringy effects in the bulk. 

In~\cref{sec:holography}, we discuss in detail the holographic interpretation of our protocol, especially in the context of the SYK model. This material is a non-trivial extension of the discussion in our first paper. We specifically revisit and clarify the role of size-momentum correspondence in the size-winding phenomena and elaborate on geometric interpretations of different teleportation regimes. 

\begin{figure}
\includegraphics[width=8.6 cm]{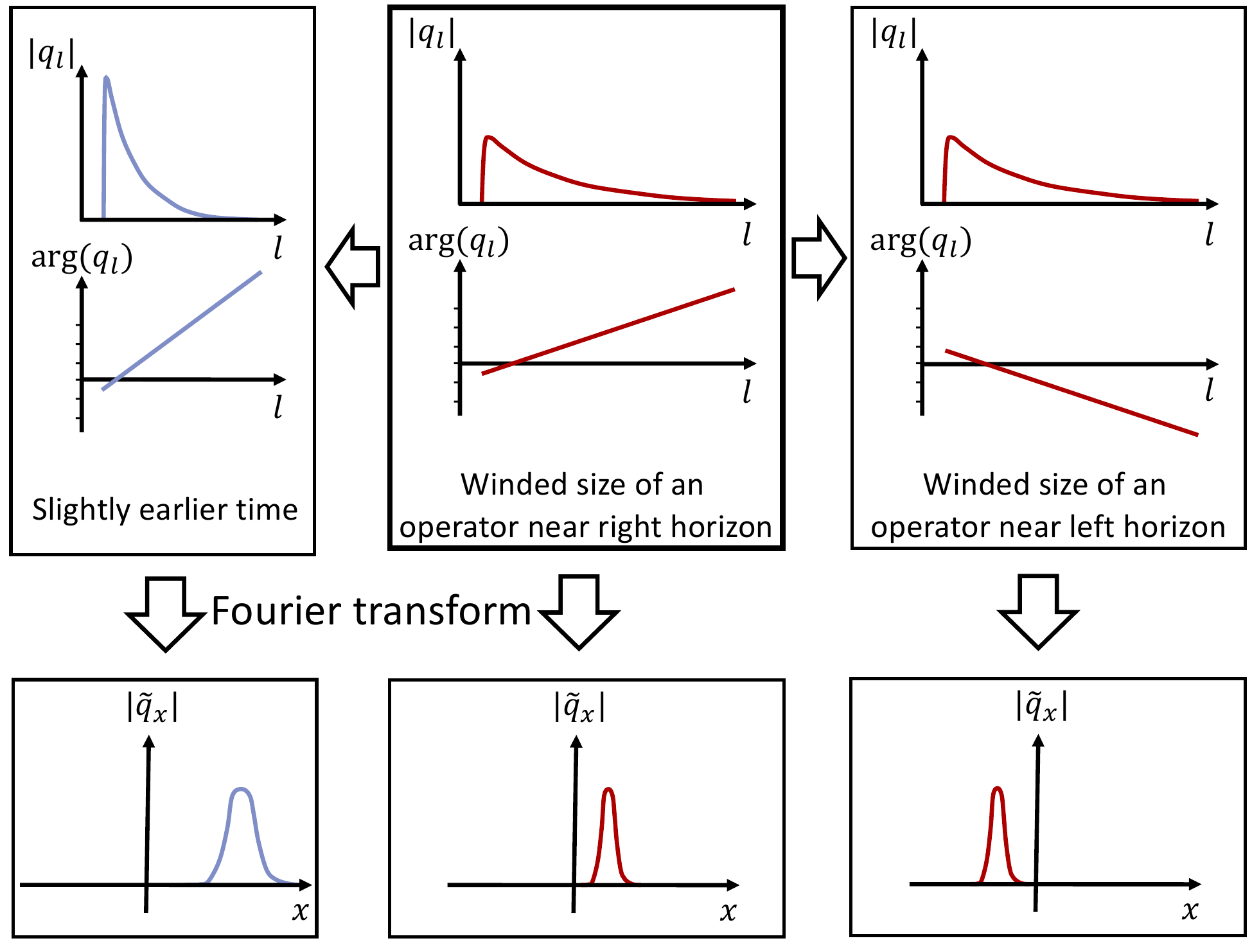}
\caption{{\bf Refinement of size-momentum duality to the level of wavefunctions.} If the expansion of the time evolved thermal Pauli (or fermion) is $\rho_\beta^{1/2} P(t) =\sum c_P P$ (or $\rho_\beta^{1/2} \psi(t) =\sum c_u \Psi_u$), then one can define the winding size distribution $q(l) = \sum_{|P|=l} c_p^2$ (or $q(l) = \sum_{|u|=l} c_u^2$), in contrast to the conventional size distribution $\sum_{|P|=l} |c_p|^2$. We argue that the $q_l$ is the boundary analog of the bulk momentum wavefunction. The plot is the schematic drawing of the winding size distribution in the SYK model, near, but slightly before the scrambling time. This is the regime that the width of the distribution is of order $n$. One can observe that the Fourier transform of the winding size distribution mimics the behavior of the position of the infalling particle (measured, e.g., from the black hole horizon). {\bf Up middle.} Magnitude and phase of winding size distribution. {\bf Bottom middle.} The Fourier transform (or bulk location) is near the origin. {\bf Up left.} Magnitude and phase of winding size distribution at a slightly earlier time. The size distribution is smaller and winds faster. {\bf Bottom left.} The Fourier transform (or bulk location) is farther from the origin. {\bf Up right.} The size distribution after acting by $e^{igV}$, with the proper value of $g$. Now the distribution is winding in the opposite direction. {\bf Bottom right.} The Fourier transform shows that the particle is on the other side of the origin, a manifestation of the fact that the infalling particle has moved from one side of the horizon to the other side.  
}
\label{fig:size_momentum}
\end{figure}

In~\cref{sec:holography} and~\cref{sec:exsyk} we will observe that size winding can be used to roughly model the emergence of bulk radial direction in the boundary system. Indeed, we will see that the winding size distribution mimics the momentum wavefunction of a particle falling into the wormhole, hence its phase describes the particle location in the bulk. In this picture, the coupling $e^{igV}$ is similar to a shift of location, moving the particle from the left region to the right one. See~\cref{fig:size_momentum} for a pictorial demonstration, and~\cref{sec:holography} and~\cite{lin2019symmetries,maldacena2017diving} for precise arguments.

The traversable wormhole is a geometrization of the Hayden-Preskill protocol \cite{maldacena2017diving} which in turn is closely related to the black hole information paradox. Recently, major progress has been made in resolving the information paradox (for a review, see \cite{pagecurvereview}). 
The central ingredient in the recent progress has been an application/refinement of the quantum extremal surface (QES) formula in a regime where quantum effects are large, and a bulk derivation of this formula from the replica trick.
This formula delineates when information can be recovered from a black hole. Applied to our context, it gives a non-trivial bound on when the traversable wormhole protocol can succeed.
We suggest a relatively simple setup where an experimental measurement of certain von Neumann (or Renyi) entropies could be used as a non-trivial experimental check of the QES formula in a regime where quantum effects are important. Our setup just involves measuring the entropy of a subset of the degrees of freedom of a one-sided system in the thermal state. This is presumably simpler experimentally than measuring the entropy in time-dependent examples.

We conclude in~\cref{sec:experiment} with a discussion of possible benchmarks for building a traversable wormhole in the lab.

In \cref{app:SYK}, we show that the large-$q$ SYK model exhibits near-perfect size winding at low temperatures, where its holographic interpretation is best understood. At finite temperatures the large-$q$ model is still tractable, and we demonstrate a rather precise match with stringy expectations.
Indeed, this is to a large extent nothing but a translation of existing results on two-point functions for traversable wormholes~\cite{maldacena2017diving, qi2019quantum} in the language of size, as we discuss in \cref{sec:holography}. 
Although perfect size winding might be specific to systems with good holographic duals, we expect to see imperfect winding even in non-gravitational systems (In~\cref{sec:subsecGGH}, we show that random non-local Hamiltonians exhibit weak size winding, which shows that traces of geometrical wormhole physics can exist in very generic quantum systems).

\subsection{Related work}
Other studies of information transfer through traversable wormholes and related notions include~\cite{freivogel2019traversable,bak2019experimental,Bao_2018,bao2019wormhole}. In particular, one small-scale experiment with trapped ions has already been carried out~\cite{Landsman_2019} based on~\cite{Yoshida2017,Yoshida2018}. This experiment implemented a probabilistic protocol and a deterministic Grover-like protocol~\cite{Yoshida2017}. In the deterministic case, the circuit in~\cite{Landsman_2019} can be related to our \cref{fig:Wormhole_Circuit}(a) if we specialize to infinite temperature, push the backward time evolution through the thermofield double, and replace~$V$ by a projector onto a Bell pair. Another recent experimental investigation looked at a similar scrambling circuit but in a qutrit setting~\cite{blok2020quantum}.

\emph{Note added}: After this work had been completed, we learned of an independent investigation of the role of operator spreading in wormhole-inspired teleportation protocols by Schuster, Kobrin, Gao, Cong, Khabiboulline, Linke, Lukin, Monroe, Yoshida, and Yao, which will appear in the same arXiv posting.

\begin{figure}
\includegraphics[width=10 cm]{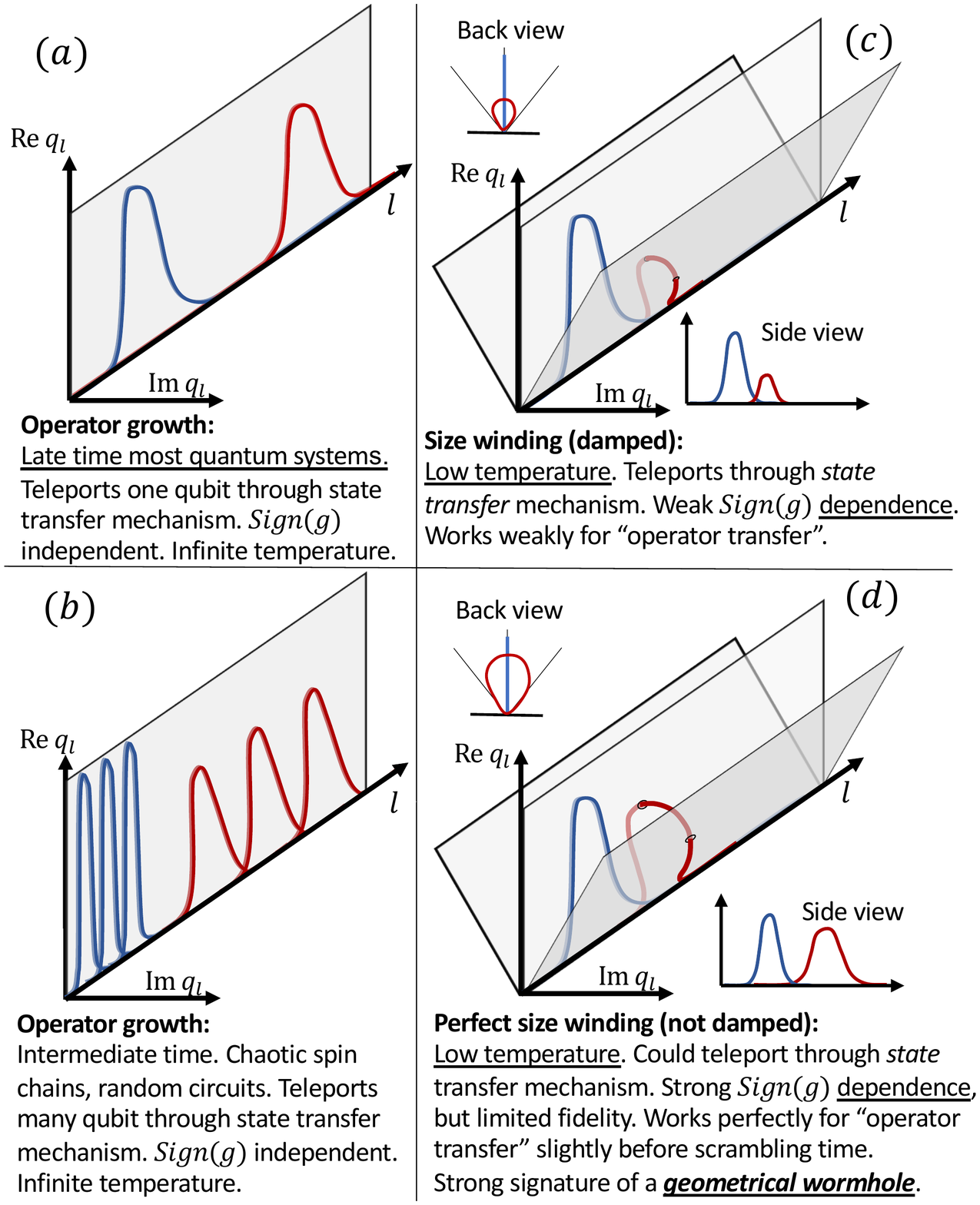}
\caption{A short summary teleportation by size, discussing different systems, different patterns of operator growth, and consequence of each growth pattern for signal transmission. {\bf Blue:} Initial operator-size distribution. {\bf Red:} Operator-size distribution of the time-evolved operator.}
\label{fig:Summary_of_everything}
\end{figure}
\subsection{Review of our first paper}\label{sec:survey}

In this section, we review definitions and concepts presented in~\cite{firstpaper}.
Readers familiar with that work may skip this summary section. 
\par 
Our starting point is the thermofield double (TFD) state on two Hilbert spaces indexed by $L$ and~$R$. We further assume that the Hamiltonians of $L$ and $R$ are the transpose of one another, and that the entire $LR$ system has a holographic gravitational dual (e.g., the Hamiltonian could be that of the SYK model). The gravitational dual of the initial TFD state is a wormhole geometry connecting $L$ and $R$ subsystems (see, for example~\cite{maldacena2003eternal}). 
\par 
The presence of the wormhole in the geometry dual to the TFD does not imply that messages can be sent between the left ($L$) and right ($R$) systems. Indeed, a message sent through the wormhole from one boundary system cannot reach the other boundary, and it instead meets its fate in the singularity. In that sense, the wormhole is \emph{not} traversable, and it cannot be used to transmit signals between the boundaries.
\par
\begin{figure}
\includegraphics[width=\textwidth]{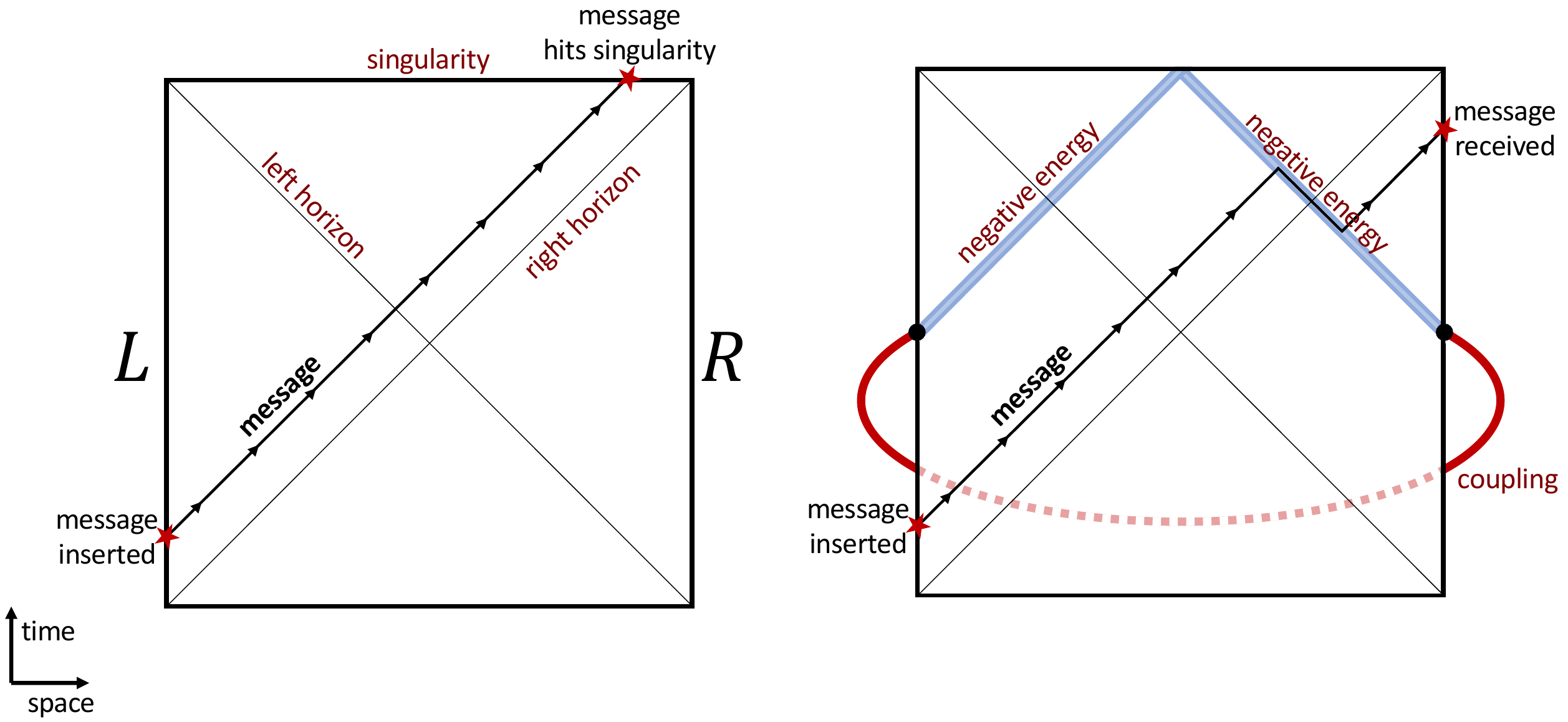}
\caption{Penrose diagram of wormholes.
\textbf{Left}: Without the coupling, a message or particle inserted at early times on the left passes through the left horizon, and hits the singularity (the top line of the diagram).
\textbf{Right}: In the presence of the left-right coupling, the message hits the negative energy shockwave (the thick blue line) created by the coupling. The effect of the collision is to rescue the message from behind the right horizon.}
\label{fig:Penrose_Diagram}
\end{figure}

 
By simultaneously acting on the two boundaries by a finely-tuned interaction at $t=0$, \citet{gao2017traversable} showed that one can render the wormhole traversable. The effect of acting with the carefully chosen, bipartite coupling proposed in~\cite{gao2017traversable} is that a negative energy shock-wave is created in the bulk, which is able to pull a signal out of the wormhole, thereby delivering it to the other boundary. See~\cref{fig:Penrose_Diagram} for an illustration. Thus, the $LR$ coupling modifies the geometry of the wormhole in such a way that a signal free-falling from one boundary can escape the singularity and reach the other boundary. 
\par 
\begin{figure}
\includegraphics[width=4.2cm]{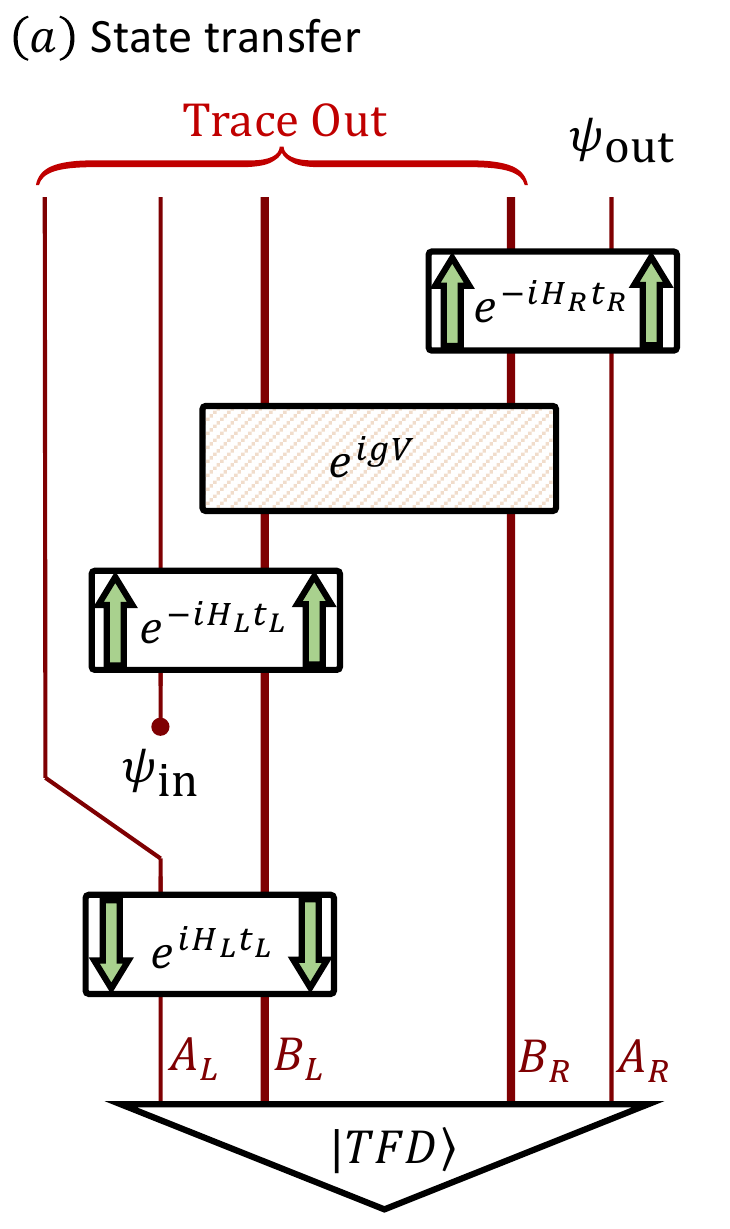}
\qquad\qquad\qquad
\includegraphics[width=4.2cm]{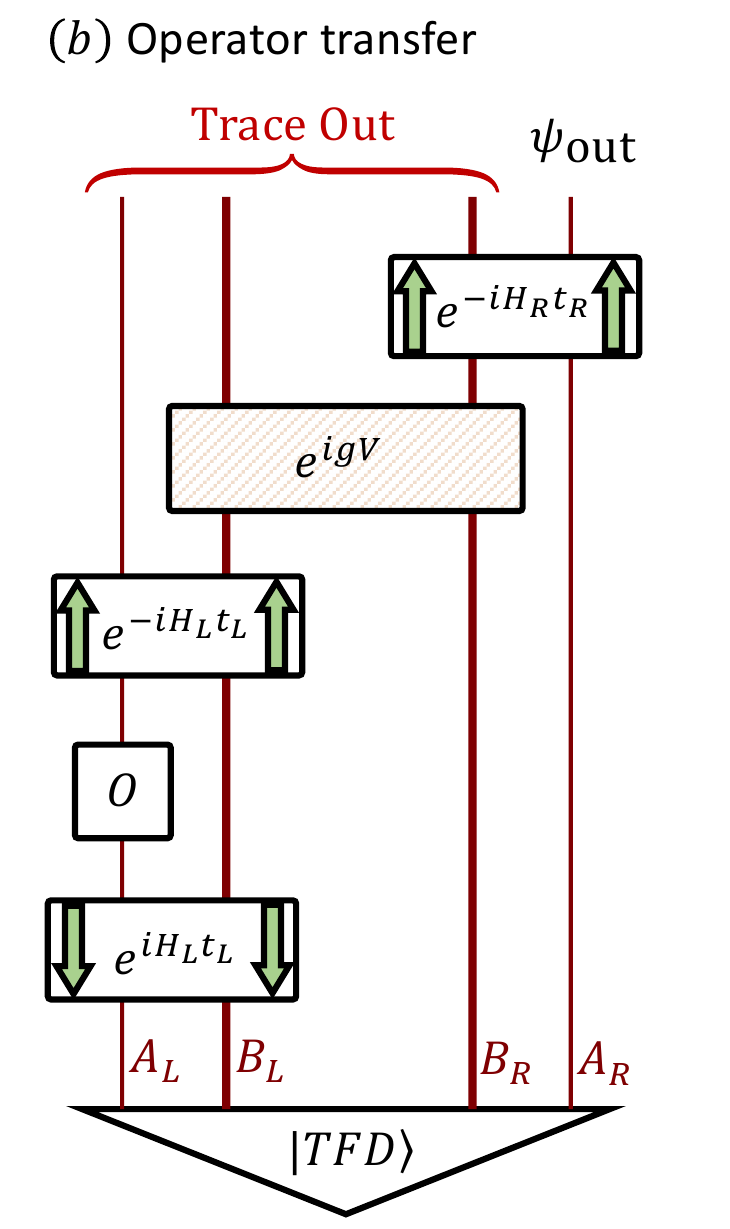}
\caption{The circuits considered in this paper, with $H_L = H_R^T$. Downward arrows indicate acting with the inverse of the time-evolution operator. In both protocols, the goal is to transmit information from the left to the right.
The \textbf{(a) state transfer} protocol calls for us to discard the left message qubits ($A_L$) and replace them with our message $\Psi_{\mathrm{in}}$. The output state on the right then defines a channel applied to the input state.
The \textbf{(b) operator transfer} protocol calls for the operator $O$ to be applied to $A_L$. Based on the choice of operator, the output state on the right is modified, similar to a perturbation-response experiment. \label{fig:Wormhole_Circuit}}
\end{figure}

In the story above, an exotic two-sided coupling produces a negative energy shockwave, which renders the $LR$ wormhole traversable, thereby allowing a message to be transmitted from one side to the other. This explanation of the signal transmission is given from a purely bulk gravitational perspective, and it does not make clear the way that information can be transmitted between the $L$ and $R$ Hilbert spaces in the dual quantum picture. In~\cite{firstpaper}, and further expanded in this paper, we explain this communication phenomenon from the quantum mechanical, boundary perspective. As discussed in~\cite{firstpaper}, the traversable wormhole communication protocol described above can be given a boundary interpretation, illustrated as a quantum circuit and shown in~\cref{fig:Wormhole_Circuit}. Moreover, the process of sending and receiving signals corresponds to a form of state transfer or quantum teleportation on the boundary~\cite{firstpaper}.
\par 
We argue that the information transmission through the boundary circuit (\cref{fig:Wormhole_Circuit}) can be understood with a proper analysis of the size distribution and spreading of the operators involved. Hence, we coined the term \emph{teleportation by size} to describe the different ways that the wormhole circuit can be used to transmit signals. 
\par
\begin{figure}
    \centering
    \includegraphics[width=13cm]{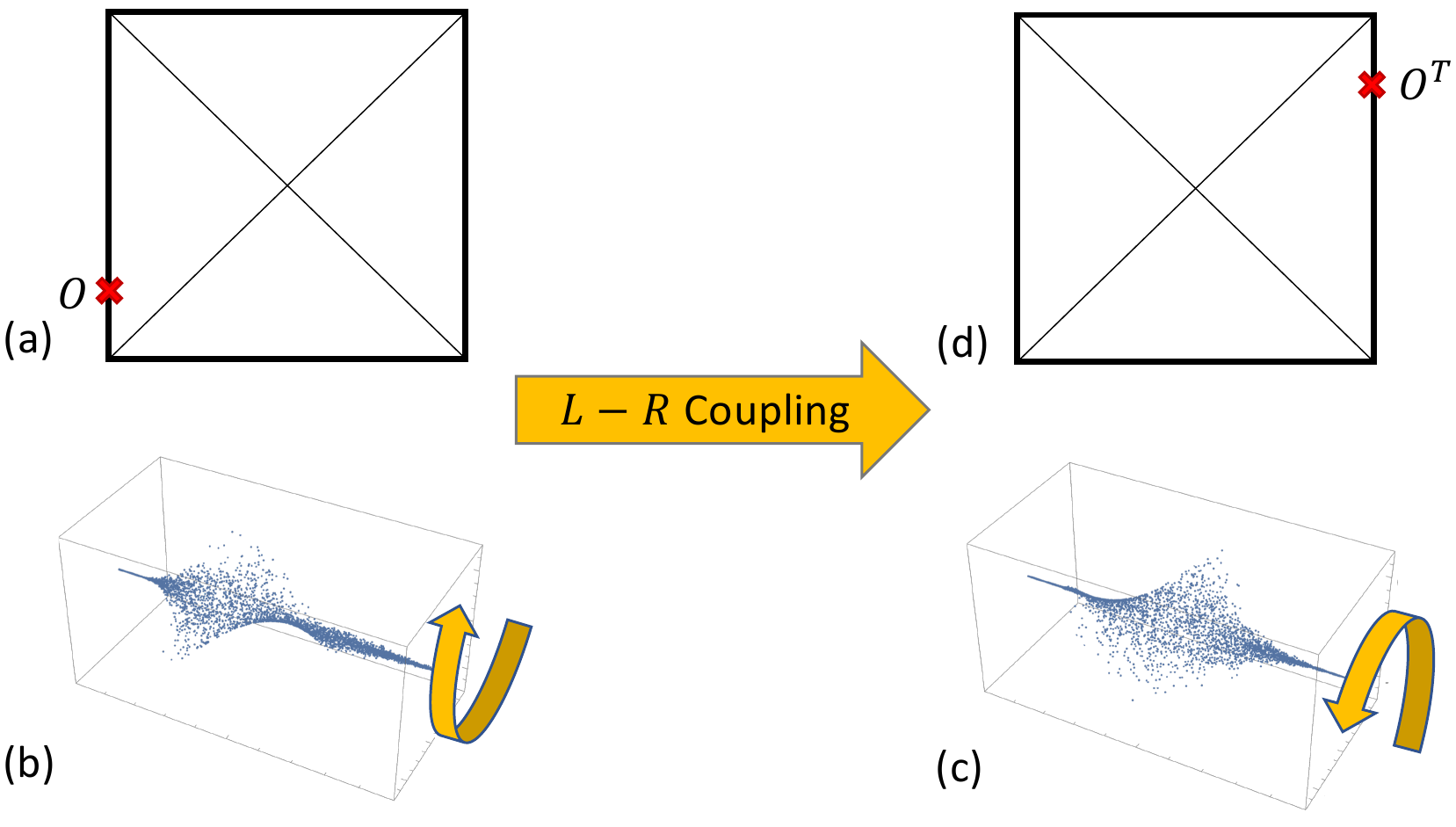}
    \caption{Traversing the wormhole from the boundary point of view. (a) An operator $O$ inserted at negative time $t$ into the left boundary. (b) The (winding) size distribution of the thermal operator $O(t)\rho_\beta^{1/2}$, which is winding in the clockwise direction. (c) The size distribution after the application of $LR$ coupling. The coupling applies a linear phase to the size distribution of the thermal operator in part (b), unwinds it, and winds it in the opposite direction. In this way, we obtain a counter-clockwise size distribution corresponding to the thermal operator $\rho_\beta^{1/2}O(t)$. (d) As we saw in~\cite{firstpaper}, winding in the opposite direction corresponds to the operator inserted on the other boundary at a positive time. Thus, the coupling maps the operator $O$ on the left to operator $O^T$ on the right.}
    \label{fig:lrwindi}
\end{figure}

 
As previously observed, there  are~\emph{two} mechanisms of transmission with the wormhole circuit:
\begin{enumerate}
    \item {\bf Late time high-temperature teleportation.} This mode of transmission is---to a large extent---unexpected from the gravitational point of view, and it does not immediately correspond to a signal traversing a semi-classical wormhole. It can be used to teleport a qubit at high-temperatures when the system is described by (potentially non-holographic) chaotic Hamiltonians. We studied this model in~\cite{firstpaper} and provided fidelity bounds for the corresponding teleportation protocol. Here, we will show how it can be used to teleport many qubits in spin-chains, Brownian circuits, and Hamiltonians with i.i.d. random matrix elements.
    \item {\bf Low temperature through-the-wormhole transmission.} This mode corresponds to the transmission of signals through a semi-classical wormhole. On the boundary, the transmission can be understood using the size distribution of thermal operators. More precisely, we argue that the size distribution of the operators should~\emph{wind} in the complex plane, i.e., obtain a phase that is linear in the size. This phenomenon---which we call~\emph{size-winding}--- is at the heart of signal transmission through the wormholes. The role of the $LR$ coupling is to undo the complex winding of the operator's size, and instead wind the size distribution in the opposite direction. In effect, this reversal of the winding direction corresponds to mapping the operator inserted in one boundary to the other boundary. See~\cref{fig:lrwindi} and~\cite{firstpaper} for more details. We explicitly show~\emph{size-winding} of thermal operators near the scrambling time for the SYK model, and we conjecture that the phenomenon can also be found in other holographic systems. Specifically, the size distribution behaves similarly to the (null) momentum wave-function in the bulk, and its winding rate is dual to the ``location'' of the infalling operator in the bulk. In~\cref{sec:holography}, we expand the arguments of~\cite{firstpaper} and further elucidate the gravitational picture as well as connections to the recent developments on ``gravitational islands''~\cite{westcoast}.
\end{enumerate}
In this paper and~\cite{firstpaper}, we assume that the total number of qubits (or fermions) on each side is~$n$, and the number of \emph{message} qubits (or fermions) that are transmitted by the state transfer or operator transfer protocols is $m$. Furthermore, the coupling is assumed to act on $k$ \emph{carrier} qubits (or fermions). Unless otherwise states, we will assume that $k=n-m$, i.e., the coupling acts on all qubits but the message qubits, or $k=n$. The difference between the performance of these cases will be insignificant in the regime of interest $m \ll n$.

\subsubsection{Size Winding}\label{subsec:swinding}
Before proceeding, let us review the definition of size winding, which is a property of a Hermitian operator $O$, the Hamiltonian $H$ of the system, the temperature $\beta$ and the time $t$ at which the operator $O = O(t)$ is evaluated at in the Heisenberg picture.
For simplicity we assume that the Hilbert space of our system is that of $n$ qubits (there is an analogous definition for fermions \cite{qi2019quantum}).
Let $\rho_\beta = e^{-\beta H}/\tr{e^{-\beta H}}$. Expand the operator $\rho_\beta^{1/2} O$ in the Pauli basis as 
\begin{equation}\label{eq:def-c}
\rho^{1/2}_\beta O = 2^{-n/2}\sum_P c_P P,
\end{equation}
where the sum runs over all $n$-qubit Paulis\footnote{From now on, we suppress the subscripts $L$ and $R$ when there is no confusion.}.
Write $\lvert P\rvert$ for the size of an $n$-qubit Pauli operator, i.e., the number of qubits the operator acts non-trivially on $|X_1 Y_2 Z_5| = 3$.
We define the \emph{winding size distribution}:
\begin{align}\label{eq:size_dist}
  q(l) := \sum_{|P|=l} c_P^2. 
\end{align}
The winding size distribution should be contrasted with the more conventional conventional size distribution, $\mathcal P(l)$, which is manifestly positive:
\begin{align}\label{eq:conv_size_dist}
  \mathcal P(l) := \sum_{|P|=l} |c_P|^2. 
\end{align}
Note that the conventional size distribution and the winding size distribution coincide at infinite temperature $\beta = 0$ since $\rho_\beta^{1/2} O(t)$ is then Hermitian operator and thus $c_P \in \RR$.
However for $\beta > 0$, $c_P$ can have arbitrary phases.
\emph{Size winding}, in its perfect form, is an ansatz for the phase of $c_P$
\eqn{c_P = e^{i \alpha |P|/n} r_P. \quad r_P \in \RR. \la{perf}}
This equation says two things. (1) the phases of $c_P$ only depend on the size $|P|$, up to $\theta_P = \alpha |P|$. There are many operators $P$ of the same size, so this is already a non-trivial statement. And (2) the phase dependence on size is linear. One subtlety of this condition is that $r_P$ could be positive or negative; the phase is thus allowed to take the form $\theta_P = \alpha |P| + \pi n(P)$, with $n(P) \in \mathbb{Z}$. The condition on the phases of the operator wavefunction \ref{perf} is fine-grained and for a chaotic system seems challenging to check at first glance. However, a necessary and sufficient condition for perfect size winding is
\eqn{q(l) = \mathcal P(l) e^{2 i\alpha l/n} . \la{nes}}
Necessity follows immediately from the definitions above. To see that this condition is sufficient, note that $|q(\ell)| \le \mathcal P(l)$ with equality iff the phase of $c_P $ depends only on $|P|$. Thus $c_P = r_P e^{i \theta(|P|)}$ for $r_P \in \RR$. Plugging in to (\ref{nes}), we get $\theta(|P|)= \alpha |P|$ which is exactly the size winding condition. We will use this criterion to show perfect winding for the SYK model in~\cref{app:SYK}. 
\par 
We expect that systems with a ``clean'' holographic dual to exhibit size winding. In this paper, we check this condition by a direct bulk calculation in nearly-AdS$_2$ gravity. 
In systems without a clear holographic dual the winding size distribution $q(l)$ will still typically obtain a complex phase. However, this phase may be a non-linear function of $l$, or $|q(l)| < P(l)$. In these cases, we say the system has imperfect winding. An example of such systems that we study in this paper is the ensemble of low temperature random Hamiltonian systems (see~\cref{sec:subsecGGH}). Imperfect winding sometimes can lead to low quality transmission of information near the scrambling time. In other words, some systems with imperfect size winding can transmit information with the exact same mechanism as the true gravitation systems transmit information through a traversable wormhole.

\section{Example Systems}\label{subsec:examples}
In this section, we study a number of paradigmatic quantum systems and analyze their ability to teleport with state-transfer or operator transfer protocols. The simplest case, the case of infinite temperature random unitary time evolution, has already been discussed in the~\cite{firstpaper} and will not be repeated here.

\subsection{Random Hamiltonians}\label{sec:subsecGGH}
In this section, we present a detailed analysis of state and operator transfer when the Hamiltonian~$H$ is coming from the Gaussian Unitary Ensemble (GUE) or the Gaussian Orthogonal Ensemble (GOE). GUE (GOE) means that each matrix element of the Hamiltonian matrix is an independent random variable coming from the complex (real) Gaussian distribution, with the extra constraint that $H$ is Hermitian (symmetric). These ensembles are not expected to have proper gravitational dual, but they can be analytically solved for all regimes of all relevant variables ---in particular, at low temperatures. Nevertheless, we will see that even random Hamiltonians have some common features with the gravitational systems and exhibit imperfect size winding. The detailed analytical calculations are complicated and tedious, and mostly computerized. The relevant techniques are reported in~\cref{sec:HaarPre} and~\cref{app:mainappGUE}.

\textbf{High temperatures regime.} The eigenvalue distribution of both GUE and GOE ensembles is the Wigner's semi-circle~\cite{liu2000statistical}, and we assume that the distribution is normalized such that the edges of the semi-circle are located at $\{-1,+1\}$. Such systems scramble ``too fast'', and at the time of the order of thermalization time. The (winding) size distribution $q_{l_0}(l)$ (define in~\cref{eq:size_dist}) of an operator with the original size of $l_0\geq 0$ can be explicitly computed using the random matrix theory techniques (See~\cref{app:appsizedist}), and the result is a distribution with two branches (at $\beta=0$):
\begin{equation}q_{l_0}(l) =  \delta_{l,l_0}\times f(it)^4 + \mathcal N(l)\times (1-f(it)^4),\end{equation}
with $\mathcal N(l)$ being the normal distribution with mean $(3/4)n$ and the standard deviation $\sqrt{2n}/3$, and $f$ being the Fourier transform of the semi-circle distribution: $f(z) := 2\,I_1(z)/z$, and $I_1(z)$ is the modified Bessel function of the first kind and of order $1$. Note that this is not similar to the growth pattern of local Hamiltonians, for which the operator size continuously flows to larger sizes (see, e.g.,~\cref{fig:brownian_circ} for the operator growth in random $2$-local Brownian circuits). Here, $q_{l_0 =0 }(l) =  \delta_{l,l_0}$, we observe that as time increases $f(it)$ goes from $1$ to $0$, moving the size of the operator from a peak at $l_0$ to a peak at $l=3n/4$. The distribution has two branches at all times, because the random Hamiltonian evolution either keeps a Pauli operator untouched, or maps it to random Pauli strings.
\par
\begin{figure}
    \centering
    \begin{subfigure}{}
        \includegraphics[width=7.0 cm]{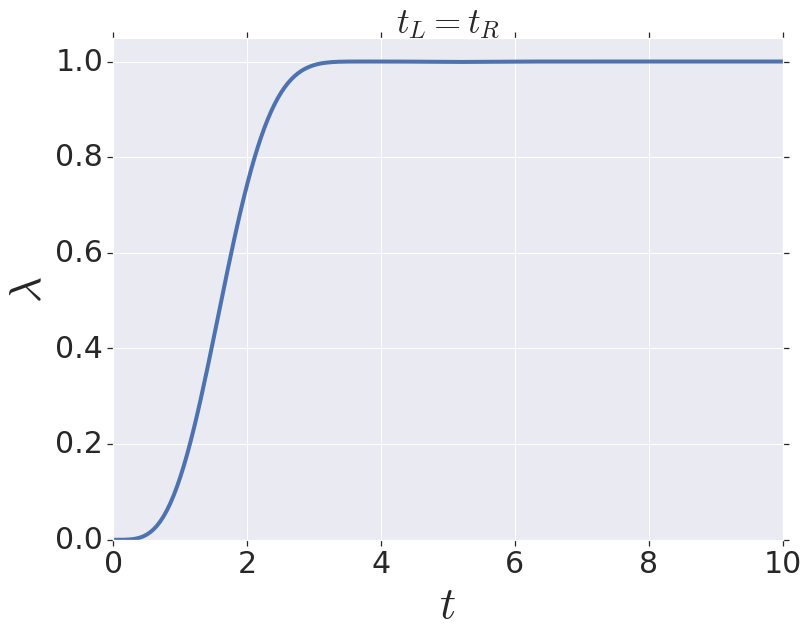}
    \end{subfigure}
    \begin{subfigure}{}
        \includegraphics[width=7.0 cm]{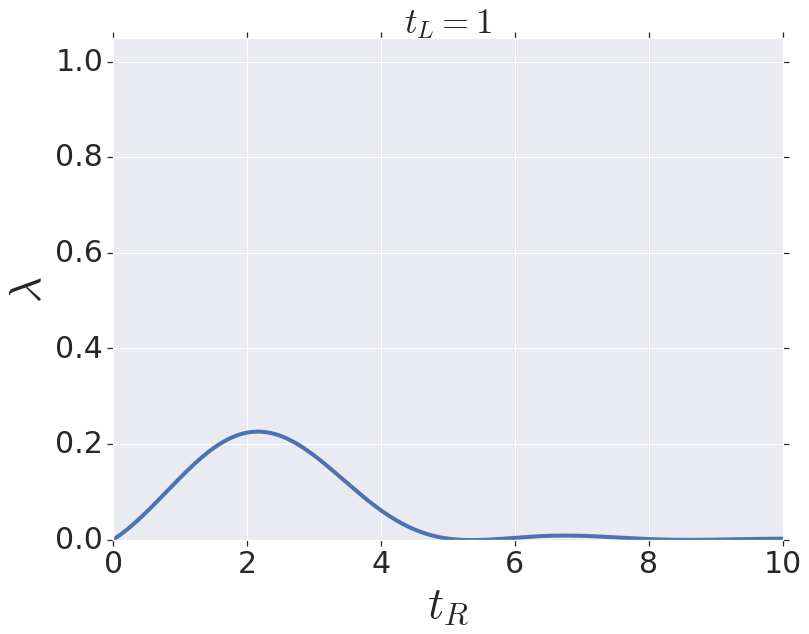}

    \end{subfigure}
    ~
    \begin{subfigure}{}
        \includegraphics[width=7.0 cm]{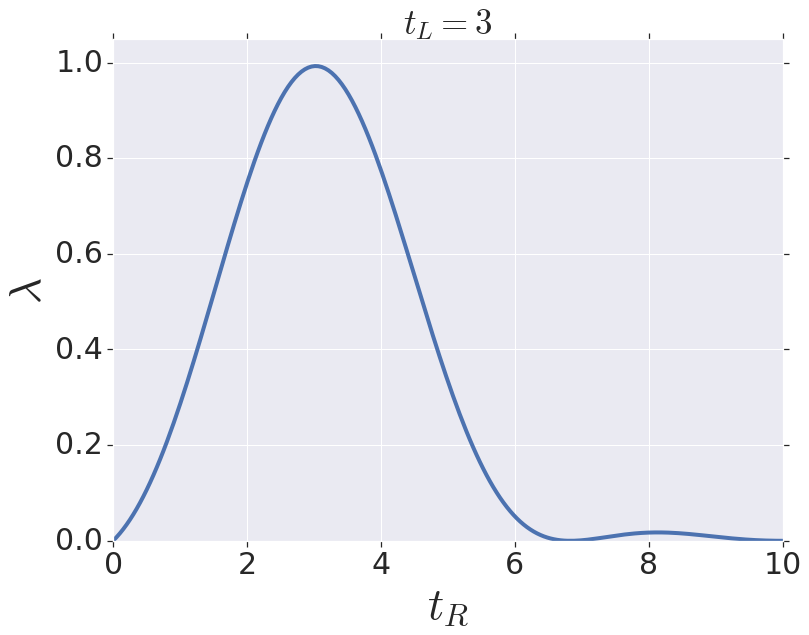}
    \end{subfigure}
    ~
    \begin{subfigure}{}
        \includegraphics[width=7.0 cm]{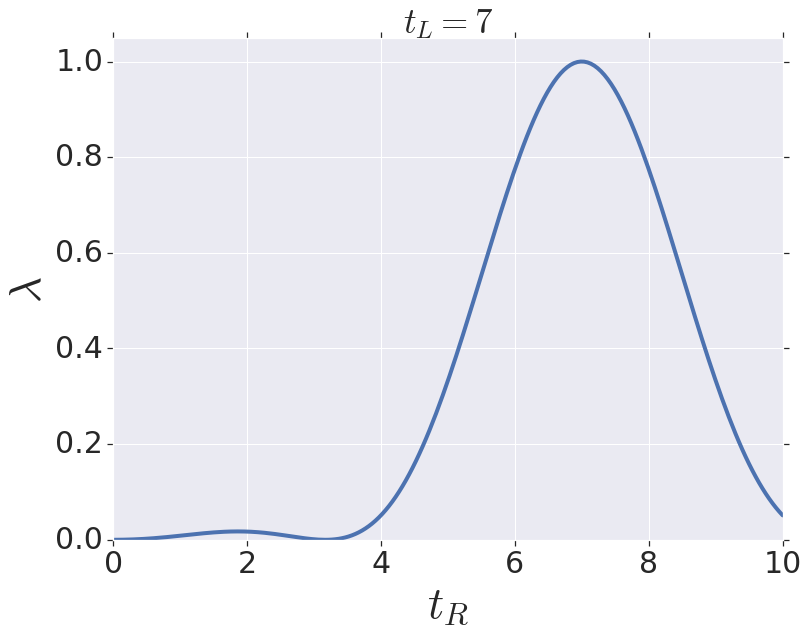}
    \end{subfigure}
\caption{Analytical results for infinite temperature state transfer protocol for sending one qubit ($m=1$), when the Hamiltonian is coming from the GOE ensemble, and at $g=\pi$. The average channel always has the form of a depolarizing channel with parameter $\lambda$, conjugated by Pauli $Y$ (see Eq.~3 in~\cite{firstpaper}). Small $\lambda$ corresponds a degraded signal, while $\lambda=1$ indicates perfect transmission. {\bf Top left.} Plot of $\lambda$ when the sending time $-t_L$ is the same as the probing time $t_R$. {\bf Top right.} Plot of $\lambda$ as a function of the probing time $t_R$, when $-t_L = -1$. A weak signal comes out at $t_R\simeq 2$. {\bf Bottom left.} Plot of $\lambda$ as a function of the probing time $t_R$, when $-t_L = -3$. A strong signal is observed at $t_R=3$. {\bf Bottom right.} Plot of $\lambda$ as a function of the probing time $t_R$, when $-t_L = -7$. After the scrambling time, the probed signal is maximized at $t_r = t_L$. }\label{fig:infgoe}
\end{figure}

Random matrix theory techniques are powerful enough to let us compute the precise average teleportation channel, even with the freedom of choosing the left and right times $t_L, t_R$ independently. See~\cref{app:mainappGUE} for the calculations, and~\cref{fig:infgoe} for an snapshot of the results.
\par
{\bf Low temperature regime and size winding.}
The winding size distribution $q_l$ can be explicitly computed at low temperatures. Again, $q_{l_0}(l)$ has two branches at the initial size $l_0$ and $l=3n/4$. The calculation techniques are explained~\cref{eq:GUE_winded_size}, and here we report the final formula for the winding size distribution: 
\begin{equation}
q_{l_0\geq 1}(l) = \delta_{l,l_0}\frac{ f(it-\beta/2)^2 f(it)^2 }{f(-\beta)^2}+
\mathcal N(l)\frac{ [f(\beta/2)^2-f(it-\beta/2)^2 f(it)^2]}{f(-\beta)^2},
\end{equation}
where $\mathcal N$ and $f$ are defined above. It is easy to see that $f(it-\beta/2)$ becomes non-real when $t$ and $\beta$ are both non-zero, giving a complex twist to $q_{l_0}$. This confirms the existence of some size winding for the GUE ensemble. Unlike the SYK model, the absolute value of size $|q(l)|$ is highly damped, indicating that the phase winding of individual coefficients is not coherent in this model. We call this ``imperfect'' size winding, which should be common in quantum systems on the general grounds of analyticity.
\par{\bf Low temperature regime, state transfer, and operator transfer.}
\begin{figure}
    \centering
    \begin{subfigure}{}
        \includegraphics[width=7.0 cm]{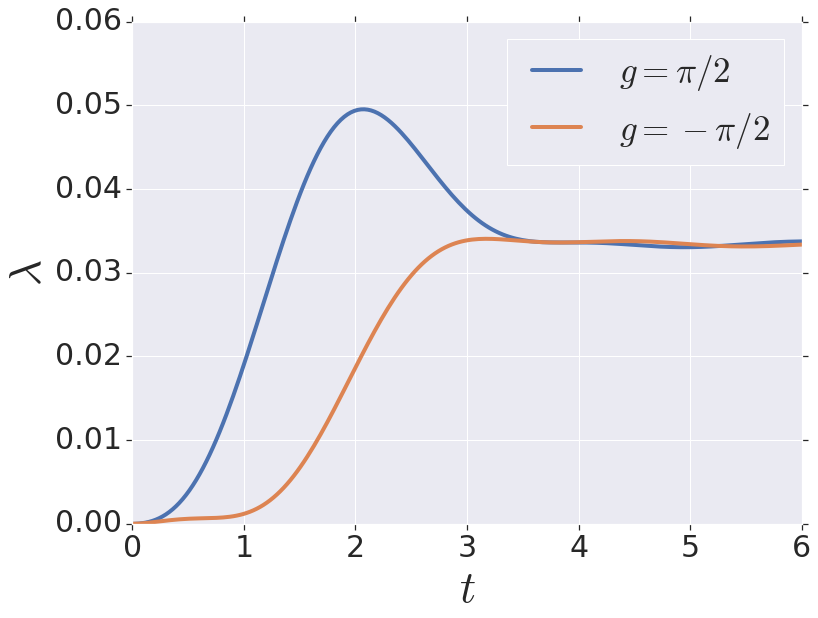}
    \end{subfigure}
    \begin{subfigure}{}
        \includegraphics[width=7.0 cm]{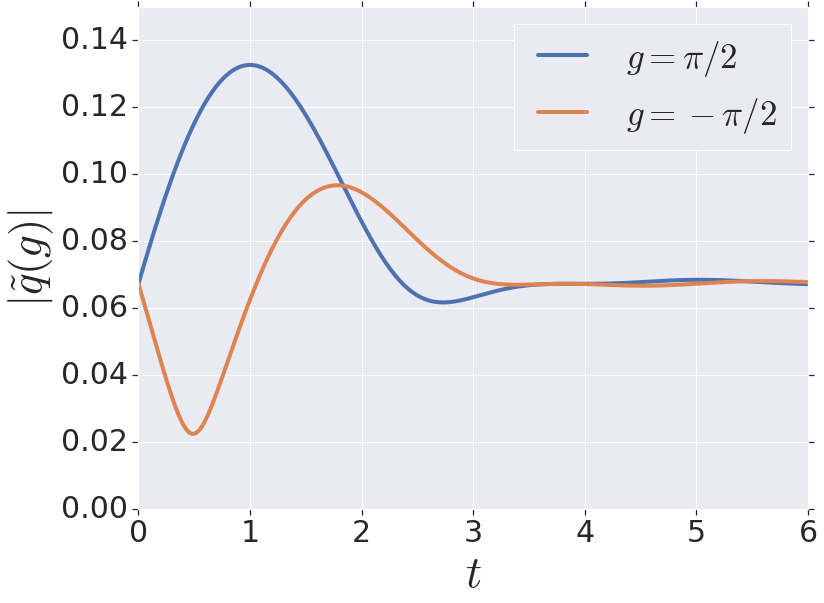}
    \end{subfigure}  
    \begin{subfigure}{}
        \includegraphics[width=7.0 cm]{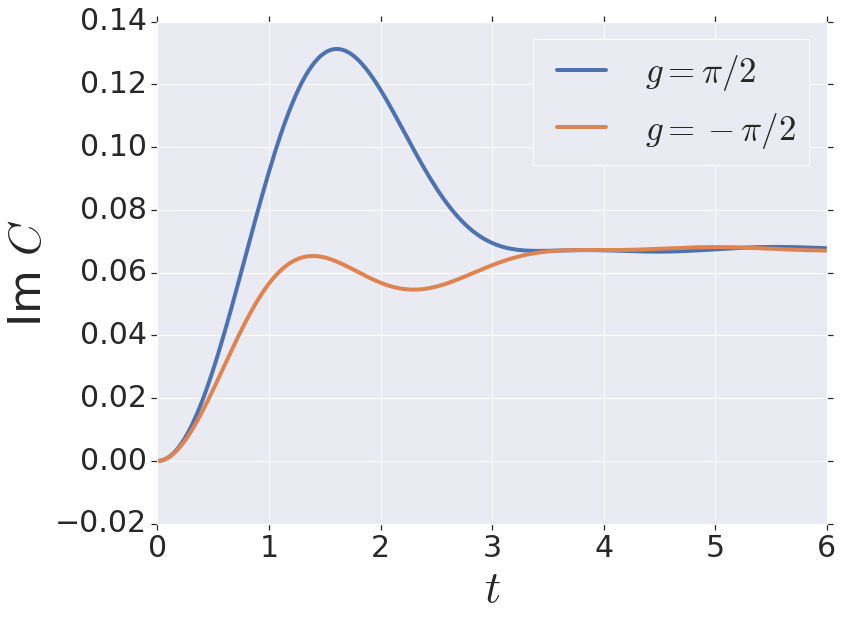}
    \end{subfigure}
    \begin{subfigure}{}
        \includegraphics[width=7.0 cm]{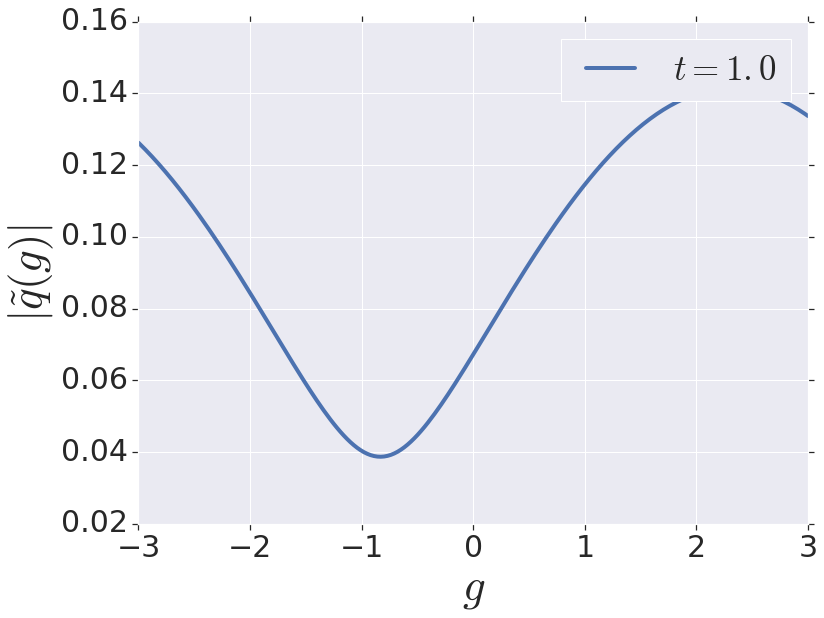}
    \end{subfigure}
\caption{Teleportation channel with random Hamiltonian at $\beta=20$. {\bf Top left}. GOE Hamiltonian depolarizing parameter $\lambda$. It clearly shows an improvement in the fidelity and quality of the channel for positive $g$. {\bf Top right}. $|\tilde q(g)|$ for GUE Hamiltonian (Recall $\tilde {q}(g) \approx e^{-ig} \bra T  P_R(t) e^{igV} P_L^T(-t) \ket T$, see Eq.~12 of~\cite{firstpaper}). Phenomenon of size winding modifies the value of the two-point function  $|\tilde q(g)|$ for early times in a $\mathrm{sign(g)}$ dependent way. {\bf Bottom left}. Imaginary part of $C= \langle e^{-igV}\phi_L(-t)e^{igV} \phi_R(t)\rangle$. $\Im \,C$ is the value of the commutator of left and right, an indicator of the causal signal. {\bf Bottom right}.  $|\tilde q(g)|$ as a function of $g$ at $t=1$. It shows a clear asymmetry around $g=0$ at low temperatures.}\label{fig:GOE_maint}
\end{figure}
Let us focus on the state transfer protocols for GUE or GOE Hamiltonian systems. 
In~\cref{app:mainappGUE} we show that it still has the form of Eq.~3 of~\cite{firstpaper}, with $\lambda$ showing a strong dependence on $\mathrm{sign}(g)$\footnote{$\mathrm{sign}(g)$ dependence is crucial from the gravitational point of view, as the wrong sign of $g$ will not make the wormhole traversable (it will send a positive energy shockwave and make the wormhole longer).}. See~\cref{fig:GOE_maint}. Note that the overall fidelity and the magnitude of the two-point function is still very small for the fidelity bounds Eq.~14 and Eq.~15 in~\cite{firstpaper} to be useful. Nevertheless, we see in~\cref{fig:GOE_maint} that there is a clear improvement of the signal for one sign of $g$. We believe that the $g\leftrightarrow-g$ asymmetry of the fidelity is partly sourced by size winding (proving this mathematically is the subject of future research). In fact, a close examination of the $\lambda$ for these systems show that it gets contributions from the real part of two-point function $\langle e^{-igV} P_L(-t)e^{igV} P_R(t)\rangle$ as well as the three-point functions $\langle P''_R(t) e^{-igV} P'_L(-t)e^{igV} P_R(t)\rangle$ with $\{P,P',P''\}=\{X,Y,Z\}$. A clear $\mathrm{sign}(g)$ dependent signature in two-point function is sufficient to produce a $g\leftrightarrow-g$ asymmetry in the fidelity as seen in~\cref{fig:GOE_maint}(a). This shows that although the bump in the fidelity is very suggestive of the signal traversing in a wormhole, one should be careful to attribute the bump to size winding alone (the situation is less challenging in the operator transfer protocols, which are directly related to the left-right 2-point function). 
\par
Indeed, the state transfer experiment above is not specifically designed to detect the physics of traversable wormholes (i.e. size winding), and it is only \emph{sensitive} to that phenomenon. This is not true when the size winding is near perfect, because in that case $|\tilde q_{l_0}(g)|$ is close to $1$ and the bound Eq.~14 in~\cite{firstpaper} is saturated, predicting a specific value for the fidelity (which might be small). 
\par
To more directly observe the signals moving in traversable wormholes, one can use operator transfer protocol in~\cref{fig:Wormhole_Circuit} and select $O = e^{i\epsilon \phi}$. Then, the response of the right hand-side density matrix attained by measuring $\phi$ operator in the leading order in $\epsilon$ is the left-right commutator $\langle [e^{-igV}\phi_L(-t)e^{igV}, \phi_R(t)]\rangle$. See~\cref{fig:GOE_maint}. The response that one gets should have a sharp bump at low temperatures, which should be stronger for one sign of $g$. Moreover, the gravity should be able to send multiple operators, hence it should have a large capacity. Although this behavior is similar to what we expect from the fidelity of the teleportation protocols, it is much easier to interpret this phenomenon as a consequence of size winding in the operator transfer experiment. This is particularly evident when we move away from completely non-local GUE/GOE Hamiltonian systems: for most $k$-local systems the size distribution of the thermal state is narrow compared to $n$ and $\langle e^{-igV} \phi_L(-t)e^{igV} \phi_R(t)\rangle\simeq e^{ig\langle V\rangle}\langle \phi_L(-t)e^{igV} \phi_R(t)\rangle$ as the coupling gives the same phase to same sized operators. This shows that the commutator is directly related to $\tilde q_{l_0}(g)$ and is highly sensitive to size winding. In practice, one needs many runs of experiments to measure this type of signal as $\epsilon$ is small.

\subsection{Spin chains}\label{sec:spchm}
\begin{figure}
\includegraphics[width=8.5 cm]{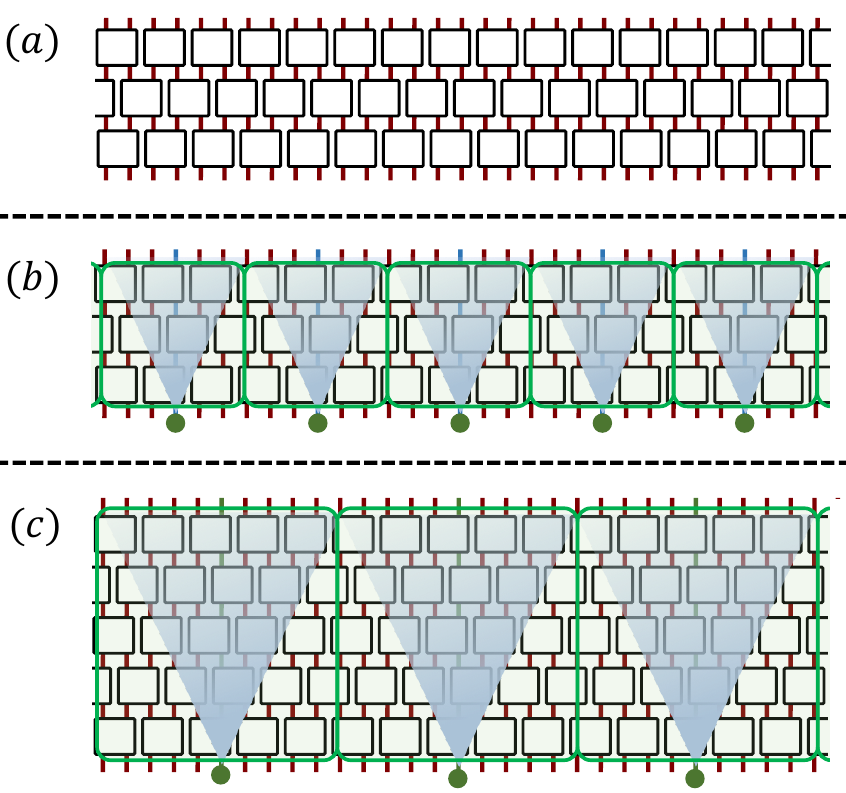}
\caption{{\bf (a)} A snapshot of the time evolution of a one-dimensional random local circuit. {\bf (b)} The circuit scrambles the degrees of freedom locally, and intuitively, it looks like a stack of parallel random unitaries (the green boxes) with the length scale given by the size of the light cone. Teleportation by size for random unitaries teach us that we can use each individual random unitary as a resource teleporting exactly one qubit (teleporting qubits indicated by dark green circles, while the $ZZ$ coupling acts on other qubits)  {\bf (c)} As time passes, the random boxes start to grow and recombine. Hence, the number of available boxes, and consequently the number of teleported qubits, will decrease.} 
\label{fig:Random_Local_Circuit}
\end{figure}

Knowing how the teleportation circuit~\cref{fig:Wormhole_Circuit}.a works when the time evolution is modeled by random unitaries we can deduce its behavior when the time evolution is a local random circuit or a that of a chaotic spin chain. Here, we primarily focus on local random circuits~(\cref{fig:Random_Local_Circuit}.a), as they have similar operator growth patterns as chaotic spin chains and are slightly easier to analyze. We only give an intuitive description of teleportation for such systems in this section and refer to~\cref{sec:precisespinchain} for a rigorous argument.
\par
Random local circuits can be constructed by considering a chain of qubits and applying randomizing unitaries between neighboring sites. See~\cref{fig:Random_Local_Circuit}(a). We start by looking at the system at relatively early times. At that time scale, the spin chain looks like a stack of many random unitary blocks in parallel (see~\cref{fig:Random_Local_Circuit}.(b)), and therefore is capable of sending many qubits through the parallel application of the random unitary teleportation discussed earlier, i.e., one qubit per block (see~\cite{firstpaper} for the case of random unitary time evolution). However, as time passes, the random blocks grow and recombine (see~\cref{fig:Random_Local_Circuit}.(c)), which reduces the total number of random blocks in the system, hence reducing the capacity of the channel\footnote{Of course, we are not discussing precise definitions of channel capacity of channels here, which could be of independent interest.}. This continues until the whole system is one large random scrambling block, which can teleport only one qubit.
\par

\par 
Indeed, the protocol cannot be used too early, as the random local circuit needs some time in order to start to scramble the system locally. The teleportation can be successful as long as (1) the total value of $g$ charge in each random block is equal to $\pi$, and (2) the size of the block is large, such that error term $\pi^2/{(\mathrm{block\,size})}$ is small (see~\cref{sec:precisespinchain}). Note that $g$ should be updated to $\pi/{(\mathrm{block\,size})}$ in real-time to satisfy (1).

\subsection{Non-local Brownian circuit models} \label{sec:brownian}
Non-local Brownian circuits, in the context of gravity, were proposed to model the black hole dynamics as fast scramblers~\cite{lashkari2013towards}. In this section, we study the operator growth in such systems, and their capacity to teleport via teleportation by size. We will mostly discuss an intuitive picture and refer the reader to~\cref{app:Brownian_Calc} for exact definitions and calculations.
\par 
Unlike the previous models that we have studied, i.e., random unitary time evolution, random GUE/GOE Hamiltonian evolution, and chaotic random circuits, this model has a proper operator growth pattern that mimics that of holographic systems such as the SYK model. On the other hand, we will observe that this model lacks size-winding which severely limits its potential for holographic teleportation. 
\par
\begin{figure}
\includegraphics[width=8.5 cm]{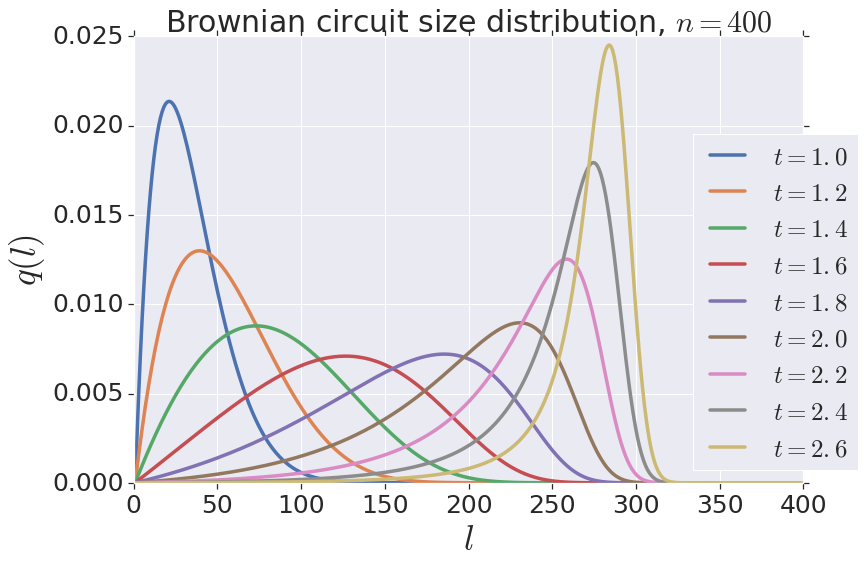}
\caption{Analytical evaluation of the size distribution of $2$-local Brownian circuits on $n=400$ qubits. The operator starts small, and its evolved size is plotted for a different range of times. At $t=1$, the operator size is still small. Then, it gets wider until $t\simeq 1.6$, and starts to reconcentrate to the final value afterward. The final width is a function of $n$, and the peak is sharper for larger $n$ (this is because at late times the time evolved Pauli looks like a random string of single-qubit Pauli operators.). However, the width of the size distribution at the intermediate times only weakly depends on $n$. See~\cref{fig:brownian_circ_2} for evidence.} 
\label{fig:brownian_circ}
\end{figure}

Here, we consider a $2$-local Hamiltonian where all terms independently and rapidly change with time. These systems only exist at infinite temperature and can be completely analyzed using It\^o calculus. This analysis indicates that the operator growth behaves as follows: (1) Operators start small, where the width of the distribution is also small. (2) Then the average size of the operators, as well as their ``width'' grow, and stay proportional to each other for a while, (3) and lastly, the operators recombine to a Gaussian distribution peaked at $3/4n$ near the scrambling time. See~\cref{fig:brownian_circ} and~\cref{fig:brownian_circ_2}. 
\par
As it is clear in~\cref{fig:brownian_circ}, the size distribution is \emph{over-dispersed} before the scrambling time, i.e., its standard deviation is large and comparable to its mean. This feature is not specific to this model and is generic in a variety of systems including the SYK model. Over-dispersion is detrimental to operator-transfer at early times for high-temperature systems. It will cause the two-point function $\tilde q_{l_0=1}(g)$ to have a norm much smaller than one, as the Fourier transform of a wide real function is highly damped. Therefore the operator transfer protocol will not work properly in the intermediate times. 
\par
For the state transfer protocols, over-dispersion suggests that the actual fidelity cannot get very large. More precisely, consider sending one qubit. For some value of $g$ we need $\tilde q_{l_0=1}(g)\approx -1$ in order to have good fidelity. However, the large width of the distribution will significantly decrease the absolute value of $\tilde q_{l_0=1}(g)$, hence, even though $\tilde q_{l_0=1}(g)$ can be a negative number it is highly damped. This shows that $F_q$ (defined in Eq.~13~\cite{firstpaper}) is much smaller than $1$, and $F$ cannot be arbitrarily close to $1$ as a consequence of Eq.~15 in~\cite{firstpaper}. Note that since $|\tilde q_{l_0=1}(g)|$ is small, the upper bound of Eq.~14 in~\cite{firstpaper} does not match $F_q$ and we cannot read the exact fidelity from the size distribution.\footnote{One can consider encoding the qubit into a code space to increase the fidelity, however, for the systems of a few hundred qubits and using simple codes our efforts have not been successful.}
\par
In the next section, we argue that as a result of size winding the size distribution $q_{l_0}(l)$ can get non-real for low temperatures. Hence, despite the over-dispersion, $|\tilde q(g)|$ can get large.

\subsection{The SYK model}\label{sec:exsyk}
Using the elegant techniques developed by Qi and Streicher~\cite{qi2019quantum}, we study size winding in the SYK model in~\cref{app:SYK}. 
For simplicity, we use the same notation, i.e., $P_{l_0=1}(l)$ and $q_{l_0=1}(l)$, for the size and winding size of the fermionic system, but keep in mind that for the fermions we use the product of Majoranas instead of Pauli operators to expand the time evolved thermal fermion (see~\cite{qi2019quantum}). The size distribution starts concentrated at a fixed value, and just before the scrambling time, it grows enough so that its width is comparable to $n$. In this time regime, we show that the winding size distribution $q_{l_0=1}(l)$ obtains a phase linear in $l$, while $|q_{l_0=1}(l)|$ remains the same the conventional size $P_{l_0=1}(l)$. Hence, all of the expansion coefficient in $q_{l_0=1}(l)$ should have a linear phase that only depends on $l$. More precisely, the equality $|q_{l_0=1}(l)|=P_{l_0=1}(l)$ shows that this phase is coherent, otherwise, the absolute value of $q_{l_0=1}(l)$ would have been damped due to phase cancellations. 
\par
In~\cref{app:SYK} we show that the number of windings per standard deviation of the size distribution is a constant and equal to $\beta J/(\pi^2 q)$. Hence, as time passes, the size distribution gets exponentially wider and the rate of size winding gets exponentially smaller. This explains why one needs a larger $g$ to make the wormhole traversable at earlier times, and the task is easier as we get closer to the scrambling time. Of course, if one waits too much, the size distribution recombines at the final equilibration value ($n/2$ for fermions and $3n/4$ for qubits) and the operator transfer through size winding will not work. 
\par
The phenomenon of size winding has a natural interpretation in terms of the size-momentum relations~\cite{susskind2019complexity,lin2019symmetries}, which we properly explain in~\cref{sec:holsizemom}. The size distribution is believed to be dual to the momentum in the bulk\footnote{Or more precisely, the derivative of the size is dual to the momentum~\cite{lin2019symmetries}, but when the momentum is growing exponentially the distinction is unimportant.}, and the exponential over-dispersion of the size distribution is indicative of the boost symmetry of the bulk, which multiplies the momentum distribution by an exponential factor. Here, we argue that the size distribution naturally contains a phase when looked at through the quantity $q_{l_0}(l)$ instead of $P_{l_0}(l)$, which act similarly to the phase of a momentum wavefunctions: its frequency determines the bulk location relative to the black hole horizon (see~\cref{fig:size_momentum} for a pictorial presentation and~\cref{sec:holography} for a more accurate interpretation). 
\par
Notice that the fermion or Pauli expansion coefficients of a certain size usually average to zero, hence $q_{l_0}(l)$ (or its better known Fourier transform, the two-point function) defined using the square of the fermion (or Pauli) is a natural quantity to observe this winding. The winding suggests some form of factorization of information on the boundary: the information about the details of the operator inserted in the bulk is encoded in the details of $c_P$ coefficients defined in~\cref{eq:def-c} (or the fermionic counterparts), while the position and momentum of the infalling signal are encoded in the phase of the winding size distribution and its absolute value, respectively.
\par
Let us follow the wormhole teleportation more accurately at the time of interest, i.e., slightly before scrambling time. At that time, the size distribution has expanded enough such that the average size is of order $n$, but possibly only a small fraction of it. The operator is highly boosted, and the momentum of its corresponding quanta is growing $\sim e^{2\pi t/\beta}$ (see~\cref{eq:avg_size}), saturating the chaos bound. But, since the number of windings per standard deviation of the size is constant and equal to $\beta J /(\pi^2 q)$, the size frequency is also decreasing by the same exponential factor $e^{-2\pi t/\beta}$, indicating that the operator is getting very close to the horizon. However, the operator is winding, e.g., in the clockwise direction. The left-right coupling simply twists the operator by an extra phase proportional to $g$ and size. As described before, if $g$ has the right value such that $e^{igV}$ accurately reverses the winding direction, the operator from the left side of the horizon will be mapped to its right side. 

As mentioned above, one needs a larger $g$ in order to teleport the particle at a slightly earlier time. This can be seen from the gravitational picture as well: at earlier times, the particle is farther from the horizon and a stronger negative energy shock-wave is needed to push it to the other side of the horizon

\section{The Holographic Interpretation}\label{sec:holography}
In the above section, we described size-winding purely from the boundary point of view. In this section, we relate these ideas to bulk concepts.
The growth of the size of an operator is a basic manifestation of chaos, and is related to a particle falling towards a black hole horizon~\cite{Roberts:2018mnp, Brown:2018kvn, qi2019quantum}. In the context of low temperature SYK, or nearly-AdS$_2$ holography, the bulk interpretation of size is particularly sharp~\cite{lin2019symmetries}. We will review this discussion, and elaborate on its applications to the traversable wormhole. Viewing size as a bulk symmetry generator gives predictions for the wavefunction of the infalling particle that traverses the wormhole which may be compared with boundary calculations. In particular, we emphasize how the symmetry viewpoint makes predictions about the entire size wavefunction and not just the average size of an operator.

\subsection{Size and momentum}\label{sec:holsizemom}
 In the traversable wormhole, the particle crossing the negative-energy shockwave experiences a (null) translation. The shockwave can, therefore, be interpreted as the generator of this translation, otherwise known as (null) momentum.
The shockwave is a direct consequence of the interaction between the two sides, which in the SYK model is simply the ``size'' operator. Thus, the size operator is simply related to null momentum~\cite{maldacena2017diving,lin2019symmetries}.

\begin{figure}
\includegraphics[width=3.3 cm]{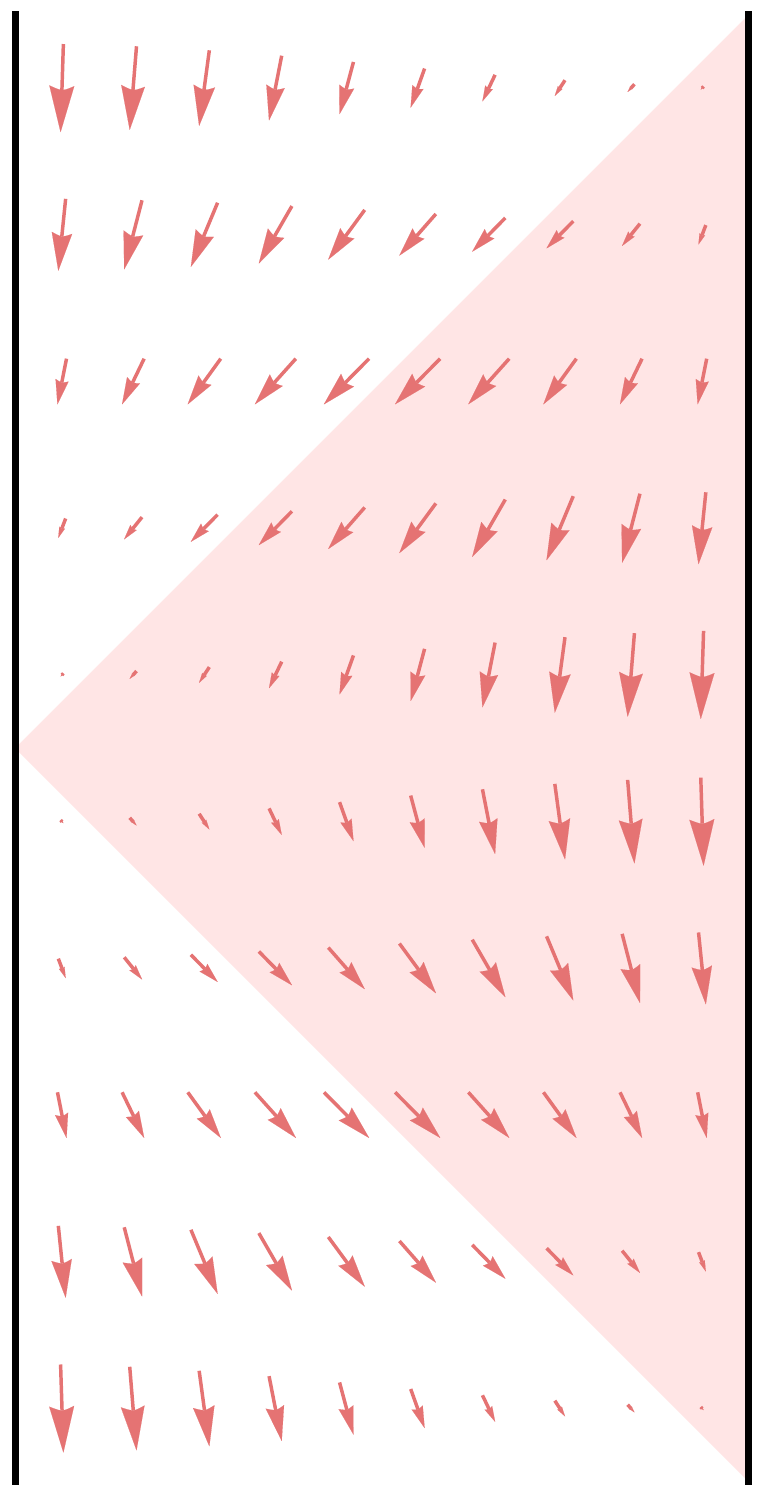}\qquad
\includegraphics[width=3.3 cm]{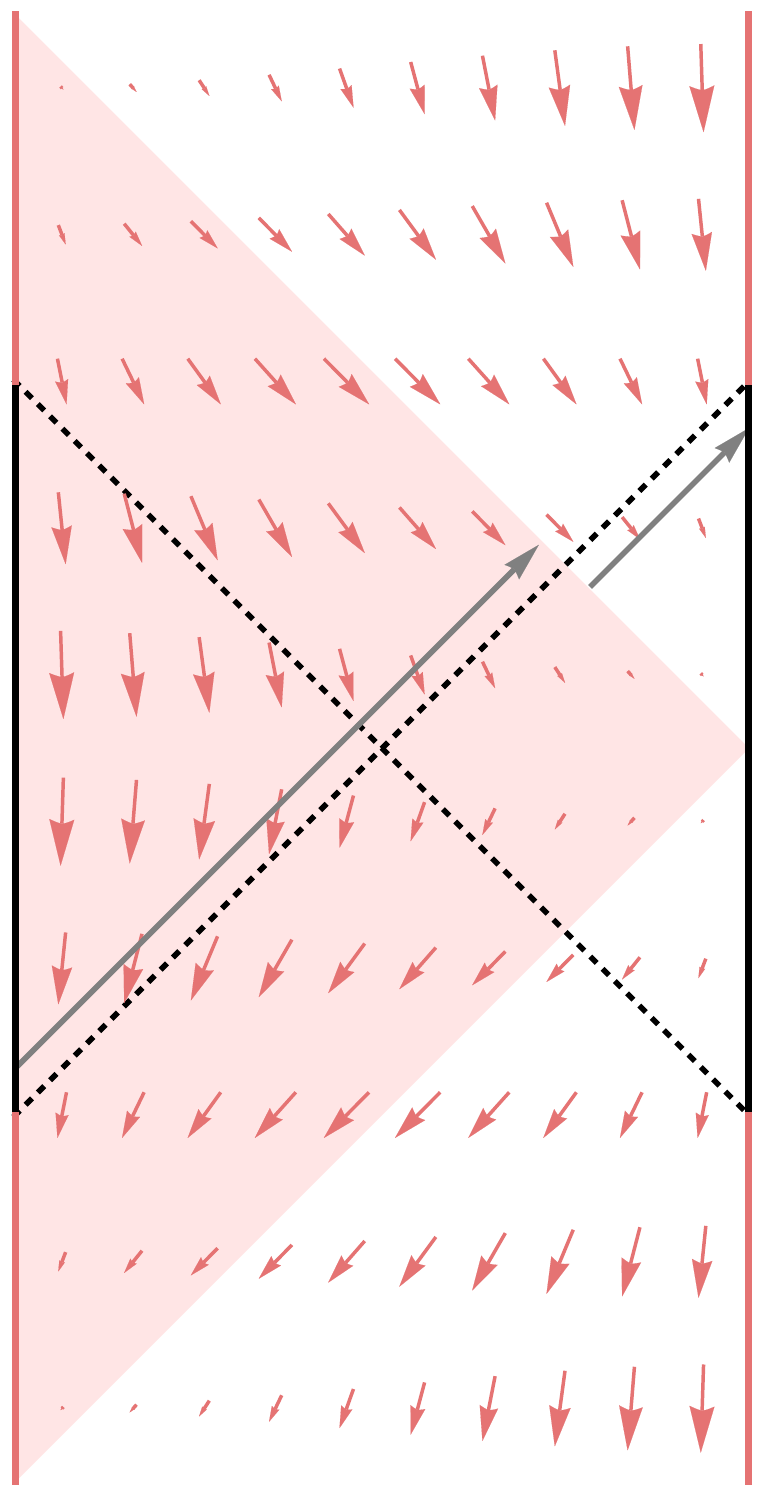}
\caption{The action of $P_+$ (left) and $P_-$ (right) in AdS$_2$. These generators act by shifting Poincare time. The particle in the right diagram would experience a null shift when leaving the Poincare patch (shaded in pink). We also indicate on the right diagram the horizon (dotted black line) and the nearly-AdS$_2$ boundaries (solid black segments).  Notice that with our conventions, physical, future-directed quanta satisfy $P_+ <0$, $P_- <0$. Here we are thinking of $P_+$ and $P_-$ as being generated by operators at $t_L = t_R = 0$. By boosting these operators, we can get generators that act at $ t_R = - t_L = t$. For large values of $t$, this will give us the null generators at the horizon.
\label{fig:pgen} }
\end{figure}

A more precise argument ~\cite{lin2019symmetries, maldacena2018eternal} can also be given.  Readers unfamiliar with nearly-AdS$_2$ may jump to the next section.
Our starting point is the approximate symmetry generators
\begin{equation}
B = H_R - H_L, \quad E=H_L + H_R + \mu V - E_0 \label{eq:approximate-sym}
\end{equation}
Here $V$ is a sum of operators on both sides $V = \sum_{i=1}^k O_i^L O_i^R$; in the SYK model, the simplest choice would be to take $O_i = \psi_i$ so that $V\propto i \sum \psi^j_L \psi^j_R$ is essentially the size operator\footnote{It exactly matches the size operator when the number of carrier fermions, $k$, is equal to $n$.}.
The operators $B$ and $E$, acting on states near the TFD, have semi-classical interpretations as the boost generator and the global energy generator (after we tune the value of $\mu$ so that the TFD is an approximate ground state of $E$), see \cite{maldacena2018eternal}. We then choose the constant $E_0$ such that $E \ket{\mathrm{TFD}} = 0$. It is also natural to consider the combinations
\begin{equation}P_\pm = -\frac12 (E\pm B).\end{equation}
The action of these generators in the NAdS$_2$ spacetime is depicted in \cref{fig:pgen}. Note that $P_\pm < 0$ in quantum field theory on AdS$_2$. For our purposes, the important point is that $e^{ia^\pm P_\pm}$ generate a null shift as the particle leaves the pink region. Furthermore, by choosing the appropriate sign of $a^\pm$, we can shift the particle backward so that it leaves the horizon. Now notice that 
\begin{equation}    -P_+ = H_R +  \mu V/2 + E_0/2, \quad -P_- = H_L + \mu V/2 + E_0/2. \end{equation}
The remarkable feature of this formula is that the action of $P_+$ $(P_-)$ is exceedingly simple on the left (right) Hilbert space\footnote{These operators generate Poincare time. If we chose $V \sim O_L O_R$ to be slightly more complicated so that $O_L$ and $O_R$ are bosonic operators, then we would have obtained the alternative Hamiltonian discussed in Section 7 of \cite{kourkoulou2017pure}.}. 

Consider a collection of Hermitian operators $\Gamma_I$, constructed by multiplying distinct Majorana fermions 
\begin{equation}
\Gamma_{I} \equiv \Gamma_{i_{1} i_{2} \ldots i_{k}}= 2^{-n/4} i^{\frac{k(k-1)}{2}} \psi_{i_{1}} \ldots \psi_{i_{k}} \quad 1 \leq i_{1}<i_{2}<\ldots<i_{k} \leq n
\end{equation} which forms an orthonormal basis for all operators acting on our Hilbert space $\tr (\Gamma_I \Gamma_J) = \delta_{IJ}$. We define a temperature-dependent wavefunction for a given operator $O$,
\begin{equation}
    \rho^{1/2}_\beta O = \sum_I c_I \Gamma_I \ . 
    \end{equation} 
We define the size winding distribution as before
\eqn{ q(s) := \sum_{|I|=s} c_I^2. \la{ qssyk} }
Just to familiarize ourselves with these definitions, let us check that
\eqn{\bra{\tfd} O_R(t)O_L^T(-t)   \ket{\tfd} &= \tr \lp \rho^{1/2} O(t)  \rho^{1/2}  O(t)  \rp\\
&= \sum_I c_I(t)^2 = \sum_s q(s,t).} 
Notice that this identity tells us that $\sum_s q(s,t)$ is independent of $t$. From now on we will assume $O = O^T$ and drop the transpose (this is true if the initial operator is a product of an odd number of fermions).
Now the generalization of this formula that we are interested in is
\begin{align}
\bra{\mathrm{TFD}}O_R(t) e^{i \mu V} O_L(-t) \ket{\mathrm{TFD}}  
&= \sum_I c_I^2(t) e^{i \mu |I|}\\
&= \sum_s q(s) e^{i \mu s}.
\label{eq:size1}
\end{align}
So we see that the Fourier transform of the size winding distribution $q$ is related to the traversable wormhole signal. On the other hand, the Fourier transform of the conventionally defined size distribution $\mathcal{P}$ is
\eqn{\bra{\mathrm{TFD}}O_R(t) e^{i \mu V} O_R(t) \ket{\mathrm{TFD}}  =  \sum_s \mathcal{P}(s,t) e^{i \mu s}.} 
If we consider a more general correlator with $t_L \ne t_R$, we can obtain $q(s)$ by setting $t_L = -t_R = t$ and $\mathcal{P}(s)$ by analytically continuing $t_L = -t + i \pi$, $t_R = t$.
We would like to compute both $q(s,t)$ and $\mathcal{P}(s,t)$ using a bulk calculation.
\eqn{
C(t) &\equiv
    \bra{\mathrm{TFD}}O_R(t) e^{i a^+ P_+} O_L(-t) \ket{\mathrm{TFD}}\\
    &= 
\bra{\mathrm{TFD}}O_R \int d{p_+} \ket{p_+}\bra{p_+}e^{i a^+ P_+} O_L \ket{\mathrm{TFD}} \\
&=
\int dp_+ \psi_R^*(p_+) \psi_L(p_+) e^{i a^+ p_+},\\
\psi_L(p_+) &= \bra{p_+} O_L \ket{\tfd}. \label{eq:mom1}
}
Here we are imagining that $O$ creates a single particle state of the quantum fields propagating in $AdS_2$. This suggests that we should identify $c_s^2 \sim \psi_R^* \psi_L$. However, the quantities on the LHS of equations \ref{eq:size1} and \ref{eq:mom1} are not quite the same. They would be the same if we could neglect the $H_R$ term in $P_+$. To understand this, let us use the boost symmetry to rewrite the correlator as
\begin{align}
    C(t) &=     \bra{\mathrm{TFD}}O_R(0) e^{-i B t} e^{i a^+ P_+} e^{i B t} O_L(0) \ket{\mathrm{TFD}}\\
    &= e^{- i a^+ E_0/2}\bra{\mathrm{TFD} }O_R(0) e^{-i a^+ \lp \mu V(t,-t) + H_R \rp } 
    O_L(0) \ket{\mathrm{TFD}}
\end{align}
We see that the time dependence is coming from the boosted size operator $V(t,-t)$. More precisely, if we are interested in times $1 \ll t/\beta$ (but shorter than the scrambling time),
\begin{align}
    C(t)  &\approx e^{-i a^+ \lp E_0/2 + \ev{H_R}\rp }  \bra{\mathrm{TFD} }O_R(0) e^{-i a^+ \lp \mu V(t,-t) \rp } 
    O_L(0) \ket{\mathrm{TFD}}
\end{align}
Using the definition of $E_0$, the global phase is just the average size of the TFD 
$E_0/2 + \ev{H_R} =  \mu \ev{V}/2 $. Thus, we conclude that for our purposes,
\begin{equation}- P_+ \approx \hf \mu \lp V -  \ev{V}\rp \end{equation}
Here $V$ is the infinite temperature size operator. At finite temperature, the relation is
\eqn{\text{thermal size} = - 2 (\delta_\beta \mu)^{-1}  P_+, \quad \delta_\beta \mu = {2 \alpha_S \cj \over \Delta} (2\pi /\beta \cj)^{2} } 
We emphasize that this equation (with the above approximations) should hold as an operator equation. 
Comparing both sides of the equation, we can equate the wavefunctions
\begin{equation}  \sum_{|I|=s} c_I^2 \approx \psi_R^*(p_+) \psi_L(p_+) \ .
\end{equation} 

The sum on the LHS runs over all operators with a given size $s = \ev{V} - 2 p_+/\mu$.
Now the wavefunctions can be evaluated in NAdS$_2$ by using conformal symmetry. Up to an overall normalization, the wavefunctions are given by \cite{maldacena2017diving}, yielding
\eqn{q(s,t) \propto {(-p_+ )^{2\Delta-1} } \exp \lp {-i 4 p_+ e^{-2 \pi t/\beta}}\rp, \quad p_+ < 0 \ .  \la{windingNAdS2}} 
Here $\Delta$ is the dimension of the operator which inserts the particle; for a single fermion in SYK $\Delta=1/q$. We expect the above formula to be valid in the low temperature limit, and for times $t\gg \beta$.
Here is the result of Maldacena-Stanford-Yang, in the regime of small $\tilde{g}$. We have defined $\mathcal{C}$ by stripping off an irrelevant multiplicative constant $\mathcal{C} \propto C$ and setting $2 \pi t /\beta \to t$ (to be restored by dimensional analysis):
\eqn{
\mathcal{C}(t_L,t_R) &=\frac{1}{\left(\cosh \frac{t_{L}+t_{R}}{2}\right)^{2 \Delta_{\phi}}}  \int_{-\infty}^{0} dp_+ \left(-p_{+}\right)^{2 \Delta_{\phi}-1} e^{-4 i p_{+}} \exp \lb i \tilde{g} p_+ {e^{(t_R-t_L)/2 } \over \cosh \frac{t_L+ t_R}{2}} \rb\\
&=   e^{\Delta_\phi(t_L-t_R)}  \int_{-\infty}^{0} dp_+ \left(-p_+\right)^{2 \Delta_{\phi}-1} \exp \lb -4 i p_{+} {\cosh \frac{t_L+ t_R}{2} \over e^{(t_R-t_L)/2 } } \rb e^{ i \tilde{g} p_+} \\
}
Now let us first specialize to $t_L = -t_R, t_R = t- i \epsilon$. Then,
\eqn{ 
\mathcal{C}(-t,t) &= e^{-2 \Delta t} \int_{-\infty}^{0} dp_+ \left(-p_+\right)^{2 \Delta_{\phi}-1} \exp \lb 4  p_{+}  e^{- t }(-i+ \epsilon)  \rb e^{ i \tilde{g} p_+}. \la{qc}}
On the other hand, to compute $\mathcal{P}(s,t)$ we set $t_L = -t + i \pi - i \epsilon, t_R = t - i \epsilon$. This yields
\eqn{\mathcal{C}(-t+i\pi,t) &=   (-e^{2t})^{\Delta} \int_{-\infty}^{0} dp_+ \left(-p_+\right)^{2 \Delta_{\phi}-1} \exp \lb  4  p_{+} e^{-t} \epsilon \rb e^{ i \tilde{g} p_+} \la{pc} .\\}
Here $\epsilon$ is a UV regulator that makes the two-point function at $t=0$ finite. This is necessary to match with any system like SYK where the conformal symmetry emerges in the infrared, see e.g., \cite{lin2019symmetries} Section 6.2 and our Appendix \ref{app: SYKGravity}.
By comparing the integrand of \nref{pc} with that in \nref{qc}, we conclude that 
\eqn{ q(s,t) = \mathcal{P}(s,t) e^{i \alpha s}, \quad \alpha \propto e^{-t}}
If we want to go beyond zeroth order in $\epsilon p_+ e^{-t}$, we had to assume that $\epsilon$ in \nref{qc} is the same as that in $\nref{pc}$. This can be justified in the large $q$ SYK model, see Appendix \ref{app: SYKGravity}.
As a further application of these formulas, we can compute the average size of a fermion. The integrand of \nref{pc} has an exponentially decaying factor $ \exp \lp -4 \epsilon (-p_+) /e^t \rp $ in $s \sim -p_+$. So we expect an average size $\propto e^t/\epsilon$.  To be more precise, we can evaluate the integral:
\eqn{
\mathcal{C} &\propto \frac{1}{\left(2 \cosh \frac{t_{L}+t_{R}}{2}+\frac{\tilde{g}}{2} e^{\frac{t_{R}-t_{L}}{2}}\right)^{2 \Delta}}.}
Then interpreting $\mathcal{C}(-t+i \pi,t)$, as a generating functional for size, and restoring factors of $2\pi/\beta$, 
\eqn{
\ev{\text{thermal size}} &=  -2(\mu\delta_\beta)^{-1} \ev{P_+} =2 i (\mu\delta_\beta)^{-1} \partial_{\tilde{g}} \log \mathcal{C}(-t+i \pi, t)|_{\tilde{g}=0}\\
&\propto {\Delta^2 \over \alpha_S \epsilon  \cj} (\beta \cj )^2 e^{2\pi t/\beta}
}
Setting $\epsilon \sim 1/\cj$ as in \cite{lin2019symmetries} gives the correct answer for the average size in the regime $\beta \ll t \ll t_*$.
Our formulas for $C(t)$ agree with the results of \cite{maldacena2017diving} in the probe limit, where gravitational backreaction is ignored. Indeed, after the scrambling time\footnote{If we were to use the exact expressions from \cite{lin2019symmetries} for $E,B$ instead of the approximate ones, this formula would be exact, but the connection to simple boundary quantities is obscure.} the semi-classical approximation breaks down and the symmetry generators no longer are given by Eq.~\ref{eq:approximate-sym}. From the boundary point of view, since our system is finite the size of the operator cannot continue to grow exponentially but must saturate.

To summarize, we learned that the momentum-space wavefunction of a particle in AdS$_2$ is closely related to the size wavefunction of the operator. Using the momentum wavefunction, we verified that simple operators satisfy size winding at times $1 \ll t/\beta$ but smaller than the scrambling time $t \ll t_*$.
The Fourier transform of the momentum wavefunction is the position wavefunction.    ``Position'' here means the coordinate conjugate to $P_+$, which is Poincare ``time'', see Figure \ref{fig:pgen}. On the edge of the pink region, this coordinate parameterizes a null shift. (This is just to say that the ``position'' conjugate to size is null or time-like; it is not a space-like position.)
The action of the two-sided coupling $e^{igV}$ in the traversable wormhole protocol simply shifts the position of the particle, allowing the particle to potentially exit the black hole.
The upshot is that in a holographic setting size winding is related to the location of the particle, e.g., whether the particle is inside or outside of the black hole horizon.

\subsection{A modified bulk protocol \la{modbulk}}
One disadvantage of the old protocol is that we must insert the signal at some particular time $t_L = - t$ on the left side. If $t$ is too large, then the back-reaction destroys the effect. If $t$ is too small, the signal does not make it across the wormhole (at least in the semi-classical limit). In this subsection, we discuss a modified protocol that should work even when $t$ is small. 

This protocol is also interesting from the bulk perspective, synthesizing \cite{kourkoulou2017pure} and \cite{lin2019symmetries}. Indeed, we have described how the instantaneous coupling acts like a null momentum generator on the particle. One could wonder whether there is a teleportation protocol that simply evolves with a symmetry generator.  

To understand this new protocol, let us first remind the reader that the microscopic definition of the approximate global energy $E \sim H_L + H_R + \mu O_L O_R$ is somewhat ambiguous. In particular, the important fact is that we deform the usual Hamiltonian $H_L + H_R$ by some operators that are highly correlated in the TFD state. The simplest choice is to take $O_L O_R \sim \sum_{i=1}^n\psi_L^i \psi_R^i$. However, there are many other choices. For example, we could consider a sum over only half the fermions on the left and right. (The price that we pay is that at a fixed temperature, using fewer fermions makes $\mu$ bigger, which means that the range of temperatures in which $E$ can be interpreted as the global energy of AdS$_2$ would be more limited.)
Another option is to consider what happens when
\begin{equation} O_L O_R \sim  \sum_{i=1}^{n/2} \psi^{2i}_L \psi^{2i-1}_{L} \psi^{2i}_R \psi^{2i-1}_R  \ . 
\end{equation} 
This might seem like a somewhat odd choice, but notice that it has the feature that we can group the Majorana fermions into qubits $Z^i_L = \psi_L^{2i} \psi_L^{2i-1}$. Now with this choice of global Hamiltonian $E$, the Poincare generator 
\begin{equation} -P_+ = H_R + \frac{\mu}{2} \sum_i Z^i_L Z^i_R \end{equation} 
Now instead of evolving with $P_+$, we may instead measure the operators $Z^i_L$, which projects the left qubits into the eigenstate $\ket{z_1, z_2, \cdots}$, and then evolve with
\eqn{ - \tilde{P}_+ = H_R + \frac{\mu}{2} \sum_i z^i_L Z^i_R  \la{eq:tildep}}
Note that this operator only acts on the right Hilbert space.

\begin{figure}
\includegraphics[width=5cm]{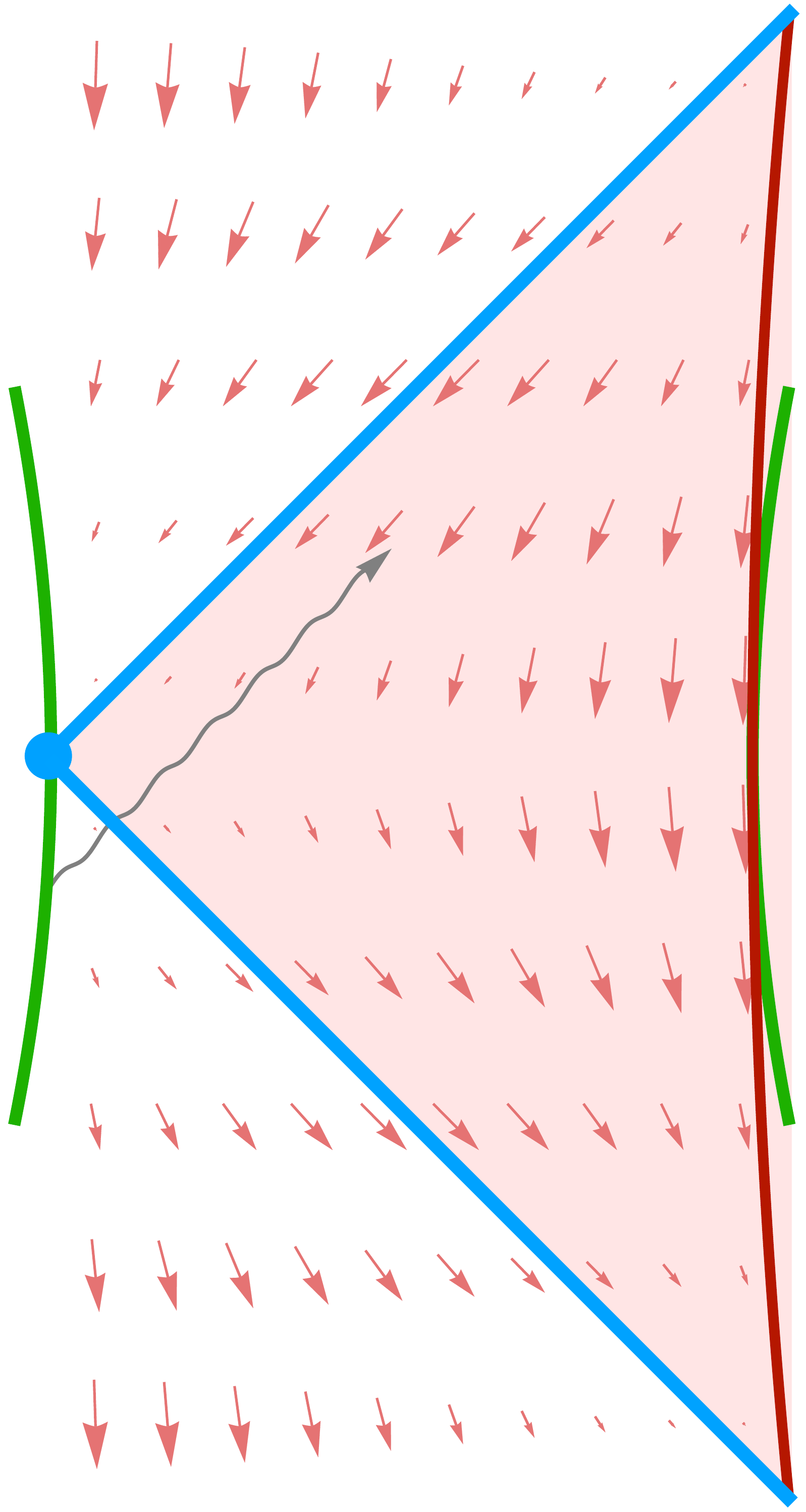}
\caption{ Teleportation at the end of the world. In this modified protocol, we insert a particle on the left at some time $t_L = -t$. Unlike the protocol in~\cite{gao2017traversable}, $t$ could be arbitrarily close to zero. The green trajectories show the standard evolution by $H_L + H_R$. A particle (wavy gray line) inserted on the left does not make it to the green boundary on the right. If we instead evolve with the symmetry generator $\tilde{P_+}$, the boundary particle on the right follows a trajectory of constant Poincare distance (dark red). The particle will then intersect the trajectory. Notice that the point on the left indicated by the blue circle is a fixed point of the Poincare symmetry. When we evolve by $P_+$, the interpretation is that the left boundary is fixed at this spacetime point. Alternatively, we may project onto an eigenbasis of the Pauli $Z_i$ operators by performing a complete measurement on the left-hand side. This inserts an EoW brane \cite{kourkoulou2017pure}. 
\label{fig:eow} }
\end{figure}

In fact, the procedure we have just described may be viewed as a ``symmetry'' based derivation of the ``state-dependent'' Hamiltonian described in \cite{kourkoulou2017pure}. There the state $e^{-\beta H }\ket{z_1, z_2 \cdots }$ was viewed as a 1-sided state with an end-of-the-world (EoW) brane, see also \cite{westcoast}. Here we are taking a 2-sided view; we expect that the act of projecting onto a state inserts a high energy shock, see Figure \ref{fig:eow}. Notice, however, that if we instead evolve with the two-sided operator $P_+$ (without the tilde), we would not inject this shockwave/EoW brane.

\subsubsection{\label{sec:entwedge}Entanglement wedge of the right side and a subset of the left}
As a sidenote let us remark that one does not need access to the full left side in order for the protocol to work. As we have already remarked, with some caveats we can use a finite fraction of the fermions. A useful concept is the entanglement wedge of the right side combined with a subset of fermions on the left side. Roughly speaking, the entanglement wedge is the bulk spacetime region where a probe operator could (in principle) be reconstructed~\cite{Czech2012,Almheiri2014,Dong2016,JLMS2016,TwirledPetz,UntwirledPetz}.

Just to make the connection with the recent island discussions \cite{westcoast}, we again consider a protocol where only a finite subset $k$ of the $n$ fermions on the left hand side is used. This means that at time $t=0$, we separate the $k$ fermions we will use in the construction of $\tilde{P}_+$ from the $n-k$ fermions that we no longer use. We consider the leftover $(n-k)/2$ qubits on the left hand side the ``garbage''.

A further motivation for studying the extremal surface is the following. As discussed in \cite{maldacena2017diving} traversable wormhole can be thought of as a geometrization of the Hayden-Preskill protocol. Alice throws her diary into one side of an old black hole. The black hole is sufficiently old that it is highly entangled with the Hawking quanta that have already been emitted. These Hawking quanta can be captured and collapsed into another black hole which is entangled with the original black hole. Under favorable conditions, there will be a wormhole connecting the two black holes. The recovery of Alice's diary on the other side of the wormhole (by the teleportation experiment) may then be viewed as a protocol to recover information from the Hawking radiation.

The Hayden-Preskill protocol suggests that after waiting a scrambling time after Alice's diary is dropped in, one just needs to include a few bits from the original black hole (together with the earlier radiation) in order to recover what was in the diary. Rephrased in a modern language, the Hayden-Preskill claim is that the entanglement wedge of the earlier radiation + a few qubits of radiation from Alice's black hole contains enough of the interior of the black hole to include her diary.
In our setup, this translates to the statement that the entanglement wedge of the right side + a small fraction of the left will include a significant portion of the interior.

In the section \ref{modbulk} above, we adopted a two-sided view, but we could also adopt a 1-sided view to make the connection to \cite{westcoast} explicit. Let us introduce some notation. We denote 1-sided pure states 

\begin{align}
\ket{B_{i \alpha} } &= e^{-\beta H/2} \ket{i}_k \otimes \ket{ \alpha}_{n-k} \\
\ket{\tilde{B}_{i \alpha} } &= e^{-\beta H/2} O(t_L) \ket{i}_k \otimes \ket{ \alpha}_{n-k}.
\end{align}
Here we have arbitrarily divided the right side into $k$ fermions and $n-k$ fermions; $i$ and $\alpha$ label the computational basis elements for the corresponding $k/2$ and $(n-k)/2$ qubits, correspondingly. This is a pure state for the right side system with an EoW brane \cite{kourkoulou2017pure}. $\ket{\tilde{B}}$ is a perturbation of $\ket{B}$ with some particle behind the horizon.
With these notations established, we return to our protocol. 
At $t=0$ we have a two sided state. After separating garbage qubits, we measure in the computational basis on the $k/2$ left qubits. 
This projects onto some definite state $\ket{i}_L$. So the combined system is in a state where the state of the $k/2$ left qubits factorizes from the garbage $\otimes$ right side. 
However, the state of  (garbage $\otimes$ right side) does not tensor factorize:
\eqn{ \sum_\alpha \ket{\alpha}^L e^{-\beta H/2 } \ket{i}^R \ket{\alpha}^R = \sum_{\alpha} \ket{\alpha}_\text{garbage} \ket{\tilde{B}_{i\alpha}}_R \la{eq:westcoast} }
We are left with a one-sided state, where the EoW brane states are entangled with an auxiliary system (the garbage/Hawking radiation).

If we model these brane states $\ket{B_{i \alpha}}$ using JT gravity $+$ EoW branes like in the West Coast model, we may use their results to compute the entanglement wedge \cite{westcoast}. 
The garbage in our setup is the analog of the ``Hawking radiation'' in the West coast model. Indeed, we may compute the entanglement wedge of the garbage. If we discard too many qubits, an ``island'' will form and the signal we want to extract will be in the garbage! On the other hand, if the garbage only contains a small number of fermions, no island will form and the entanglement wedge of the right hand side will contain the signal. 

A potentially confusing point is that in the Hayden-Preskill setup, the garbage is the analog of the black hole (after a few Hawking quanta have escaped) whereas in the analogy with the West coast model, the garbage plays the role of the radiation. There is no contradiction here. If we adopt the Hayden-Preskill interpretation of the traversable wormhole, the garbage is really the black hole; the connection to the West Coast model is just a device for computing the entanglement wedge.
Just as in \cite{maldacena2017diving}, we can say that the radiation system in the Hayden-Preskill setup has been processed and collapsed into its own black hole, which is why we interpret it as a black hole from the West coast point of view.

The West coast model has no propagating quantum fields, which is qualitatively dissimilar to SYK. A better model in the SYK context would be to study the entanglement wedge of JT gravity with free fermions in the bulk.
The boundary of the entanglement wedge is determined by the quantum extremal surface (QES) prescription (in $1+1$ dimensions, the surface is just a point), which amounts to balancing the gradient of the dilaton $\phi$ against the entanglement entropy of $m$ quantum fields in the bulk, where the $m$ fields correspond to the $m$ fermions we are including from the left. (For a review of the QES prescription in a similar context, see \cite{pagecurvereview, Almheiri:2019qdq}.)

To understand the contribution of the matter entropy, note that if we include a fraction of the fermions on the left-hand side, we can reconstruct the bulk fields dual to these fermions in a small wedge that is very close to the left boundary. More accurately, the state of any subset of fermionic qubits is basically unchanged over a timescale of order $\sim 1/J$ (when the interaction between becomes important). So using the HKLL prescription \cite{hkll}, we can reconstruct fields in a small wedge near the boundary. This means that the bulk matter entropy $S_m$ will get a contribution from $m$ fields in the subregion given by the union of a small interval near the boundary (the size of which is set by the UV cutoff $\epsilon \sim 1/J$) and the interval to the right of the quantum extremal surface.

If $m$ is zero, then the extremal surface is just at the horizon, where the dilaton is minimized. However, the entanglement entropy of the $m$ fields will shift the extremal surface further to the left, see Figure \ref{fig:ew}.
Some details are given in Appendix \ref{eeads}, more computations and a comparison to the SYK model will be reported in a future work \cite{tba}.

\begin{figure}
\includegraphics[width=15 cm]{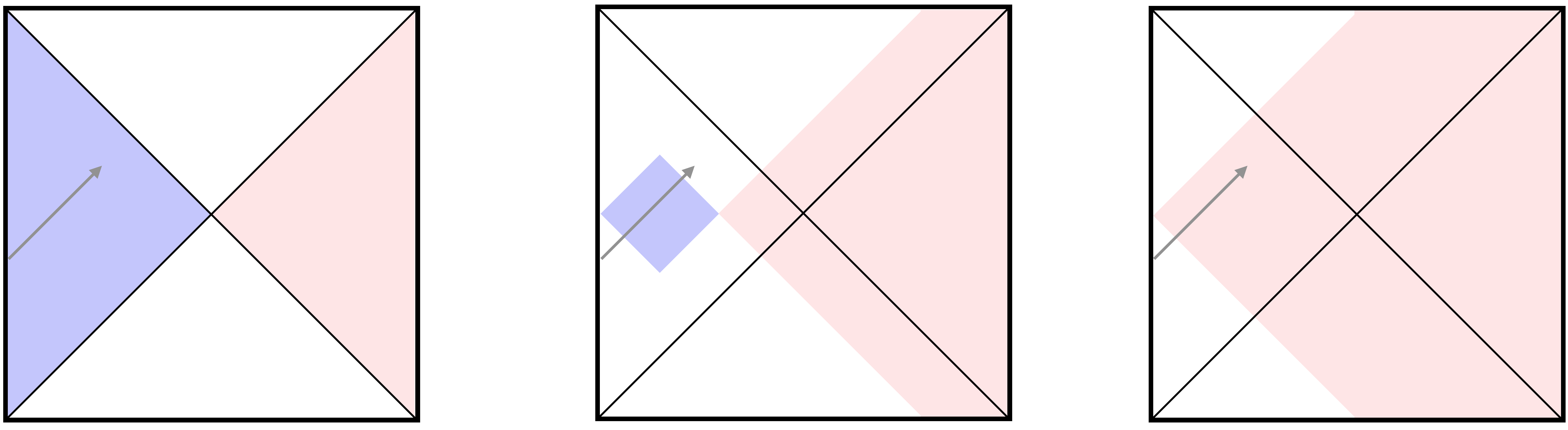}
\caption{ The entanglement wedge of the garbage is displayed in blue. The entanglement wedge of the right system + $m$ qubits on the left is displayed in pink. {\bf Left:} when $m=0$, the garbage contains the whole left side, so the quantum extremal surface is just the horizon. {\bf Middle:} as we decrease the size of the garbage, the island shrinks. {\bf Right:} $m=n$. The entanglement wedge of the pink region extends all the way to the left boundary. 
\label{fig:ew} }
\end{figure}

\FloatBarrier

\section{Discussion and Outlook}\label{sec:experiment}
We have shown, first in \cite{firstpaper} and then in more detail in this paper, that there are two distinct mechanisms by which the traversable-wormhole-inspired quantum circuit can be used to transmit information: 
\begin{itemize}
    \item Mechanism 1: a generic information teleportation phenomenon with no clear geometric interpretation, which works at high temperatures and times larger than the scrambling time.
    \item Mechanism 2: a signal transmission phenomenon that works via size-winding, and which occurs at low temperatures and near the scrambling time.
\end{itemize}
While both mechanisms can transmit signals, only for mechanism 2 can the transmitted signal be interpreted as a particle traversing a holographic wormhole. 

The two mechanisms above can be distinguished by measuring the winding size distribution, which, as we saw in Eq.~12 of our earlier paper~\cite{firstpaper}, is the Fourier transform of the left-right two-point function (as a function of the coupling $g$). Therefore, the procedure for measuring the winding size distribution is not necessarily more difficult than the measurement of the original holographic teleportation signal---one could, in principle, reconstruct it from the left-right two-point function. Therefore, one can obtain something akin to a benchmark for `quantum gravity in the lab': any holographic teleportation protocol must necessarily display a signature of size winding.

In fact, we suggest something further: the existence of size winding alone could potentially provide a geometrical picture for teleportation. One can interpret the winding size as a momentum wavefunction of a one dimensional particle. Indeed, in the low-temperature SYK model (where the holographic dual is understood) we observed that the winding size distribution is dual to the (null) momentum wavefunction of the infalling particle. Yet, even if we are not aware of the existence of a bulk dual, we could heuristically define the winding size distribution to be the wavefunction of some one dimensional particle, thereby obtaining a geometric picture. We leave this as an avenue for further study.

Finally, let us speculate on two other related benchmarks that could be used for `quantum gravity in the lab'. 

The first benchmark is the Lyapunov exponent of the OTOC~\cite{garttner2017measuring,li2017measuring,landsman2019verified,joshi2020quantum}, which determines the exponential rate of growth of the average size. Since this quantity only depends on the size distribution, it is not immediately sensitive to the phases of the operator wavefunction, and it could be considered complementary to size winding. 

A second related benchmark could be to verify the quantum extremal surface (QES) formula for the entropy of a subsystem by measuring the entropy and comparing with the holographic prediction.
In practice it may be easier to measure various Renyi entropies (using the replica trick) than measuring the von Neumann entropy. Moreover, the Renyi entropies can also be computed using a more sophisticated bulk calculation. Although at first glance this benchmark seems very different from the traversable wormhole experiment, the two are in fact closely connected. As described in \cref{sec:entwedge}, the QES associated to the joint system of one side and a few qubits on the other must necessarily encompass the infalling particle within the black hole for the traversable wormhole protocol to work.

In summary, we have seen that traversable wormholes provide a promising model system in which to study quantum gravity in the lab. 

\ssection{Acknowledgements}
We thank Patrick Hayden, Bryce Kobrin, Richard Kueng, Misha Lukin, Chris Monroe, Geoff Penington, John Preskill, Xiaoliang Qi, Thomas Schuster, Douglas Stanford, Alexandre Streicher, Zhenbin Yang, and Norman Yao for fruitful discussions. We also thank Iris Cong, Emil Khabiboulline, Harry Levine, Misha Lukin, Hannes Pichler, and Cris Zanoci for collaboration on related work. H.G.~is supported by the Simons Foundation through It from Qubit collaboration.
H.L.~is supported in part by an NDSEG fellowship.
G.S.~is supported by DOE award Quantum Error Correction and Spacetime Geometry DE-SC0018407, the Simons Foundation via It From Qubit, and the IQIM at Caltech (NSF Grant PHY-1733907).
L.S.~is supported by NSF Award Number 1316699.
B.S.~acknowledges support from the Simons Foundation via It From Qubit and from the Department of Energy via the GeoFlow consortium. M.W. is supported by an NWO Veni grant no.~680-47-459. S.N. is supported by the Walter Burke Institute for Theoretical Physics and IQIM at Caltech.
The work of G.S.~was performed before joining Amazon Web Services.

\bibliography{cecc}
\clearpage
\appendix
\begin{widetext}

\section{Preliminaries on Pauli Operators}\label{sec:pauli}
In this section we review the algebra of $n$-qubit Pauli operators and recall some useful identities.
Consider the Hilbert space $(\CC^2)^{\ot n}$ of an $n$ qubit system.
For any integer vector $\vec v=(\vec p,\vec q)\in\ZZ^{2n}$ we can define a corresponding \emph{Pauli operator} (also known as \emph{Weyl operator}) by
\begin{align}\label{eq:def pauli}
  P_{\vec v}=i^{-\vec p \cdot \vec q } \, Z^{p_1}X^{q_1} \ot \cdots \ot Z^{p_n}X^{q_n}.
\end{align}
The Pauli operators~$P_{\vec v}$ for $\vec v\in\{0,1\}^{2n}$ form a basis of the space of $n$-qubit operators.
However, we caution that $P_{\vec v}$ depends on $\vec v$ modulo $4$ and is well-defined modulo $2$ only up to a sign.
Namely,
\begin{align}\label{eq:mod2mod4}
  P_{\vec v +2\vec w} = (-1)^{[\vec v,\vec w]}P_{\vec v},
\end{align}
where $[\cdot,\cdot]$ is the `symplectic form' defined by $[(\vec p,\vec q),(\vec p',\vec q')] = \vec p\cdot \vec q' - \vec q\cdot \vec p'$.
Using this form, the commutation relation of the Pauli operators can be succinctly written as
\begin{align}
  P_{\vec v} P_{\vec w} = (-1)^{[\vec v,\vec w]} P_{\vec w} P_{\vec v}
\end{align}
and multiplication is given by
\begin{align}\label{eq:group_str}
  P_{\vec v} P_{\vec w} =i^{[\vec v,\vec w]} P_{\vec v+ \vec w},
\end{align}
where the addition $\vec v+\vec w$ in $P_{\vec v+ \vec w}$ must be carried out modulo~$4$ and can only be reduced to the range~$\{0,1\}$ by carefully applying~\cref{eq:mod2mod4}.
Finally, we note that the transpose of a Pauli operator is given by
\begin{align}\label{eq:Pauli_Transpose}
P_{\vec v}^T = (-1)^{\vec p\cdot\vec q} P_{\vec v},
\end{align}
since transposing only impacts the $Y$ operators.

With these facts in mind, let us momentarily discuss the size of Pauli operators.
The \emph{size} of a Pauli operator~$P_{\vec v}$, which we denote by $\abs{P_{\vv}} = \abs{\vec v}$, is defined as the number of single-qubit Paulis in \cref{eq:def pauli} that are not proportional to the identity operator.
If $\vec v \in \{0,1\}^{2n}$ then this can be calculated as $\vec p \cdot \vec p + \vec q \cdot \vec q - \vec p \cdot \vec q$, where the last term ensures we do not double count $Y$ operators.
Using the properties above, we arrive at an identity which holds for all $\vec v\in\ZZ^{2n}$ and will frequently be used:
\begin{align}\label{eq:Ytranspose_Pauli}
  Y^{\ot n} P_{\vec v}^T Y^{\ot n}
= (-1)^{\vec p \cdot \vec p + \vec q \cdot \vec q} P_{\vec v}^T
= (-1)^{\abs{P_\vec v}} P_{\vec v}.
\end{align}

\section{Preliminaries on Haar Averages}\label{sec:HaarPre}
In this section we recall how to compute averages of the form $\EE[\Tr[U^{\ot r} A U^{\dagger,\ot r} B]]$ where $U$ is a random unitary $d\times d$ matrix chosen from the Haar measure.
Any average over the unitary group that involves the same number~$r$ of matrix elements of $U$ (or $U^T$) as of $U^\dagger$ (or $\bar U$) can be cast in this form for a suitable choice of~$A$ and~$B$.

We first introduce some notation.
Consider the permutation group $S_r$ of $r$ elements.
For any Hilbert space $\Hil$, the permutation group $S_r$ acts naturally on $\Hil^{\ot r}$ by permuting the $r$~replicas, i.e.,
\begin{align}
  R_\pi \ket{\psi_1} \ket{\psi_2} \cdots \ket{\psi_r} = \ket{\psi_{\pi^{-1}(1)}} \ot \ket{\psi_{\pi^{-1}(2)}} \ot \cdots \ot \ket{\psi_{\pi^{-1}(t)}}.
\end{align}
where $\ket{\psi_i} \in \Hil$.
If $\Hil=(\CC^2)^{\ot n}$ is the Hilbert space of $n$ qubits then $R_\pi$ is a tensor power operator with respect to the $n$ qubits and their replicas:
\begin{align}
  R_\pi = r_\pi^{\ot n}, \quad \text{where } r_\pi \ket {j_1, j_2,\ldots, j_r} = \ket {j_{\pi^{-1}(1)},j_{\pi^{-1}(2)}\cdots j_{\pi^{-1}(r)}} \text{ and }j_k\in \{0,1\}.
\end{align}

Now consider an arbitrary $r$-th moment of the form
\begin{align}\label{eq:r moment}
  \EE[U^{\ot r} A U^{\dagger,\ot r}],
\end{align}
where $U$ is chosen from the Haar measure on~$\U(d)$ and $A$ is an arbitrary operator on~$(\CC^d)^{\ot r}$.
Since \cref{eq:r moment} commutes with all $U^{\ot r}$ for $U\in\U(d)$, Schur-Weyl duality implies it is a linear combination of the operators $R_\pi$ for $\pi\in S_r$.
For example,
\begin{align}
  \EE[U A U^\dagger] = \Tr[A] \frac I d,
\end{align}
since for $r=1$ there is only the identity permutation.
Similarly,
\begin{align}
  \EE[U^{\ot 2} A U^{\dagger,\ot 2}] = \frac{d \Tr[A] - \Tr[AF]}{d(d^2-1)} I + \frac{d \Tr[AF] - \Tr[A]}{d(d^2-1)} F,
\end{align}
where $F=R_{(1\,2)}$ denotes the flip or swap operator which permutes the two replicas.

Higher moments can be conveniently calculated using the Weingarten calculus developed in~\cite{collins2003moments}.
To state the result, define the \emph{length} $\ell(\pi)$ of a permutation~$\pi\in S_r$ as the number of disjoint cycles in~$\pi$ (including fixed points).
Then $\abs\pi := r - \ell(\pi)$ is the minimum number of transpositions needed to write~$\pi$, and it is easy to see that $\dist(\pi,\sigma) := \abs{\pi\sigma^{-1}}$ defines a metric on~$S_r$.
Now, the results of~\cite{collins2003moments,collins2006integration} imply the following general formula for an $r$-th moment:
\begin{align}\label{eq:haar average operator}
  \EE[U^{\ot r} A U^{\dagger,\ot r}]
= \sum_{\sigma,\tau\in S_r} \Wg(\sigma\tau, d) \Tr[A R_\tau] \, R_\sigma,
\end{align}
where $Wg(\pi, d)$ is the \emph{Weingarten function}.
We will not define the Weingarten function here but only state its limit as $d\to\infty$:
\begin{align}
  d^{r+\abs{\pi}} \Wg(\pi,d) \to \Moeb(\pi),
\end{align}
where $\Moeb$ is the \emph{M\"obius function} defined using the Catalan numbers~$c_n = \frac{2n!}{n!(n+1)!}$ as
\begin{align}
  \Moeb(\pi)
= \prod_{i=1}^{\ell(\pi)} (-1)^{\abs{C_i}-1} c_{\abs{C_i}-1}
= (-1)^{\abs{\pi}} \prod_{i=1}^{\ell(\pi)} c_{\abs{C_i}-1},
\end{align}
with $\{C_1,\dots,C_{\ell(\pi)}\}$ the cycle decomposition of the permutation~$\pi$.
Clearly, \cref{eq:haar average operator} implies that
\begin{align}\label{eq:haar average}
  \EE[\Tr[U^{\ot r} A U^{\dagger,\ot r} B]]
= \sum_{\sigma,\tau\in S_r} \Wg(\sigma\tau, d) \Tr[A R_\tau] \, \Tr[B R_\sigma].
\end{align}
In most cases of interest for us, the matrices $A$ and $B$ in~\cref{eq:haar average} are close to tensor product operators of the form $A = \otimes_{i=1}^n a_i$, $B = \otimes_{i=1}^n b_i$,.
In this case, $\tr[A R_\sigma] = \prod_{i=1}^n\tr[a_i r_\sigma]$ and $\tr[B R_\sigma] = \prod_{i=1}^n\tr[b_i r_\sigma]$, which are easy to calculate.

\section{Examples}\label{app:examples}


\subsection{Chaotic spin chains and random local circuits}\label{sec:precisespinchain}
In this appendix, we revisit the random local circuit discussion of~\cref{sec:spchm} in a more quantitative way. 
\par
As stated before, chaotic spin chains and random local circuits provide prime examples for the phenomenon of high temperature state transfer, which could be experimentally realized with the current technology. Again, we primarily discuss the local random circuit models, and point out that chaotic spin chains should behave similarly.  
\par
 Consider a one-dimensional chaotic spin chain (higher-dimensional extensions are straightforward), where the $A$ qubits (message qubits) are spread among the $B$ qubits (carrier qubits), such that the distance of any two $A$ qubits is larger than the $2v_B t$, where $v_B$ is the butterfly velocity and $t$ is the time of the evolution. We assume the random circuit evolution (or the chaotic Hamiltonian evolution) turns a Pauli operator into a collection of random strings on its light-cone (Note that this should be exactly true for random circuit models). Now, we discuss the size distribution. Start with operator $P_A$ at $t=0$, with the initial size $l_0 = |P_A|$. At later times, each non-identity single Pauli operator in $P_A$ expands to a Pauli string of average size equal to $2\times 3v_Bt/4$ (which is $3/4$ of the light-cone size $2 v_Bt$). If $v_Bt\gg1$ such that the variance of the size is negligible, we will have
\begin{equation} q_{l_0}(l) = \delta(l,l_0\times 3v_Bt/2),
\end{equation} see~\cref{fig:Summary_of_everything}.(b). This shows that (See Eq. 12 of~\cite{firstpaper} for the definition of $\tilde q_{l_0}(g)$)  
\begin{equation}
    \tilde q_{l_0}(g)=e^{-il_0(4g/3)3v_Bt/2n}=e^{-il_0(g/n)2v_Bt}.
    \end{equation}
As long as $(g/n)2v_Bt$, i.e., the total $g$ charge on the light cone, is an odd multiple of $\pi$ the norm of $\tilde q_{l_0}(g)$ is equal to one and it has an alternating sign as a function of $l_0$. Hence, this system can teleport all of the qubits in $A_L$ with high fidelity according to Eq.~14 of~\cite{firstpaper}. We point out that the standard deviation of the size distribution is proportional to $\sqrt{l_0 v_B t}$, which should be much smaller than $v_B t$, the distance between peaks in the size distribution (see~\cref{fig:Summary_of_everything}.(b)). Therefore the $m$-qubit fidelity is large only if $m\ll \sqrt{v_Bt}$.

\subsection{Random non-local Hamiltonian in GUE or GOE ensemble}\label{app:mainappGUE}
Now, we focus on the actual Hamiltonian evaluations. We start with non-local completely random Hamiltonians. The benefit of using such systems is that we have complete analytical control over the calculations, and one can easily go to the infinite system size and arbitrary parameters. The calculations are usually tedious, computerized, and use the techniques described in~\cref{sec:HaarPre}. In~\cref{app:GOE_GUE}, we compute the state transfer average output state as well as some two-point function. These quantities are generically complex, and we only report the final results. Size distribution, which is calculated in~\cref{app:appsizedist}, is more manageable and we spend more time explaining the details of its calculations. Therefore, we refer the reader interested to learn the techniques of our calculations to~\cref{app:appsizedist}.
\subsubsection{Average evolution with GUE or GOE Hamiltonians}\label{app:GOE_GUE}
Here we report the formulas for the infinite and finite temperature GOE set-up. We believe that there is no significant difference between the GUE and GOE, except that for GOE we have the luxury of having $H_L=H_R=H$. Again, we are going to assume that there are $n$ qubits on each side, and the coupling act on all $n$ of the qubits, and there is only one qubit to send ($m=1$).~\footnote{We will ignore the insignificant $1/n$ corrections caused by action of the coupling on the message qubits.} 
\par
Let us relax the condition of acting on left and right at the same time, and assume that the left qubit is inserted at time $-t_L$ and the observation time is $+t_R$. The one qubit teleportation channel is given by:
\begin{equation}\label{eq:tel_signal_GOE}
    \Psi_{S_L} \rightarrow [1-\lambda(\beta, t_L,t_R)] \tau + \lambda(\beta, t_L,t_R) Y\Psi Y  = Y \Delta_{\lambda(\beta,t_L,t_R)}( \Psi_{S_L}) Y,
\end{equation}
where we explicitly relaxed the condition of acting by the same times on the left and right, and $\lambda(\beta,t_L,t_R)$ is the parameter of the depolarizing channel. In the $n\to \infty$ limit, $\lambda(\beta,t_L,t_R)$ is given by the following formula:
\begin{multline}\label{eq:MASTER_GOE}
    \lambda(\beta,t_L,t_r)=\\
    \Re\Bigg[{\frac12(1-e^{ig})f(-\beta/2+i(t_R- t_L)) f(i( t_R- t_L))f(-\beta/2))}+
    {f(it_L)f(it_R)f(-\beta/2+it_L)}\times\\
    \bigg(
    \frac12(1-e^{ig})f(-\beta/2+it_R)
  +(1-e^{ig}) f(i t_L)f(it_R) f(-\beta/2-it_L)\\
  -(1-e^{ig}) f(it_L) f(-\beta/2+i(t_R-t_L))
 -(1-\cos(g))f(it_R)f(-\beta/2)
    \bigg)\Bigg]/f(-\beta)
\end{multline}
As in the main text, $f(z) = 2\,I_1(z)/z$ is the modified Bessel function of the first kind of order one. This is a relatively complicated formula, but it is valid for all $g$, $\beta$, $t_L$, $t_R$ and completely characterizes the teleportation channel  through~\cref{eq:tel_signal_GOE}. As one decreases $n$, then the first corrections to the formula above are obtained by the following substitutions of the $g$ dependent coefficients:
\begin{align}
\text{For finite $n$:}\quad &(1-e^{ig}) \rightarrow \cos{(g/n)}^n\left[\cos{(n/k)}^n-e^{ig}\right], \\
\text{and}\quad  &(1-\cos(g))\rightarrow \cos(g/n)^{n}\left[\cos(g/n)^n-\cos(g)\right]. \end{align}
For very small $n$, e.g., $n=6$ or $n= 7$, there are more corrections due to
(1) the distribution of eigenvalues starting to deviate from the semi-circle distribution, and (2) the details of the eigenvalue statistics becoming relevant, as the separation between the time scale dictated by the inverse eigenvalue spacing and the time scale of interest $t=O(1)$ gets less significant.
\par
In the $n\rightarrow \infty$ limit, we can focus on to the infinite temperature limit, $\lambda(t_L,t_R) := \lambda(\beta=0, t_L,t_R)$:
\begin{multline}\label{eq:formula_GOE_inftemp}
    \lambda(t_L,t_R)=\\
    \frac{1-\cos(g)}{2}\Big[f(-i t_R+i t_L)^2+2f(i t_L)^4 f(i t_R)^2-f(i t_L)^2f(i t_R)^2 -2 f(i t_L)^3 f(i t_R) f(-i t_R+i t_L)  \Big],
\end{multline}
which for $t_L=t_R$ simplifies to:
\begin{align}\label{eq:formula_GOE_sametime}
    \lambda(t)=\frac{1-\cos(g)}{2}\Big[1 +2f(i t)^6 -3 f(i t)^4\Big],
\end{align}
\par
One can see that the same time teleportation signal $\lambda(t,t)$ starts from $0$, and plateaus at the value of $(1-\cos(g))/2$, which is the same as the random unitary result.
See~\cref{fig:same_time_GOE}. The plateau starts at time around $t=2.7$, which is around the same time as the scrambling time or the thermalization time (which are both more or less the same for GOE non-local Hamiltonians).
\begin{figure}
    \centering
    \includegraphics[width=0.6\textwidth]{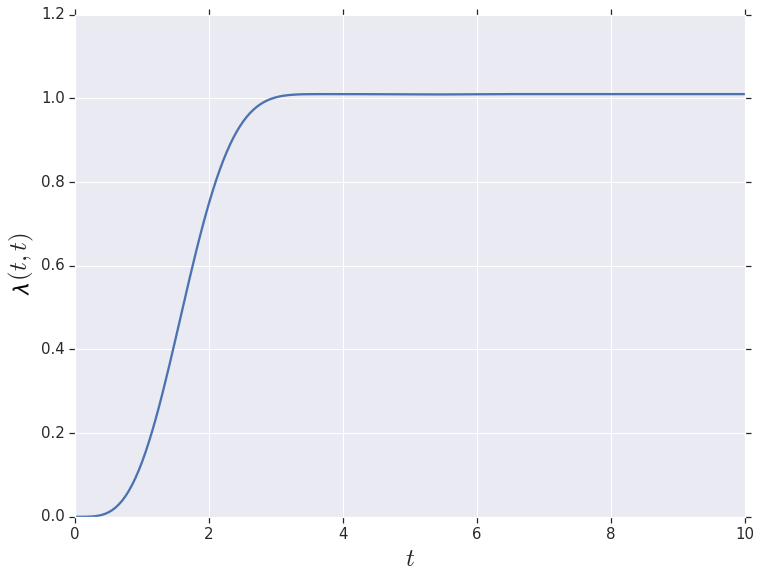}
\caption{Plot of the teleportation signal $\lambda(t,t)$ as defined in~\cref{eq:tel_signal_GOE}, for Hamiltonian coming form the GOE ensemble. This is plotted for $g=\pi$.}\label{fig:same_time_GOE}
\end{figure}
\par
It is interesting to look at the problem $t_L \neq t_R$. Indeed, we can fix the time that we send the signal ($t_L$ fixed) and vary the observation time $t_R$. See~\cref{fig:infgoe}. If one sends the signal before the scrambling time ($t_L\lesssim 3$), the signal is weaker, but still comes out at near scrambling time. However, for times larger than scrambling time the signal comes out at $t_R=t_L$.

We also report the results of two-point function calculations:
\begin{itemize}
\item \emph{two-point function calculations I}.
We can compute the following two-point function for GUE/GOE Hamiltonians:
\begin{align}\label{eq:2pt1}
\bra T P_{R}(t) e^{igV}  P_{L}(-t)\ket T,
\end{align}
in the limit of very large system size $n\rightarrow \infty$, and with the assumption that $P$ is a single-qubit Pauli.
This can be evaluated using Eq.~12 of~\cite{firstpaper} and the result that we will derive in next section (\cref{eq:GUE_winded_size}) to:
\begin{align}\label{eq:GUE_winded_size_twopt}
\frac{e^{ig} f(it-\beta/2)^2 f(-it)^2 + \left[f(-\beta/2)^2-f(it-\beta/2)^2f(-it)^2\right]}{f(-\beta)}.
\end{align}
\item\emph{two-point function calculations II}.
We analyze a different two-point function:
\begin{align}\label{eq:2pt2}
\bra T e^{-igV} P_{R}(t_R) e^{igV}  P_{L}(-t_L)\ket T,
\end{align}
for the GOE ensemble. For most $k$ local systems the size distribution of the thermal state has a small variance, and therefore the thermofield double state is close to being an eigenvalue of $e^{igV}$. This means that~\cref{eq:2pt2} is simply proportional to the simpler two-point function~\cref{eq:2pt1}. This is not true for non-local random Hamiltonians as their thermal state has two branches, one at the identity matrix and one a narrow Gaussian at $3n/4$. Hence, it is not possible to easily infer~\cref{eq:2pt2} from~\cref{eq:2pt1}. With direct calculations,~\cref{eq:2pt2} evaluates to:
\begin{align}
  &\frac{1}{f(-\beta)}\Bigg[2f(-\beta/2+it_L)f(-it_L)f(it_R)f(-it_R)f(-\beta/2) \\
  &-f(-\beta/2-it_R+it_L)f(-\beta/2)f(+it_R-it_L)\\&
  -f(-\beta/2+it_L)f(-it_L)f(-\beta/2+it_R)f(-it_R) \\
  &+f(-\beta/2-it_R+it_L)f(-\beta/2+it_R-it_L)\\&
     -e^ {i g} f(-\beta/2+it_L)f(-it_L)f(+it_R)f(-it_R)f(-\beta/2) \\
  &   + e^ {i g} f(-\beta/2+it_L)f(-it_L)f(-\beta/2+it_R)f(-it_R)  \\
  &  -e^{-i g} f(-\beta/2+it_L)f(-it_L)f(+it_R)f(-it_R)f(-\beta/2) \\
  &  +e^{-i g} f(-\beta/2-it_R+it_L)f(-\beta/2)f(+it_R-it_L)\Bigg].
\end{align}
Again, the proof of this formula requires lengthy computerized calculations involving Brauer algebra contractions.
\end{itemize}
\subsubsection{Size distribution}\label{app:appsizedist}
In this section, we present a more detailed derivation of the size distribution for the case of random Hamiltonian time evolution. Let us momentarily focus on the GUE ensemble, and near the end we mention the required modifications for GOE.
We start with a calculation of the following form:
\begin{align}\label{eq:definition_s}
    s(\alpha, \beta, \vu , \vv) =\frac{1}{d^2} \EE_{H \sim \mathrm{GUE}} {\left(\tr{\left[e^{\alpha H}\,P_\vu \,e^{\beta H}\, P_\vv\right]}\right)^2}.
\end{align}
We can easily see that the size distribution of $\rho_\beta^{1/2} e^{itH}=e^{itH-\beta/2H}$ can be read from the $s$ function by proper substitution of its parameters.
\par
We can diagonalize $H$, $H=U D U^\dagger$, where $U$ is from the Haar measure on the unitary group, hence the name Gaussian Unitary Ensemble. One can check that the eigenvalues of the diagonal matrix $D$ have a semi-circle distribution centered at $0$ with radius $1$. In large the $n$ limit, the details of the semicircle distribution would be unimportant for the times of $O(1)$ that we study the problem for, and it can be assumed to be a smooth semi-circle. The detailed eigenvalue distribution only manifests itself at exponential time, and is a subject of important investigations in the study of holographic dualities (see e.g. \cite{cotler2017black,saad2018semiclassical}).
In drawings,~\cref{eq:definition_s} is:
\begin{center}
\includegraphics[width=11.2 cm]{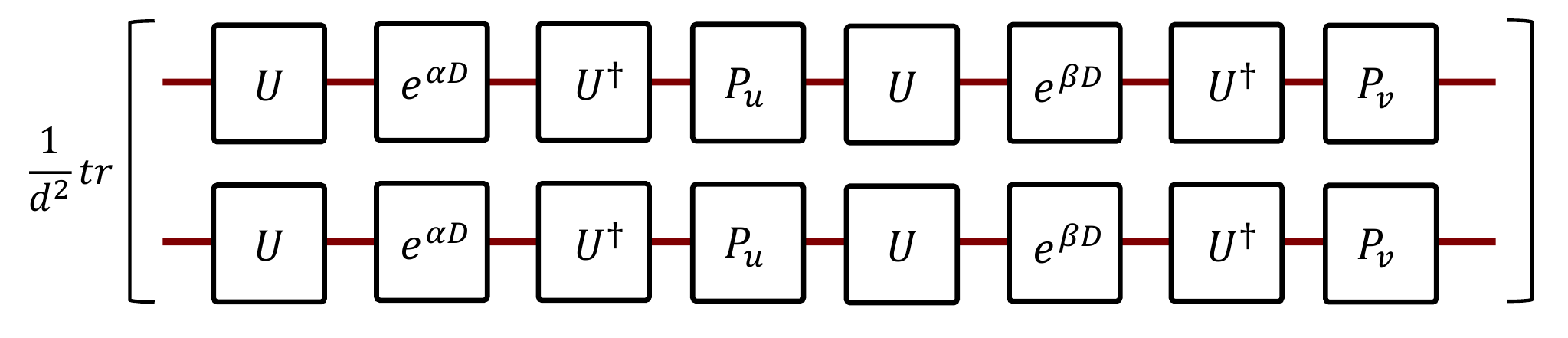}
\end{center}
With simple manipulations, this can be re-drawn as:
\begin{center}
\includegraphics[height=5 cm]{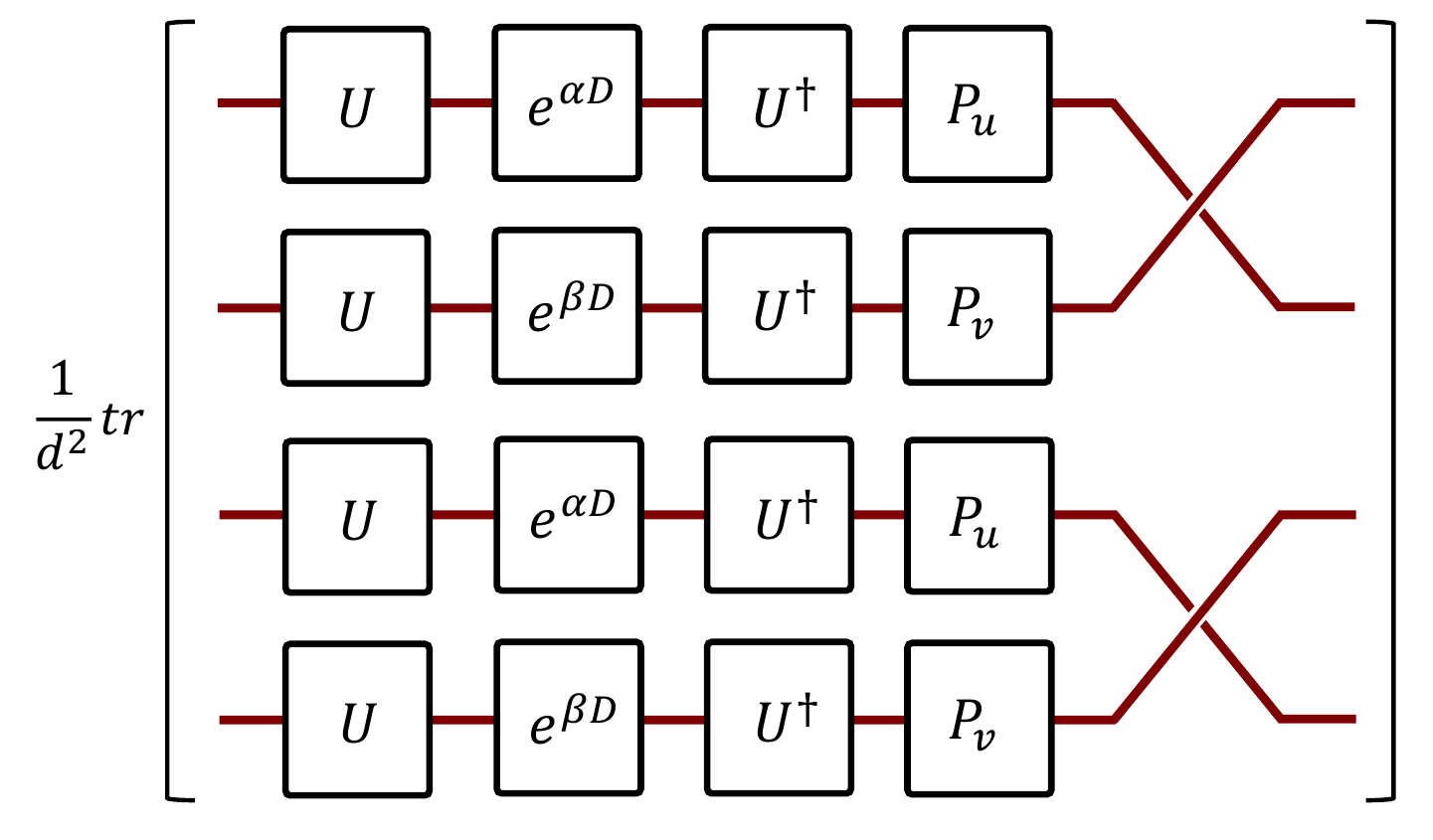}
\end{center}
Define the permutation $\kappa := \PFour{2}{1}{4}{3}$ as the permutation that swaps the first and second Hilbert spaces, and also the third and fourth (it has two swaps). Then, this shows that the quantity of interest is $\frac{1}{d^2} \tr{\left(U^{\ot 4}\left[e^{\alpha D}\ot e^{\beta D}\ot e^{\alpha D}\ot e^{\beta D}\right]U^{\dagger,\ot 4} R_\kappa\right) }$. Now, using the equation~\cref{eq:haar average}, we obtain the following formula for~\cref{eq:definition_s}:
\begin{center}
\includegraphics[height=5 cm]{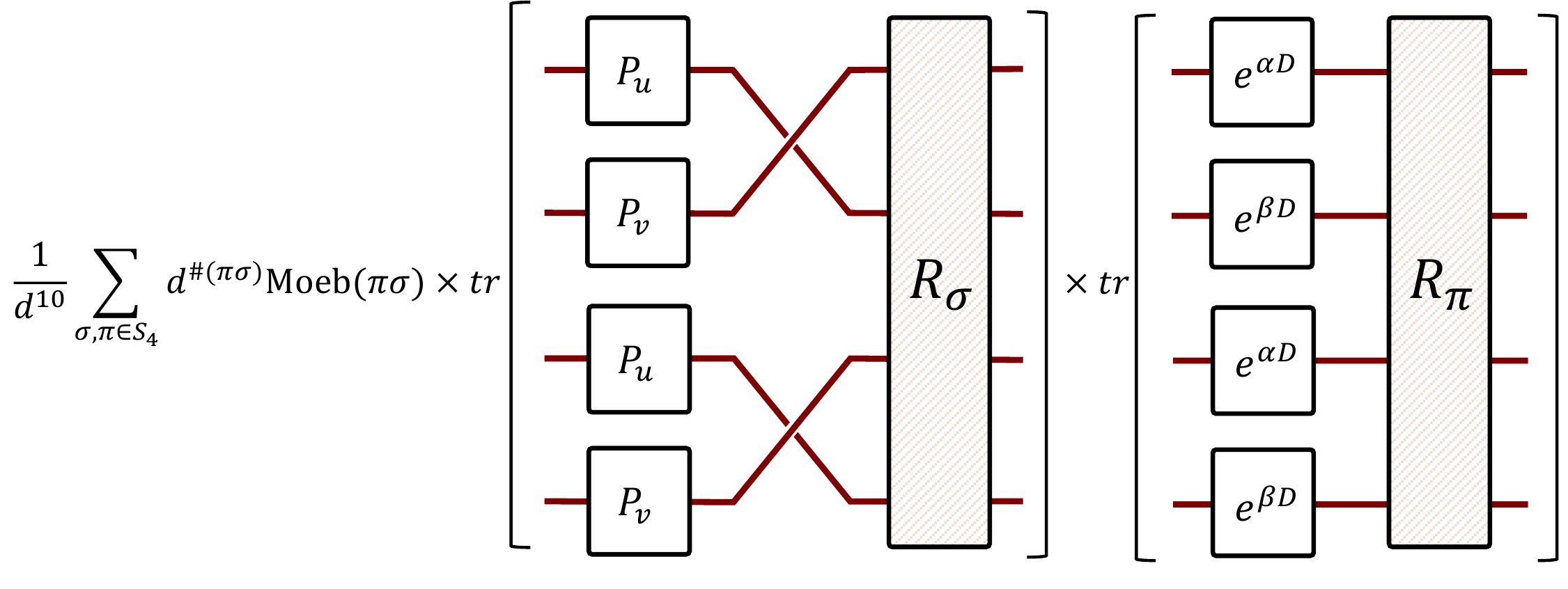}
\end{center}
There are $4! \times 4!$ terms in the sum, but most of them are subleading in $d=2^{n}$ (in fact subleading in $2^{2n}$, as the corrections to the Weingarten formulas are of order $d^2$ (see~\cite{collins2006integration}), and can be neglected.
\par
The calculation of each summand term is straightforward. For instance, if one uses $\sigma = \PFour{2}{3}{1}{4}$, and $\pi = \PFour{1}{2}{4}{3}$, then the corresponding term will be:

\begin{center}
\includegraphics[height=5 cm]{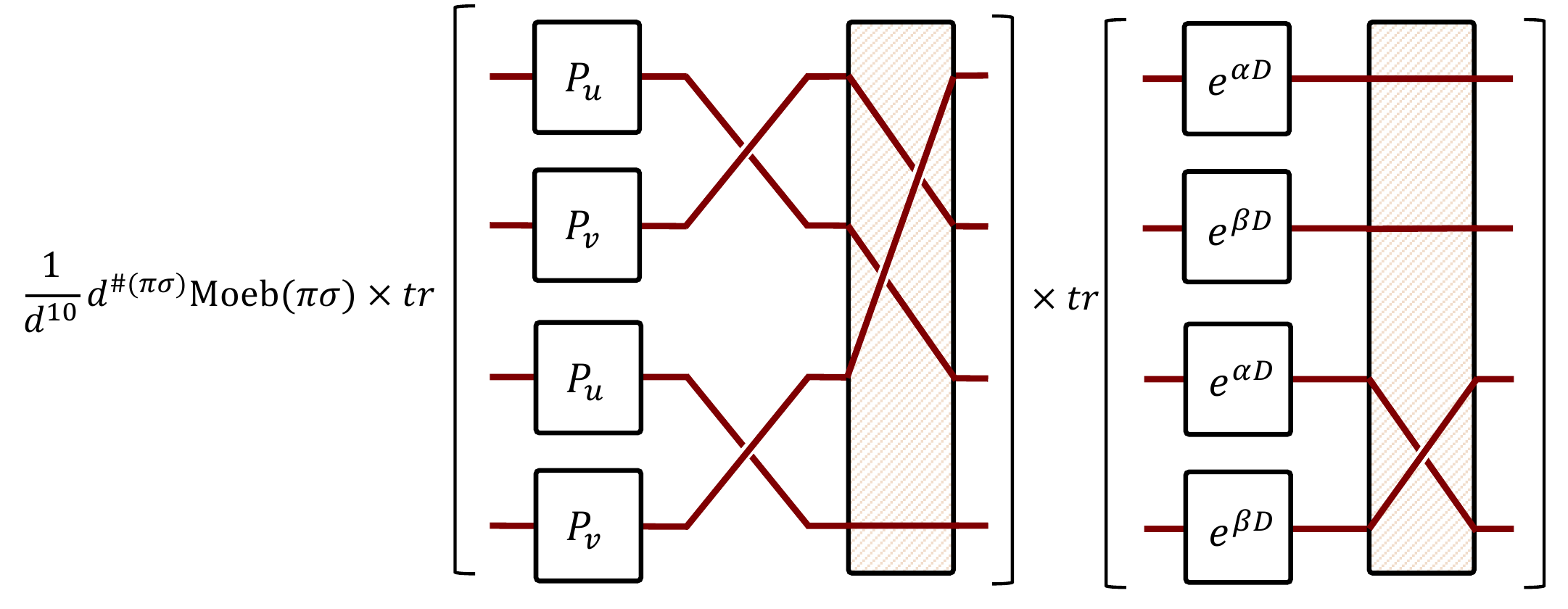}
\end{center}
which is equal to
\begin{equation}\frac{1}{d^{10}} d^{\ell(\pi\sigma)} \Moeb(\pi\sigma)\left(\tr[P_\vu P_\vu P_\vv ]\tr[ P_\vv ]\right)\left(\tr[e^{\alpha D}]\tr[e^{\beta D}]\tr[e^{(\alpha+\beta) D}]\right). \end{equation}
This can be evaluated. One is able to read off the order of this term by replacing the $\tr$ with the normalized $\Tr$:
\begin{equation}\frac{1}{d^{10}} d^{\ell(\pi\sigma)} \Moeb(\pi\sigma)\left(d^2 \Tr[P_\vu P_\vu P_\vv ]\Tr[ P_\vv ]\right)\left(d^3\Tr[e^{\alpha D}]\Tr[e^{\beta D}]\Tr[e^{(\alpha+\beta) D}]\right), \end{equation}
We define $\Tr (e^{\alpha D})=\Tr(e^{\alpha H}) = f(\alpha)$, with $f(z)$ is Fourier transform of the semi-circle distribution:
\begin{align}\label{eq:deff}
    f(z) = 2\,I_1(z)/z,
\end{align}
and $I_1(z)$ is the modified Bessel function of the first kind of order $1$.
Note that each $\Tr$ is of $O(d^0)$, so the whole term is $O(d^{-5} d^{\ell (\pi \sigma)})$. We have $\pi \sigma = \PFour{2}{3}{4}{1}$, so $\ell(\pi \sigma) =1$, the whole term is of order $d^{-4}$.
\par
Similarly for general $\pi$ and $\sigma$ we have that the corresponding term is of the order of \begin{equation}d^{-10} d^{\ell(\pi\sigma)} \times d^{\ell(\kappa \sigma)}\times d^{\ell(\pi)}=d^{2-[\dist(e,\pi)+\dist(\pi,\sigma^{-1})+\dist(\sigma^{-1},\kappa)]}.\end{equation}
From the triangle inequality for the permutation distance metric, we have that $\dist(e,\pi)+\dist(\pi,\sigma^{-1})+\dist(\sigma^{-1},\kappa)\geq \dist(e,\kappa) = 2$. Hence, no term can be of positive order in $d$.
\par
The complete calculation of $s(\alpha,\beta, \vu,\vv)$ function to the leading order is tedious and can be computerized. We report the final result, leading order in $d$, as calculated by computer:

\begin{align}\label{eq:complete_formula}
    s(\alpha, \beta, \vu,\vv) :=
       \begin{cases}
    f(\alpha+\beta)^2,               & \text{if $\vu=\vv = 0$},\\
    f(\alpha)^2f(\beta)^2,               & \text{if $\vu=\vv\neq 0$},\\
    \frac{1}{d^2} [f(2\alpha+2\beta) -f(\alpha+\beta)^2], & \text{if $\{\vu= 0, \vv \neq 0\}$ or $\{\vu\neq 0, \vv = 0\}$},\\
   \frac{1}{d^2} [f(2\alpha)f(\beta)^2 + f(2\beta)f(\alpha)^2 \\
   \ \ \ \ +f(\alpha+\beta)^2 - 3f(\alpha)^2f(\beta)^2 ], & \text{if $\vu\neq 0,\vv\neq 0, \vu\neq \vu, [\vu,\vv]=0$},\\
    \frac{1}{d^2} [f(\alpha)^2f(\beta)^2+f(\alpha+\beta)^2\\
    \ \ \ \ -f(2\alpha)f(\beta)^2-f(2\beta)f(\alpha)^2 ], & \text{if $\vu\neq 0,\vv\neq 0, \vu\neq \vu, [\vu,\vv]\neq 0$}.
  \end{cases}
\end{align}
It is easy to read the \emph{winding} size distribution of $\rho_\beta^{1/2} P_\vu(t)$, defined in~\cref{eq:size_dist}. If
\begin{multline}
\rho_\beta^{1/2} P_\vu(t) = \frac{e^{(it-\beta/2)H}P_\vu e^{-itH}}{\sqrt{d f(-\beta)}}=\frac{1}{\sqrt d} \sum_\vv c_\vv P_\vv  \Rightarrow \\
\EE\, (c_\vv)^2= \frac{1}{d^2f(-\beta)} \EE{\left(\tr{\left[e^{(it-\beta/2) H}\,P_\vu \,e^{-itH}\, P_\vv\right]}\right)^2}=\frac{s(it-\beta/2,-it,\vu,\vv)}{f(-\beta)}.
\end{multline}
This means that for large values of $n$, the winding size distribution is given by:
\begin{multline}\label{eq:GUE_winded_size}
q_{|\vu|}(l) =\\
\frac{\delta_{l,|\vu|} f(it-\frac{\beta}{2})^2 f(-it)^2 + \Norm(l;\mu=\frac{3n}{4},\sigma =\sqrt{\frac{3n}{4}}) \left[f(-\frac{\beta}{2})^2 -f(it-\frac{\beta}{2})^2f(-it)^2\right]}{f(-\beta)}.
\end{multline}
One can see that the function $f$ is real on the real and imaginary line, but becomes complex on the other parts of the complex plane. Therefore, the winding size distribution is real for $\beta=0$, but it acquires a phase for non-zero values of $t$ and $\beta$.
\par
The calculations of GOE ensemble are similar. The main difference that one should use the elements of the Brauer algebra instead of permutations, and use the Weingarten calculus for the orthogonal group. A complete and easy-to-read discussion of these techniques can be found in~\cite{collins2006integration}. The eigenvalue distribution of the GOE ensemble is the same as GUE, and follows the semi-circle law~\cite{liu2000statistical}.

\subsection{2-local Brownian circuits}\label{app:Brownian_Calc}
In this section, we focus on a different model of time evolution. Let us consider the model of fast scrambling studied in~\cite{lashkari2013towards}, a $2$-local Brownian circuit. The main technical tool here is the It\^o calculus.
\par
We assume that the system is at infinite temperature, and the time evolution is given by a rapidly changing Hamiltonian. Suppose that at each instance of time, an operator evolves by the following rule:
\begin{align}\label{eq:brownian_gov}
O_t\rightarrow O_{t+dt} =e^{i dH} O_t e^{-i dH}.
\end{align}
The infinitesimal time evolution $dH$ is a two local Hamiltonian which has the following form:
\begin{equation} dH =\frac{1}{\sqrt {8 n}} \sum_{\vv} dh_{\vv} {P_\vv}, \end{equation}
where $\vv$ is the vector indicating the Pauli term in the Hamiltonian.
We assume that $dh_{\vv}$ is only non-zero for terms that have $2$ Paulis, for which it has mean $0$ and $\EE\, dh_{\vv}^2 = dt$.
We would like to read off the operator growth structure of such systems. First, we need to expand $O_t$ in the Pauli basis:
\begin{equation} O_t = \sum_{\vu} c_\vu(t) P_\vu.\end{equation}
As discussed before, $c_\vu(t)$ averages to $0$, and we are interested in $q_\vu(t) = \EE\, c_\vu(t)^2$. To access $\EE c_\vu(t)^2$, look at $O_t \ot O_t$:
\begin{equation} O_t\ot O_t = \sum_{\vu\vv} c_\vu(t) c_\vv(t) P_\vu \ot P_\vv.\end{equation}
It is not hard to see that if $O_{t=0}$ is a single Pauli operator, then after averaging only the diagonal term ($\vu=\vv$) survive in the above sum. So we get:
\begin{equation}\EE \,O_t\ot O_t = \sum_{\vu} (\EE\, c_\vu(t)^2) P_\vu \ot P_\vu = \sum_{\vu} q_\vu(t) \,P_\vu \ot P_\vu. \end{equation}
Now, we have:
\begin{multline}
   \EE \,O_{t+d t} \ot O_{t+d t} =
   \EE \Big(\left[ e^{i dH}\ot e^{i dH}\right] \left(O_{t} \ot O_{t}\right)\left[e^{-i dH}\ot e^{-i dH}\right]\Big) = \\
   \sum_{\vu} q_\vu(t) \,\EE \, \Big(\left[ e^{i dH}\ot e^{i dH}\right] P_\vu \ot P_\vu \left[e^{-i dH}\ot e^{-i dH}\right]\Big).
\end{multline}
Using the rules of It\^o calculus,
\begin{multline}
  \EE\,\Big(\left[ e^{i dH}\ot e^{i dH}\right] P_\vu \ot P_\vu \left[e^{-i dH}\ot e^{-i dH}\right]\Big) = P_\vu \ot P_\vu +\\
   \EE\,\Bigg(\left[dH\ot I +I \ot dH\right]  P_\vu \ot P_\vu  \left[dH\ot I +I \ot dH\right]   -\frac12\Big\{\left[dH\ot I +I \ot dH\right]^2   , P_\vu \ot P_\vu \Big\}\Bigg) =\\
   P_\vu \ot P_\vu +
   \frac {\EE\, (dh_\vv^2)}{8 n}\sum_{\vv}
   \Bigg(\left[P_\vv\ot I +I \ot P_\vv\right]  P_\vu \ot P_\vu  \left[P_\vv\ot I +I \ot P_\vv\right]   \\
   -\frac12\Big\{\left[P_\vv\ot I +I \ot P_\vv\right]^2   , P_\vu \ot P_\vu \Big\}\Bigg) =\\
     P_\vu \ot P_\vu + \frac{dt}{2 n} \sum_{\vv, P_\vv\,\,\text{is a 2 Pauli operator}}{\left(P_{\vv+\vu}\ot P_{\vv+\vu} -P_{\vu}\ot P_{\vu}\right)\times\delta(P_\vu  \& P_\vv\text{ anti-commute})},
\end{multline}
where we use the notation that $\delta(P_\vu \& P_\vv\text{ anti-commute})$ is equal to one if $P_\vu$ and $P_\vv$ anti-commute and zero otherwise.
Hence, we get the following formula by change of variables in the first sum:
\begin{equation}\frac{d}{dt}\EE \left(O_t\ot O_t \right) = \frac{1}{2 n} \sum_{\vu}\sum_{\vv, P_\vv\,\,\text{is a 2 Pauli operator}}{P_\vu \otimes P_\vu \times \left[ q_{\vu+\vv}(t)-q_{\vu}(t)\right]\times \delta(P_\vu \& P_\vv\text{ anti-commute})}.\end{equation}
This gives the following simple update rule for $q_\vu(t)$:
\begin{equation}\frac d {dt }q_\vu(t) =\frac{1}{2 n} \sum_{\vv, P_\vv\,\,\text{is a 2-Pauli operator}}{\left[ q_{\vu+\vv}(t)-q_{\vu}(t)\right]\times \delta(P_\vu \& P_\vv\text{ anti-commute})}.\end{equation}
Let us define, like before, the size on operators of certain size by $q_l(t)=\sum_{|\vu|=l} q_\vu(t)$. Using some combinatoric equalities one arrives at $q_l(t)$:
\begin{multline}
  \frac d{dt}q_l(t)
= n\Bigg[
3\bigg(\frac l n-\frac 1 n\bigg)\bigg(1-\frac l n  +
\frac 1 n \bigg) q_{l-1}(t) +\\ \frac l n \bigg(\frac l n +\frac 1 n\bigg)q_{l+1}(t)-\bigg[3\frac l n\bigg(1-\frac ln \bigg)+ \frac l n \bigg(\frac l n-\frac 1 n\bigg)\bigg] q_{l}(t)\Bigg].
\end{multline}
If one defines $\Phi(x,t) = q_{nx}(t)$, and $\Delta x = 1/n$, we have for $x=k \Delta x$, ($k \in\NN$),
\begin{multline}
    \frac d{dt}\Phi(x,t) =
    \bigg[3\big(x-\Delta x \big)\big(1-x +\Delta x \big) \Phi(x-\Delta x,t) + \\
    x \big(x +\Delta x\big)\Phi(x+\Delta x,t) -\big[
    3 x\big(1-x \big)+ x \big(x-\Delta x\big)
    \big] \Phi(x,t)\bigg]/{\Delta x}.
\end{multline}
See~\cref{fig:brownian_circ_2} and~\cref{fig:brownian_circ} for solutions.
\par
From~\cref{fig:brownian_circ_2} one can see that the operator size distribution $\Phi$ is to a great extent independent of $n$, and follows the same form as long as one modifies $t$ by adding a logarithmic term in $n$ into $t$. In other words, the function $\hat \Phi(x,t)=\Phi(x,t+\log(n)/3)$ seems to be $n$-independent in the large-$n$ limit and at times slightly smaller than the scrambling time. It is important to point out that for very late times the distribution becomes Gaussian with standard deviation proportional to $\sqrt{n}$, hence the diagrams are much sharper for larger $n$ and late times. 

\begin{figure}
    \centering
    \begin{subfigure}{}
        \includegraphics[width= 5.6 cm]{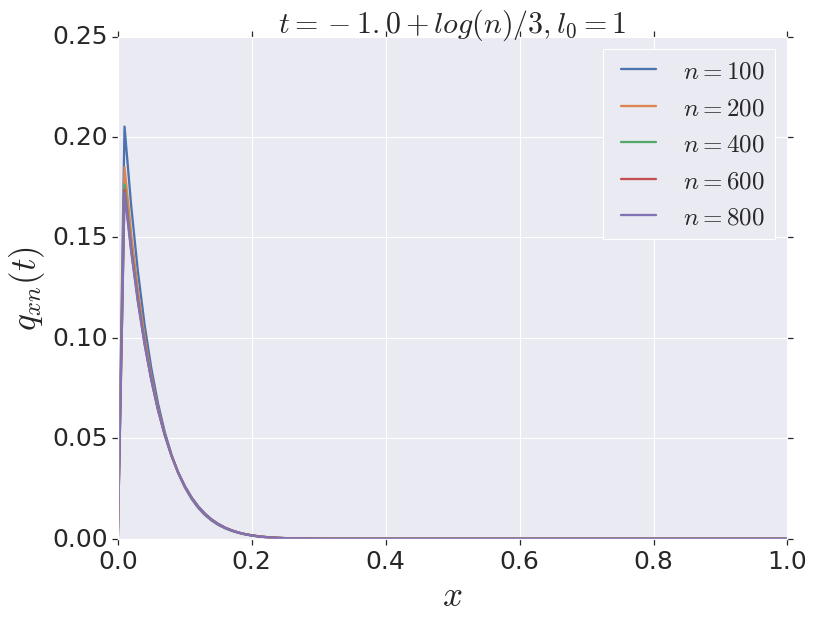}
    \end{subfigure}
    \begin{subfigure}{}
        \includegraphics[width= 5.6 cm]{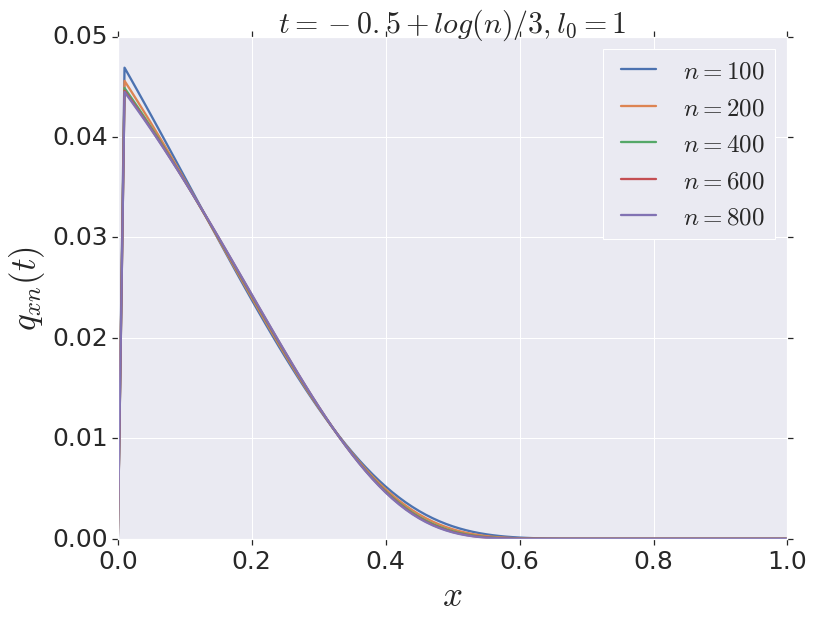}
    \end{subfigure}
    \begin{subfigure}{}
        \includegraphics[width= 5.6 cm]{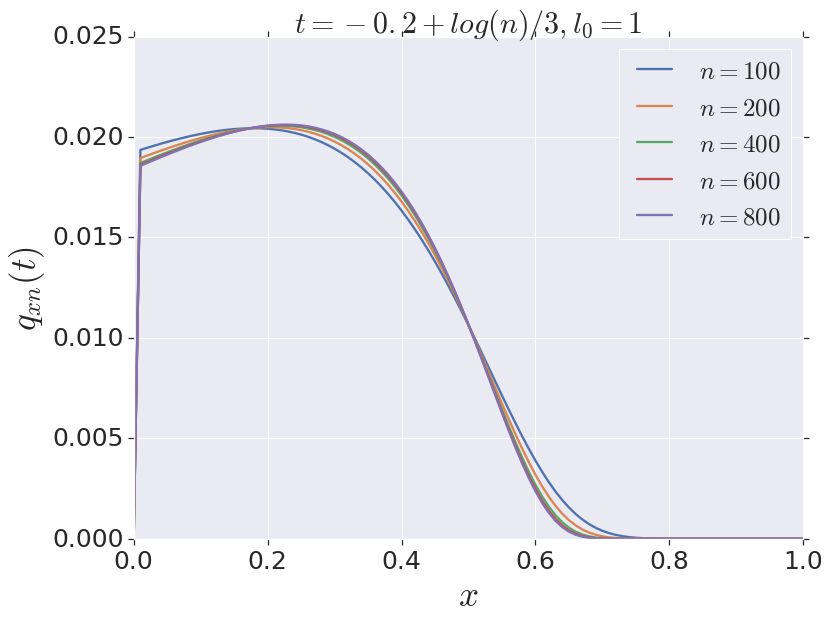}
    \end{subfigure}
    \begin{subfigure}{}
        \includegraphics[width= 5.6 cm]{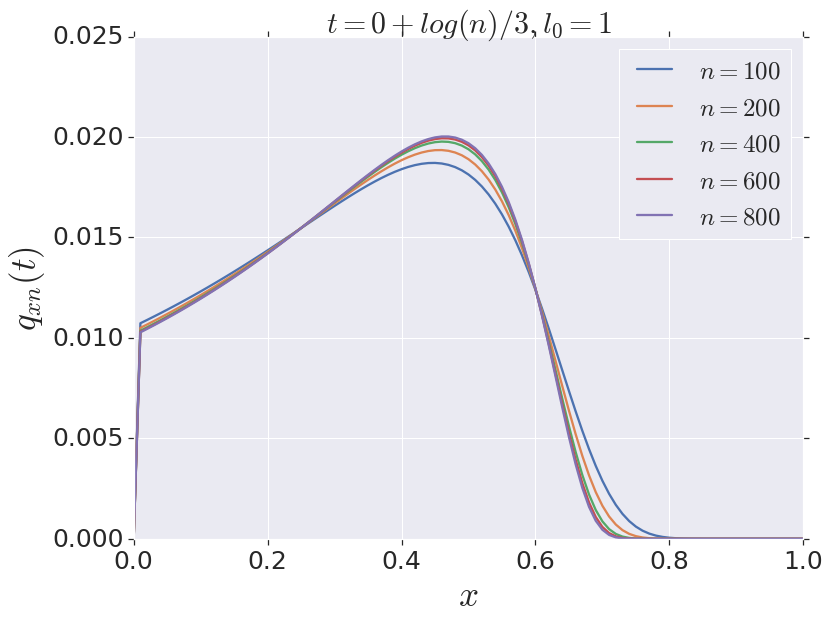}
    \end{subfigure}
    \begin{subfigure}{}
        \includegraphics[width= 5.6 cm]{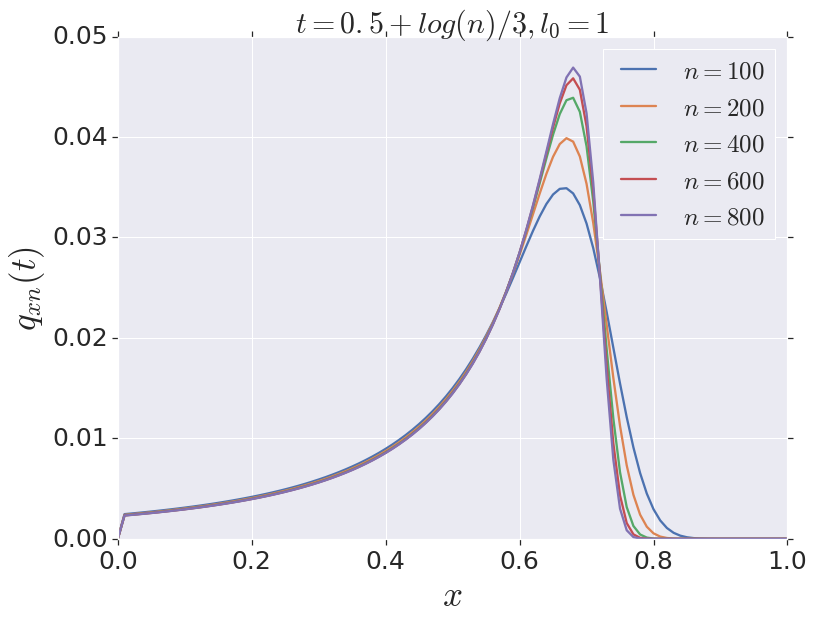}
    \end{subfigure}
    \begin{subfigure}{}
        \includegraphics[width= 5.6 cm]{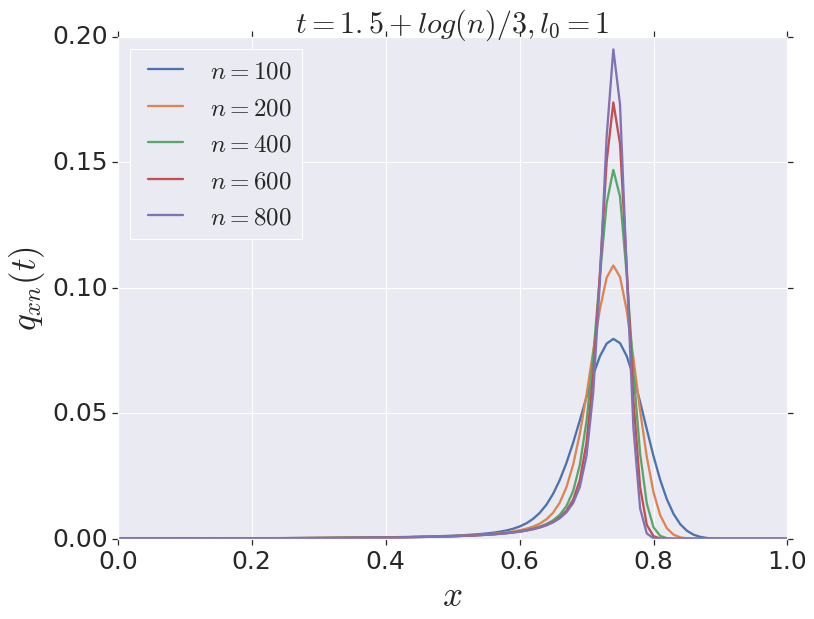}
    \end{subfigure}
\caption{Brownian circuit operator growth as function of the number of qubits and the initial time. }\label{fig:brownian_circ_2}
\end{figure}

\subsection{Size winding in the SYK model from Qi-Streicher formalism}\label{app:SYK}
In this section, we use the machinery developed by Qi and Streicher~\cite{qi2019quantum} to study the SYK model in the large $q$ limit. We demonstrate that in that in some limit (to be specified later), size winding happens in a perfectly linear and \emph{detailed} way.
\par
In~\cite{qi2019quantum} the authors study the size operator defined with the conventional absolute value notation (see~\cite{qi2019quantum}, section 3.1).
Now, we contrast their definition and what we defined as the winding size distribution. Let $\rho_\beta^{1/2} \Psi_1(t) = \sum_{u}c_u \Psi_u$, where $\Psi_1$ is a single fermion operator and the sum is over the index set $u\in \{0,1 \}^n$, with each $u_i$ indicating if the compound fermionic operator $\Psi_u$ has a fermion on site $i$. 
For the convenience of the reader, we copy definitions from earlier today.
The conventional size distribution of the operators is implicitly defined in~\cite{qi2019quantum} by:
\begin{equation}\mathcal P_l\Big[\rho_\beta^{1/2} \Psi(t)\Big] = \sum_{|u|=l} |c_u|^2.\end{equation}
The \emph{winding} size distribution is defined without the absolute value:
\begin{equation}q_l\Big[\rho_\beta^{1/2} \Psi(t)\Big] = \sum_{|u|=l} c_u^2. \end{equation}
\par
\emph{Size winding}, in its perfect form, is the following ansatz for the operator wavefunction:
  \eqn{c_P = e^{i \alpha |P|/n} r_P. \quad r_P \in \RR. }
Our goal in this appendix is to check from an SYK calculation that size winding holds by verifying that 
\eqn{q(\ell) = \mathcal{P}(\ell) e^{i \gamma \ell} }
for some $\gamma$.
Qi and Streicher define the \emph{growth distribution} $K_n^\beta[\Psi(t)]$, such that the size distribution of  $P_n[\Psi(t)\rho_\beta^{1/2}]$ is given as the convolution of the size distribution of $\rho_\beta^{1/2}$ and the growth distribution:
\begin{align}\label{eq:conv_form}
\mathcal P_n\Big[\rho_\beta^{1/2}\Psi(t)\Big] = \sum_m \mathcal P_{n-m}\left[\rho_\beta^{1/2}\right]K_m^\beta[\Psi(t)].
\end{align}

For teleportation purposes, we are interested in the winding size distribution $q(l)$ defined in~\cref{eq:size_dist}. 
Indeed, as $\rho_\beta$ is Hermitian, there is no difference between $q(l)$ and $P_m$ for the thermal state. 
By explicitly computing the two-point function, the authors of~\cite{qi2019quantum} derived the form of the growth distribution.
\par
The distribution of $K_l^\beta$ in~\cite{qi2019quantum} can be written as a negative binomial distribution. A negative binomial distribution, $\NB$, is a distribution with two parameters $r$ and $p$, and is defined as:
\begin{equation} \NB(l;r,p) := (1-p)^r \binom{l+r-1}{l} p^l.\end{equation}
 This distribution is generically useful for fitting overdispersed data, as a $\NB$ can have a standard deviation that as large as its mean. The mean and variance of a $\NB$ are given by:
 \begin{equation}\text{Mean: } \frac{pr}{1-p}\qquad \text{Standard deviation: }\frac{\sqrt{pr}}{1-p} . \end{equation}
 For the case of the SYK model,~\cite{qi2019quantum} show that
\begin{equation} K_{l=\delta_\beta(1+qm)}^\beta = \NB \left(m; 2/q, \frac{\left[\frac{J}{\alpha}\sinh(\alpha t)\right]^2}{1+\left[\frac{J}{\alpha}\sinh(\alpha t)\right]^2}\right),\end{equation}
with $\alpha$ being the first positive root of,
\begin{align}\label{eq:alpha_gamma_sol_0}
\alpha= J \sin (\gamma), \quad \text{and} \, \, \sin(\alpha \beta/2 +2\gamma) = \sin(\alpha \beta/2),
\end{align}
and \begin{equation}\delta_\beta =\left(\frac{\alpha}{J}\right)^{2/q}.\end{equation} For very large $\beta$ (small temperature) these can be expanded as 
\begin{align}\label{eq:low_temp_algam}
    \alpha = \frac{\pi}{\beta}-\frac{2\pi}{\beta^2J}+O(1/\beta^3),\quad \gamma = \frac{\pi}{\beta J} - \frac{2\pi}{\beta^2J^2}+O(1/\beta^3),
\end{align}
and the average of the growth distribution $K_l^\beta$ is
\begin{align}\label{eq:avg_size}
\Delta n =  \delta_\beta\left(1+q\frac{2/q}{1-\frac{\sinh(\alpha t)^2}{1+\sinh(\alpha t)^2}}\right)= \delta_\beta \left(1+2\left[\frac{J}{\alpha}\sinh(\alpha t)\right]^2\right),
\end{align}
with the standard deviation $\sigma( n) \approx  \sqrt{\frac{q}{2}} \Delta n$.
\par
In the SYK model, the width of the size distribution of the thermal state is small, of order $\sqrt n$. We are interested in studying the system at times slightly before the scrambling time, when the formalism of~\cite{qi2019quantum} is valid and the width of the size distribution of $\Psi(t)\rho_\beta^{1/2}$ is a small fraction of $n$. In this regime, we have $1\ll \sinh(\alpha t) \approx e^{\alpha t}$, and we get,
\begin{align}
\label{eq:limit_growth}
     K_{l=\delta_\beta(1+qm)}^\beta &= \NB \left(m; 2/q, \frac{1}{1+4\frac{\alpha^2}{J^2} e^{-2\alpha t}}\right) \\
     &\approx \left(4\frac{\alpha^2}{J^2}e^{-2\alpha t}\right)^{2/q}\times \binom{m+2/q-1}{m} \left[1-4\frac{\alpha^2}{J^2} e^{-2\alpha t}\right]^{m}.
\end{align}
\par
In this limit, it is clear that $K_l^\beta$ dictates the size distribution $P_m$, as the thermal state acts as a delta function in the convolution formula~\cref{eq:conv_form}.
\par
Now, we would like to find the growth distribution for the winding size distribution $q_l$, which we call~\emph{winding growth distribution} $\tilde q_l^\beta$. We can see that similar to $K_l^\beta$, $\tilde q_l^\beta$ dictates the $q_l$ distribution at the time regime of interest (slightly before the scrambling rime). To compute the winding size distribution, one can analytically continue Eq. 4.9 of~\cite{qi2019quantum} and set $\tau_1 = \beta/4 - i t, \tau_2=-\beta/4-i t$ to derive a formula for $\tilde q_l^\beta$, which with minor simplifications lead to:
\begin{align}\label{eq:streiqier}
    \sum_{l=0}^m \tilde q_l^\beta e^{-\mu  l} =
    \frac{e^{-\mu \delta_\beta }G}{\left[1-(1-e^{-\mu \delta_\beta q })\frac{ \sin(it\alpha_\mu)\sin(\alpha_\mu\beta/2+it\alpha_\mu)}{\sin(\,\,\gamma_\mu\,\,\,)\sin(\alpha_\mu\beta/2+\gamma_\mu\,\,\,\,\,)}\right]^{2/q}}.
\end{align}
Here we have defined $G:=\left[\sin(\gamma_\mu)/\sin(\alpha_\mu \beta/2+\gamma_\mu)\right]^{2/q}$; $\alpha_\mu$ and $\gamma_\mu$ are functions of $\mu$ through the analogue of~\cref{eq:alpha_gamma_sol_0}
\begin{align}\label{eq:alpha_gamma_sol}
\alpha_\mu= J \sin (\gamma_\mu), \quad \text{and} \, \, \sin(\alpha_\mu \beta/2 +2\gamma_\mu) = e^{-\mu \delta_\beta q}\sin(\alpha_\mu\beta/2);
\end{align}
and $\alpha_\mu$ is the smallest positive root of the above equation.
\par
One can simply expand~\cref{eq:streiqier} in terms of binomial coefficients to obtain the following relation for the winding growth distribution:
\begin{align}\label{eq:streiqierExpand}
    \sum_{l=0}^m \tilde q_l^\beta e^{-\mu  l} =
    \frac{G}{[1-h_\mu(t)]^{2/q}}
    \sum_{m=0}^{\infty}\binom{-2/q}{m}\left(\frac{h_\mu(t)}{1-h_\mu(t)}\right)^m e^{-\mu \delta_\beta(1+ qm)},
\end{align}
where
\begin{equation}h_\mu(t) :=\frac{ \sin(it\alpha_\mu)\sin(\alpha_\mu\beta/2+it\alpha_\mu)}{\sin(\gamma_\mu)\sin(\alpha_\mu\beta/2+\gamma_\mu)} \ . \end{equation}
We are interested at times where $\tilde q_l^\beta$ is a very wide distribution, with width of $O(n)$. To read such low frequency structure from $\sum_{l=0}^m \tilde q_l^\beta e^{-\mu  l} $, we have to look at $\mu$ of $O(1/n)$. In that regime, $G=\delta_\beta$, and  $\alpha_\mu$ and $\gamma_\mu$ are frozen to be very close their value for $\mu=0$ and given by~\cref{eq:alpha_gamma_sol_0}. With this approximation, we establish that:
\begin{multline}
\tilde q_{l=\delta_\beta(1+qm)}^\beta = \frac{\delta_\beta}{[1-h_0(t)]^{2/q}}\times \binom{m+2/q-1}{m} \left(\frac{h_0(t)}{h_0(t)-1}\right)^{m} =\\
\delta_\beta \NB \left(m; 2/q,\frac{h_0(t)}{h_0(t)-1} \right) .
\end{multline}
From this equation, it is clear that there should exist a linear phase in $\hat q(l)$: it is the phase of $\frac{h_0(t)}{h_0(t)-1}$ that accumulates as $l$ increases. We can compute $h_0(t)$ for very small temperatures using expansion in~\cref{eq:low_temp_algam}:
\begin{equation}h_0(t) = \frac{ \sin(it\alpha_\mu)\sin(\alpha_\mu\beta/2+it\alpha_\mu)}{\sin(\,\,\gamma_\mu\,\,\,)\sin(\alpha_\mu\beta/2+\gamma_\mu\,\,\,\,\,)}\approx i \sinh (\alpha t)\left(\frac{\beta J}{\pi} \cosh (\alpha t) + i \sinh (\alpha t)  \right)\end{equation}
We already concluded that in the time regime of interest, $\sinh (\alpha t)\approx \cosh(\alpha t)\approx e^{\alpha t}/2$, and therefore,
\begin{equation}h_0(t) \approx \frac{e^{2\alpha t}}{4} \left(i\frac J\alpha -1 \right)\end{equation}
Now, we can compute the quantity $\frac{h_0(t)}{h_0(t)-1}$:
\begin{align}\label{eq:limited_size_winding}
\frac{1}{1-1/h_0(t)} \approx e^{ 1/h_0(t)} =\left[ 1-4\frac{\alpha^2}{J^2} e^{-2\alpha t} \right]\exp{\left(-i4\frac{\alpha}{J}e^{-2\alpha t}\right)}.
\end{align}
Let us compare the absolute value of the winding growth distribution $\left|\tilde q_l^\beta\right|$ with the ordinary growth distribution $K_l^\beta$. At the times of interest,
\begin{align}
\frac{\delta_\beta}{|1-h_0(t)|^{2/q}} &\approx \frac{\delta_\beta}{\left(\frac{J}{\alpha}\frac{e^{2\alpha t}}{4} \right)^{2/q}}=\left(4\frac{\alpha^2}{J^2}e^{-2\alpha t}\right)^{2/q},\quad \text{and,}\\
\left|\frac{h_0(t)}{1-h_0(t)}\right| &\approx \left[ 1-4\left(\frac{\alpha}{J}\right)^2 e^{-2\alpha t} \right].
\end{align}
Comparing this and~\cref{eq:limit_growth} one can verify that the distributions of $\left|\tilde q_l^\beta\right|$ and $K_l^\beta$ are identical. This indicates that the winding size distribution is indeed the ordinary size distribution with a linearly growing phase.
\par
The wavelength of the size winding can be read from~\cref{eq:limited_size_winding}. The size distribution winds by $4\frac{\alpha}{J}e^{-2\alpha t}$ for each $\delta_\beta q$ increase in size. This gives the following wavelength for the size winding $\lambda_S$:
\begin{equation}\lambda_s =\delta_\beta \frac{\pi}{2} \frac{J}{\alpha}q e^{2\alpha t}\approx \pi q \frac{\alpha}{J}\Delta n \approx \frac{\pi^2 q}{\beta J} \Delta n  \Longrightarrow \Delta n/\lambda_s \approx \frac{\beta J}{\pi^2 q}\end{equation}
Hence, the winding size distribution has $\beta J/(\pi^2 q)$ windings per standard deviation of the size distribution. This can get large as one decreases the temperature. We point out that the initial phase of the size distribution, i.e., the phase of $\tilde q_{(l=0)}^\beta$ is equal to $-\frac{\pi}{q}$.
\par
We note that size winding without significant damping of the of the absolute value implies a strong form of winding, which we call \emph{detailed} winding. For simplicity, let us go back to the qubit picture and assume size winding without damping. Because $q_l = e^{i\theta} K_l$, we have that
\begin{equation}\sum_{|P|=l} c_P^2 =e^{i\theta} \sum_{|P|=l} |c_P|^2. \end{equation}
This shows that for all $P$ with $|P|=l$, $c_P = \pm e^{i\theta/2}|c_P|$. One can reasonably argue that all $c_P$'s have the same sign, say the plus sign, and therefore we have $c_P = e^{i\theta/2}|c_P|$. This shows that not only the size distribution winds in  a linear way, but all of the individual Pauli (or fermion) coefficients collectively wind by phase proportional to their size. This is a quite strong condition, which we believe is a strong sign of having a holographic dual. Indeed, if one considers the low temperature GUE ensemble, then the size winding happens, but as the size winds the absolute value of winding size damps, an indication that detailed winding is missing in that model.

\section{Stringy Effects in the Large-\texorpdfstring{$q$}{} SYK Model}\label{app: SYKGravity}

Here we analyze the traversable wormhole at finite $\beta J$, infinite $q^2 \ll N$ (and finite, instantaneous coupling between the two sides). In this section we use $N$ as the total number of Majorana fermions in order to be consistent with the quantum gravity literature. We will also first discuss the case where $g/N$ is small. This section starts in parallel  with~\cref{app:SYK}, but the approximations and techniques will diverge. In particular, we will compare the SYK results with expectations from string theory. For the convenience of the reader, we restate Qi-Streicher calculations.
  We take their twisted two-point function and analytically continue 
 
  \eqn{\tau_1 = \beta/4 + i t_r,\\ 
  \tau_2 = -\beta/4 - i t_l,\\
  \mu = -i g/N.}
  For simplicity, let us set $t_l = -t_r =t$.  
Using their twisted two-point function, we get
  \eqn{\mathcal{G} = {e^{ig/N } G(\beta/2) \over \lb 1- z \sinh(\alpha t) \sinh(\alpha t -  i \alpha \beta/2)  \rb^{2/q}  }, \\
  z= -\lp {1 - e^{ig/N} \over (\alpha/\cj)^2}  \rp G(\beta/2)^{q/2} \approx {i g \cj \over N  \alpha} .\la{streicher}}
The approximation in the last line is that $g/N$ is small. This equation is derived by finding a classical solution of the large-$q$ effective action, which has the form of a Liouville action. The boundary conditions determine $\alpha$ in terms of the parameters $(g/N),\beta \mathcal{J}$. 
 Note that when the real part of $\alpha t \gg 1$, we can replace the factor in the denominator:
  \eqn{
  G \approx e^{ig/N} { G(\beta/2) \over \lb  1- \tilde{z} e^{2\alpha t} \rb^{2/q} }, \quad \tilde{z} = {z \over 4} e^{- i\alpha \beta/2} \approx { g \cj \over 4 N \alpha} e^{i \pi/2 (1-v)}.}
  For small $g$, $\tilde{z}$ is a complex number that is proportional to $g$. The phase of $\tilde{z}$ is important. Note that the phase is non-zero. 
  Let us write it once more in the small $g/N$ approximation, and using the Maldacena-Stanford notation $\alpha = \pi v/\beta$, where $v$ is the fraction of the maximum Lyapunov exponent $\lambda = 2\pi v/\beta$:
  \eqn{  G \approx \lb  {  \pi v/(\beta \cj) \over   1 -  g {\beta \cj \over 4 N \pi v} e^{i\pi(1-v)/2}  e^{2 \pi v t/\beta}} \rb^{2/q}  \la{stringyG} }
 Here we see that $g$ should be positive. This formula is supposed to be valid in a limit where $g/N \to 0$ but $g e^{2\alpha t}/N $ is held finite. 

  Notice that at very low temperatures, $v=1$ so we expect that the Schwarzian computations with probe particles of ~\cite{maldacena2017diving} to be valid. More explicitly, if we set $t_l = - t_r = t$, then  $G(t) \sim { \lp 1 - \# g  G_N e^t \rp^{-2/q} }$ which is indeed their probe particle formulas.

  One feature of this formula is that there is a pole at some $t$ (roughly the scrambling time). In the probe limit, this is supposed to be smoothed out by stringy effects. Here we see indeed that finite temperature effects can smooth out the pole.
  In particular, when for finite temperature, $0<v <1$ and we see that the non-trivial phase in the prefactor shifts the pole away from the real $t$ axis. Notice also for finite $v$, we get an imaginary part to this correlation function immediately. So the discussion is very similar to Maldacena-Stanford-Yang, where stringy effects give rise to a (small) immediate signal between the two-sides.

One interesting aspect about this formula is that it looks very much like we are acting with some non-unitary deformation of the symmetry generator $P_+$:
  \eqn{C = \ev{\phi_L e^{-i a^+(1-i \epsilon) P_+ } \phi_R} }
  This looks morally like the stringy effects discussed in ~\cite{maldacena2017diving}, but it seems simpler than what we would have naively expected. In particular, it seems like there is no distortion. 

To complete this discussion, we should solve for $\alpha$ or equivalently for $v$. 
The boundary conditions to the Liouville equation give
  \eqn{\alpha =  \cj \sin \gamma, \quad \sin (\alpha \beta/2 + 2 \gamma) = e^{i g/N} \sin \lp \alpha\beta \over 2 \rp
	  \la{bdcond}
}
To leading order in small $g/N$, we can in fact ignore $g$ and just use the solution discussed in Maldacena-Stanford. This solution has the asymptotics
\eqn{\alpha = {\pi v \over \beta}, \quad v = 1 - {2 \over \beta \cj} + \mathcal{O}\lp  {1 \over \beta \cj}  \rp^2, \quad v = {\beta \cj  \over \pi} - {(\beta \cj)^3 \over 8\pi} }
If we go to very high temperatures, $v\to 0$, and the phase becomes purely imaginary. Expanding the denominator in \ref{stringyG}, we get
\eqn{\text{Im } C \approx {2 g \over qN} \sinh^2(\cj t)  }

An important point is that we have no reason to trust these calculations at or beyond the scrambling time, when backreaction is important. To be a little bit more precise, we can consider the probe limit in gravity. Since a fermion adds a finite amount of energy of order $\sim \cj$, we should smear the wavefunctions slightly. The analysis of \cite{maldacena2017diving} shows that the probe limit is only good in the regime
$\Delta \hat{g} G_N e^t \ll \hat{g}^{1/2 + q/4}.$

\subsection{Higher order in \texorpdfstring{$g/N$}{} effects}
We can in principle extend the above discussion to higher order in $g/N$. This is a scaling limit, where $N \to \infty$, $g \to \infty$, with $g/N$ fixed but small. So we will only sum $1/N$ effects which are multiplied by the appropriate powers of $g$.

Let us first work out the perturbative in $g/N$ corrections to $\alpha$. We can go to very low temperatures first where $v=1$. We would now like to solve this for non-trivial $g$. Expanding perturbatively around the $g = 0$ solution,
\eqn{\alpha \approx \gamma, \quad  
\lb 2 - \lp {\alpha \beta \over 2} + 2\gamma - {\pi \over 2} \rp^2 \rb  \approx e^{-\mu} \lb 2 - \lp {\alpha \beta \over 2} - {\pi \over 2} \rp^2  \rb 
}
So the leading order correction to $\alpha$ is
\eqn{\alpha ={\pi \over \beta} -i {g \over \pi N}, \quad G(\beta/2)^{q/2} = \alpha/\cj, }
\eqn{ G = {G(\beta/2) \over \lb 1 - {g \beta \cj \over \pi N} e^{2\alpha t} \rb^{2/q}} .}
 
We have already discussed how finite temperature effects remove the divergence for real values of $t$. Here it seems that finite $g/N$ effects can do the same; at the very least a complex value of $\alpha$ will shift the pole. 
So both $\tilde{z}$ and $\alpha$ will be complex and can smooth out the pole.

One question is what the maximum imaginary value of this correlator is.
The maximum value is interesting because it is one way of quantifying how well the signal gets transmitted. 
This seems to be sensitive to the imaginary contribution to $\alpha$. Note that if we stay within this approximation, this question is somewhat  artificial because backreaction could presumably change the answer a lot.

For a more general value of $\beta \cj$ and $g$, we could numerically solve for $\alpha$ using \nref{bdcond}. In general, it seems like there will be many solutions to \nref{bdcond}. We should find the solution with minimum action. For small $g/N$, we can probably ignore this by first finding the solution at $g=0$ (the usual chaos exponent), and then using this as a starting point for a numerical search.

\subsection{Comparison with gravity}

Here we will review the stringy traversable wormhole calculation of Maldacena-Stanford-Yang\footnote{We thank Douglas Stanford for discussions and useful suggestions.} \cite{maldacena2017diving}. We start by considering the gravitational scattering of a particle with wavefunction $\psi$ against $n$ particles, each in state $\chi$,

\be
\langle \phi_L (O_L O_R)^n  \phi_R \rangle =\int d p_+ \psi^*_L(p_+) \psi_R(p_+)  \left[    \int d q_- e^{ i S } 
	\chi_L^*(q_-) \chi_R(q_-)
	\right]^n\label{havesummed}.
\ee
Here we have in mind a picture where the $\chi$ particles start out on the left side and propagate to the right, and the $\psi$ particles start on the right and propagate to the left side.
Then summing over $n$ gives us the correlator
\eqn{\tilde{C} \equiv \ev{\phi_L e^{igV} \phi_R}= \int dp \, \psi^*_L(p) \psi_R(p) \exp\lb i g \int dq_- \, e^{iS} \chi^*_R(q_-) \chi_L(q_-)\rb \ . }
In the probe limit, we assume that the shockwave action $S$ is small so that the amplitude is approximately $e^{iS} \approx 1 + iS$. 

If we are interested in a very low temperature setup, we can ignore all stringy effects. Then the gravitational shockwave amplitude is determined by the on-shell action $S =  G_N e^t p_+ q_-$. 
Furthermore, the momentum wavefunctions are determined by conformal symmetry:
\eqn{\psi^*_L(p_+) \psi_R(p_+) & = { 1 \over \Gamma( 2 \Delta) } {1 \over (- p_+) } ( 2i p_+)^{ 2 \Delta } e^{-4 i  p_+}  \Theta(-p_+),}
and similarly for the momentum wavefunctions $q_-$.
This allows us to easily compute the average momentum of the shockwave, which is just a number that depends on $\Delta$.

This allows us to simplify the exponent
\eqn{i g V \approx  { i g \over 2^{ 2 \Delta+1} } \lp p_+ G_N e^t  \Delta + 2\rp .}
We have arrived at the {\bf probe particle formula:}
\eqn{ \la{ResFi} 
C_{\rm probe}  & =  e^{ - i { g  \over 2^{ 2 \Delta }}  } \tilde C = 
\int dp_+ \psi_L(p_+)^* \psi_R(p_+) e^{ - i a^+ p_+ }
  \\
 &  =  
   \langle \phi_L  e^{ - i a^+ \hat P_+} \phi_R \rangle =  { 1 \over ( 2 + {a^+ \over 2 } )^{ 2 \Delta } } 
    ~,~~~~~~~~~ a^+ = -   \Delta   { g \over 2^{ 2 \Delta+1 } } G_N e^t.
}

Now in string theory, the shockwave amplitude gets modified 
\eqn{iS =  -G_{N} \lp - i e^t p_+ q_- \rp^{v}= i G_N e^{i \pi (1-v)/2} \lp p_+ q_- e^{t}  \rp^v .} Here $0< v < 1$ and is related to the stringy chaos exponent. In the probe limit,
\bea \la{StrCase2} 
 C^{\rm stringy}_{\rm probe}
 &=&  \int dp_+   \psi^*(p_+) \psi(p_+)  \exp\left[ - i  g  G_N   
 \int  d q_-    ( - i p_+ q_- e^t)^v \psi^*(q_-) \psi(q_-)
\right]
\eea
Now note that when stringy effects are important, e.g., in the SYK model at finite temperature, the wavefunctions can also change since at finite temperature they are no longer determined by conformal symmetry. In general, we do not know how to determine $\psi_v(p_+)$, but in the large-q limit we will have a specific proposal:
\eqn{ \text{d}p_+ \psi^*_{L,v}(p_+) \psi_{R,v}(p_+) = \text{d}p_+^v \psi^*_{L,1}(p_+^v) \psi_{R,1}(p_+^v).}
\def\fq{\mathfrak{q}_-}
\def\fp{\mathfrak{p}_+}
So defining $\fp= p_+^v$, we have
\eqn{ 
C^{\rm stringy}_{\rm probe}
 &=  \int d\fp  \psi^*(\fp) \psi(\fp)  \exp\left[   
 - i  g  G_N\int  d \fq    \,    \fp \fq e^{v(t-i\pi/2) }  \psi^*(\fq) \psi(\fq) \right]
}
This gives the simple answer:
\eqn{C = { \lb 2 + {a^+ e^{i \pi (1-v)/2} \over 2} \rb }^{-2\Delta}, \quad a^+ \propto -g G_N e^{vt}  }
which agrees precisely with the large $q$ calculation.

Note that we derived the analog of the stringy probe formula. However, once we determine the S-matrix and the wavefunctions, we can use the string ansatz to go beyond the probe approximation. In other words, the ansatz we have written down gives some predictions for $1/N$ corrections to the 4-pt function. It would be nice to check these against SYK results.

Some of the formulas in this section were also derived in \cite{gao2019traversable}. The main novelty here is that a simple change in the wavefunctions allows for a quantitative match with stringy formulas.

\section{Entanglement Entropy in Nearly-\texorpdfstring{$\operatorname{AdS}_2$}{AdS\_2} Gravity \la{eeads}}
Here we compute some entanglement entropies in the global vacuum state of AdS$_2$.
Global AdS$_2$ is topologically a strip. In general, we should specify boundary conditions for the strip. We will consider standard ``reflecting'' boundary conditions. We can use a ``doubling'' trick to relate this state to the state on a topological cylinder. (This is similar to a trick familiar from studying the world-sheet of string theory which relates the open string to the closed string). We can imagine joining that the left moving modes live on one side of the strip and the right moving modes move on the other side. Then at each boundary, we join the left-movers to the right-movers.
If the quantum fields are conformal, we have a simple expression for the entanglement entropy of a single interval $[\theta_1, \theta_2]$ on a cylinder
\eqn{S \sim {c \over 6} \log \lb {\sin^2 ( \theta_1 - \theta_2)/2 \over \epsilon_1 \Omega_1  \epsilon_2 \Omega_2}\rb \ . }
Here $\Omega$ is the warp factor, and $\epsilon$ is some small UV regulator, which we will ignore (in the generalized entropy, it contributes to the renormalization of $S_0$.) Now for AdS$_2$ in global coordinates, $\Omega = \sin \sigma$. Notice that if we compute the entropy of an interval which includes the asymptotic left side of AdS$_2$ (at $\sigma = 0$) and ends at some other value $\sigma$, we should set $\theta_1 =- \theta_2 = \sigma$. Then we find that the entropy of that interval is independent of the endpoint $\sigma$, as required by the AdS$_2$ isometries. However, the entropy of a single interval with endpoints $\sigma_1$ and $\sigma_2$ in AdS$_2$ is a 2-interval computation on the cylinder. For a general 2D CFT, the entropy of 2 intervals is not known explicitly. However, if we take the OPE limit $\sigma_1 \to 0$, we get (up to a divergent constant) that the entropy is just the sum of the 2 intervals. Hence, 
\eqn{S_m \sim {c \over 6} \lp  \log {\sin^2 \lp 2\sigma_2/2 \rp \over \sin^2 \sigma_2 } + \log {\sin^2 \lp 2\sigma_1/2 \rp \over \sin^2 \sigma_1 } \rp  = 0.}
So in this limit, we get an answer that is independent of $\sigma_2$.

The leading correction is non-universal; it depends on the CFT. If we consider $c$ free Dirac fermions \cite{Casini:2005rm}, the entanglement entropy of the two-interval region $
\left[x_{1}, x_{2}\right] \cup\left[x_{3}, x_{4}\right]
$ with metric $d s^{2}=\Omega^{-2} d x d \bar{x},$ is
\eqn{
S_{\text {fermions }}
&=\frac{c}{6} \log \left[\frac{\left|x_{21} x_{32} x_{43} x_{41}\right|^{2}}{\left|x_{31} x_{42}\right|^{2} \Omega_{1} \Omega_{2} \Omega_{3} \Omega_{4}}\right]  \\
&\sim 
{c\over 6} \log \lb {\sin^2 \left(\frac{1}{2} \left(\theta _1-\theta _2\right)\right) \sin^2 \left(\frac{1}{2} \left(\theta _2-\theta _3\right)\right) \sin^2 \left(\frac{1}{2} \left(\theta _1-\theta _4\right)\right) \sin^2 \left(\frac{1}{2} \left(\theta _3-\theta _4\right)\right)  \over \sin^2 \left(\frac{1}{2} \left(\theta _2-\theta _4\right)\right) \sin^2 \left(\frac{1}{2} \left(\theta _1-\theta _3\right)\right) \Omega_1 \Omega_2 \Omega_3 \Omega_4} \rb\\
&\sim 
{c\over 6} \log \lb {\sin^2 \hf \lp \sigma_1 - \sigma_2 \rp \over \sin^2 \hf \lp  \sigma_1 + \sigma_2 \rp } \rb
}
Now expanding around $\sigma_1 = 0$, we get
\eqn{S_\text{fermions} \sim -{c \sigma_1 \over 3 \tan (\sigma_2/2) }. }
 If we further expand $\sigma_2$ around the bifurcate horizon $\sigma_2 = \pi/2$, we get $S_\text{fermions} \sim c \sigma_1 (\sigma_2- \pi/2)/3 $.
 As expected, the entropy will decrease if we move $\sigma_2$ closer to $\sigma_1$. A wide variety of bulk setups can be analyzed using these approximations which we intend to report elsewhere.

\end{widetext}
\end{document}